\documentclass[aps,twocolumn,prd]{revtex4}

\usepackage{amsmath}
\usepackage[dvips]{graphicx}
\usepackage{amssymb}
\usepackage{verbatim}
\usepackage{mathrsfs}
\usepackage{color,psfrag}

\usepackage{graphicx}
\usepackage{latexsym}
\usepackage{amsmath}
\usepackage{amssymb}
\usepackage{mathrsfs}
\usepackage{graphicx,psfrag,epsfig}
\usepackage[vcentermath]{youngtab}
\usepackage{verbatim}
\usepackage{amsfonts}
\usepackage{verbatim}
\usepackage{dsfont}

\def\KKLMMT{\mathbb{K}{\rm L}\mathbb{M}{\rm T}}

\newcommand\ba{\begin{array}}
\newcommand\ea{\end{array}}
\newcommand\ben{\begin{equation}}
\newcommand\een{\end{equation}}
\newcommand\bea{\begin{eqnarray}}
\newcommand\eea{\end{eqnarray}}
\newcommand{\al}{\alpha}

\newcommand\fverb{\setbox\fverbbox=\hbox\bgroup\verb}
\newcommand\fverbdo{\egroup\medskip\noindent%
			\fbox{\unhbox\fverbbox}\ }
\newcommand\fverbit{\egroup\item[\fbox{\unhbox\fverbbox}]}
\newbox\fverbbox

\newcommand{\llangle}{\ensuremath{\big{\langle}\!\big{\langle}}}
\newcommand{\rrangle}{\ensuremath{\big{\rangle}\!\big{\rangle}}}

\newcommand{\dslash}{d \hspace{-0.8ex}\rule[1.2ex]{0.9ex}{.1ex}}

\newcommand{\ssumint}{{\small\Sigma} \hspace{-2ex}\int}

\newcommand{\suminnt}{\sum \hspace{-3ex}\int}
\newcommand{\suminntt}{{\scriptstyle\sum} \hspace{-1.5ex}{\textstyle\int}}
\newcommand{\paral}{\textrm{\tiny $\parallel$}}

\begin{document}

\title{String Vertex Operators and Cosmic Strings}
\date{June 23, 2011}
\author{Dimitri Skliros}
\email{dimitri.skliros@nottingham.ac.uk}
\affiliation{Department of Physics and Astronomy, University of Sussex, Brighton, East Sussex BN1 9QH, UK} 
\affiliation{Cripps Centre for Particle Theory, University of Nottingham,  University Park, Nottingham, NG7 2RD, UK} 
\author{Mark Hindmarsh}
\email{m.b.hindmarsh@sussex.ac.uk}
\affiliation{Department of Physics and Astronomy, University of Sussex, Brighton, East Sussex BN1 9QH, UK} 

\begin{abstract} 
We construct complete sets of (open and closed string) covariant coherent state and mass eigenstate vertex operators in bosonic string theory. 
This construction can be used to study the evolution of fundamental cosmic strings as predicted by string theory, and is expected to serve as a self-contained prototype toy model on which realistic cosmic superstring vertex operators can be based on. It is also expected to be useful for other applications where massive string vertex operators are of interest. We pay particular attention to all the normalization constants, so that  these vertices lead directly to unitary $S$-matrix elements.
\end{abstract}

\maketitle
\tableofcontents
\section{Introduction}
\parindent = 2em
The study of cosmic strings, see e.g.~\cite{Hindmarsh11,CopelandPogosianVachaspati11} and references therein, has flourished in recent years, following the discovery \cite{MajumdarDavis02} (but see also \cite{BarnabyBerndsenCLineStoica05}) \cite{JonesStoicaTye02,SarangiTye02,Halyo03}, that such objects may be produced in string models of the early universe, thus providing an observational signature for Superstring theory \cite{CopelandMyersPolchinski04,DvaliVilenkin04,BeckerBeckerKrause06}. 

The possibility that superstrings of cosmological size may have been produced in the early universe was first contemplated by Witten \cite{Witten85} who (based on current knowledge of the time) concluded that had they been produced they would either not be observable (they would be produced before inflation and diluted away by the cosmological expansion), they would be unstable (they would disintegrate into smaller strings long before reaching cosmological scales in the case of Type I strings, or in the case of Heterotic String theory they would arise as boundaries of domain walls whose tension would cause the strings to collapse), and in any case they would nevertheless be excluded by experimental constraints, requiring string tensions, $G\mu\sim 10^{-3}$ (with $G$ the four-dimensional Newton's constant and $\mu=1/(2\pi\alpha')$ the fundamental string tension), while it was clear that strings with $G\mu>10^{-5}$ had already been ruled out. 
Much has changed since then, the discovery of dualities \cite{Polchinski_v2} and D-branes \cite{Polchinski95,Polchinski96} having completely revolutionized our understanding of string theory.

These new developments opened many new avenues for model building \cite{GiddingsKachruPolchinski02} and string cosmology, such as the brane inflation scenario \cite{DvaliTye99,Burgess01,Alexander01,DvaliShafiSolganik01,ShiuTye01,JonesStoicaTye02} in the context of large extra dimensions \cite{Arkani-HamedDimopoulosDvali98,AntoniadisArkani-HamedDimopoulosDvali98,Arkani-HamedDimopoulosDvali99}, where macroscopic strings have been found to be produced \cite{MajumdarDavis02,SarangiTye02,Jones03,DvaliVilenkin04,BarnabyBerndsenCLineStoica05} with string tensions in the range \cite{SarangiTye02}, $10^{-12}\leq G\mu\leq 10^{-6}.$ In these scenarios it is difficult to obtain a sufficient amount of inflation \cite{Quevedo02,KKLMMT} and in \cite{KKLMMT} this problem is evaded by considering instead a warped compactification \cite{RandallSundrum99a,RandallSundrum99b}, a concrete example of which is the well known $\mathbb{K}$L$\mathbb{M}$T scenario \cite{KKLT,KKLMMT}. 
It has since been realized \cite{CopelandMyersPolchinski04} that it is possible in these theories to construct macroscopic non-BPS as well as BPS strings which are stable \cite{Sen99} and potentially observable.

Unfortunately, no completely satisfactory string model of the early universe exists yet: although all moduli are stabilized in the brane inflation scenario \cite{KKLT,KKLMMT} it suffers from a reheating problem where all the reheating energy arising from the D3/$\overline{\rm D}3$-annihilation goes into a massless U(1) gauge field that lives on the stabilizing $\overline{\rm D}3$-brane instead of going into the standard model fields, for a brief elaboration on this see \cite{CopelandKibble09}. Furthermore, in the context of large extra dimensions there is no known mechanism to stabilize the moduli. Nevertheless, these drawbacks may be specific to the models considered to date and it is plausible that in more general constructions these problematic features are absent.

For an overview of cosmic strings in the pre- and post-``second superstring revolution'' era see \cite{Hindmarsh11} (as well as the older more extensive reviews \cite{HindmarshKibble95,VilenkinShellard94}) and \cite{Polchinski04,DavisKibble05,Sakellariadou07,Sakellariadou09,MyersWyman09,CopelandKibble10,Ringeval10,CopelandPogosianVachaspati11} respectively, and for an excellent review which also contains many introductory remarks and computational details associated to the latter see \cite{Majumdar05}.

\subsection{Brane Inflation}
It is possibly rather natural to suspect there to have been a multitude, or a gas, of D-branes of various dimensionalities in the early universe. The branes of higher dimensionality will annihilate first and produce lower dimensional branes and branes that are present today. As an example, in the most concrete (almost viable) scenario, namely the $\KKLMMT$ scenario \cite{KKLMMT}, one studies the relative motion of a remaining D3-brane and $\overline{D}3$-brane, which are separated in the transverse space in a throat of a Calabi-Yau (CY) three-fold. There is a U(1) gauge symmetry on each of these branes. 
Cosmological inflation is driven by the attractive interaction potential associated to the D3- and $\overline{D}3$-brane which approach each other in the higher dimensional bulk space. 

The two branes eventually collide and annihilate via tachyon condensation, see e.g.~\cite{Sen05}.  Due to the Kibble mechanism \cite{Kibble76}, when a U(1) gauge symmetry  becomes broken during the evolution of the universe, defects (and in particular cosmic strings) will be produced. The crucial observation of \cite{SarangiTye02} was that the low energy string dynamics at the end of brane inflation is described by U(1) symmetry breaking in the tachyon field, and therefore one expects the formation of defects which are seen as cosmic strings by observers on the (or one of the) remaining higher dimensional branes. It has been argued that the production of other defects such as monopoles and domain walls is suppressed \cite{JonesStoicaTye02}. These defects are identified \cite{Sen98b,Witten98} with $D1$-branes, which follows from computing the conserved charges. Nevertheless, both D-strings and F-strings are expected to arise \cite{DvaliVilenkin04,CopelandMyersPolchinski04} in this process, even though the standard language of string creation associated to a spontaneous breaking of a U(1) symmetry is not appropriate for F-strings (unless $g_s\gg1$). The standard model particles of strong and weak interactions correspond to open string modes confined to a remaining D-brane with 3 large non-compact dimensions, and the closed string modes (e.g.~the graviton, radions and massive excitations) correspond to bulk modes.

The presence of cosmic strings is likely to be a fairly generic feature of any string model of the early universe and in the present article we shall assume that such a model can be found and focus instead on the cosmic strings themselves. We shall focus in particular on the fundamental cosmic strings which have an exact perturbative (in the string coupling $g_s=e^{\langle\Phi\rangle}$ and the fundamental string length squared $\alpha'$) description in terms of vertex operators.

\subsection{Cosmic String Evolution}

The basic properties which collectively determine the evolution are string inter-commutations and reconnections \cite{Shellard87,Polchinski88b,JacksonJonesPolchinski05,AchucarroPutter06,AchucarroVerbiest10}, quantum or classical string decay \cite{DaiPolchinski89,MitchellTurokWilkinsonJetzer89,MitchellSunborgTurok90,PreskillVilenkin93,IengoRusso02,IengoRusso03,ChialvaIengoRusso03,ChialvaIengo04,Iengo06,GutperleKrym06,Chialva09} and the presence of junctions \cite{CopelandKibbleSteer06,CopelandKibbleSteer07,CopelandFirouzjahiKibbleSteer08,DavisRajamanoharanSakellariadou08,FirouzjahiKaroubyKhosraviBrandenberger09,AvgoustidisCopeland10}, and possible instabilities \cite{Witten85,PreskillVilenkin93,CopelandMyersPolchinski04}. Collectively, these properties and cosmological considerations (such as the expansion rate of the universe, density inhomogeneities, and so on) determine the various observational signatures from cosmic strings. 

An initial distribution of long strings is formed via the Kibble mechanism, the shape of any one such string resembling a random walk. The expansion of the universe stretches these strings which intercommute and reconnect producing kinks (i.e.~points on the string at which the spacetime embedding tangent vectors associated to left and right-movers are discontinuous). Any one of these kinks then separates into two kinks running along the string in opposite directions. When left- and right-moving modes meet on any given section of a string gravitational radiation is produced. There will also be long strings that self-intercommute and produce loops which subsequently are expected to decay into smaller loops, massive, and massless (including gravitational) radiation.

There is general consensus on the large scale evolution of cosmic strings. Here the string network evolves towards a scaling regime, a regime in which the characteristic length scale of the configuration evolves towards a constant relative to the horizon size \cite{HindmarshKibble95,VilenkinShellard94}. Recently, there has also been some progress in understanding the small scale structure \cite{PolchinskiRocha06,HindmarshStuckeyBevis09,CopelandKibble09,Kawasaki:2010iy,LorenzRingevalSakellariadou10}, see also \cite{SiemensOlumVilenkin02}. Here one of the most important questions is: what is the typical size at which loops are produced from long string. There has been large disagreement in the literature with estimates differing by over fifty orders of magnitude. This is an important question and further investigation is required. Another very important question which is also related to the previous one is: what is the importance of gravitational backreaction on the evolution of cosmic strings, see also below.

\subsection{Gravitational Radiation and Backreaction}

Cusps are generic in loops \cite{KibbleTurok82} and are expected to lead to very strong gravitational wave signals \cite{DamourVilenkin00,DamourVilenkin01}, although the presence of extra dimensions is likely to weaken the detected signal \cite{O'CallaghanChadburnGeshnizjaniGregoryZavala10}. Cusps on strings with junctions have also been argued to be generic in \cite{DavisRajamanoharanSakellariadou08}. Recent evidence \cite{BinetruyBoheHertogSteer10} suggests that kinks on strings with junctions also provide a very strong gravitational wave signal -- the signal from kinks on string loops \emph{with} junctions is found to be stronger than the signal due to cusps. It is very important to test the robustness of all these computations to gravitational backreaction effects. In fact, it is likely that gravitational backreaction can be important for even order of magnitude estimates \cite{QuashnockSpergel90}, and developing the necessary tools that enable one to study this problem systematically has been one of the main purposes of the present article.

Furthermore, and most importantly, it has been suggested \cite{BennettBouchet88,QuashnockSpergel90,Hindmarsh90} that gravitational backreaction sets the scale for the smallest relevant structures in cosmic string evolution, as well as the long sought-after loop production scale. It is therefore of vital importance to understand gravitational backreaction and develop the necessary tools where such questions can be addressed most naturally  -- in perturbative string theory backreaction effects can be taken into account very naturally, as first pointed out in \cite{WilkinsonTurokMitchell90}.

\subsection{Massive Radiation}

Apart from the possibility of gravitational backreaction playing a significant role in string evolution, a string theory computation is also required when there is the possibility of massive closed string states being emitted -- this might be expected to occur close to cusps and kinks and this massive radiation would presumably be invisible or difficult to calculate in the effective field theory \footnote{DS would like to acknowledge an important discussion with Andrew Strominger concerning the relevance of a vertex operator formulation of cosmic strings as opposed to an effective low energy description.}. That massive radiation may dominate over gravitational radiation was suggested in \cite{VincentHindmarshSakellariadou97,VincentAntunesHindmarsh98}, and this was motivated by the observation that loops seemed to be produced at the smallest scales, see also \cite{BennettBouchet89,AllenShellard90}, namely at the numerical simulation cutoff scale which is identified with the string width, although their conclusions relied on extrapolation of numerical results beyond the region of validity. There have also been some interesting results on massive radiation in a quantum-mechanical setup in the case of mass eigenstate vertex operators, see \cite{IengoRusso06} and references therein. Whether a significant amount of massive radiation is emitted is certainly still an open question -- this can be addressed in the vertex operator construction of the current article which is expected to provide a definite answer to this question. If one is interested in the emission of arbitrarily massive radiation one may proceed along the lines of \cite{MitchellTurokWilkinsonJetzer89,MitchellSunborgTurok90,IengoRusso02,IengoRusso03,ChialvaIengoRusso03,ChialvaIengo04,Iengo06,GutperleKrym06,Chialva09}. 

\subsection{Observational Signatures}

Signals that have been confirmed to arise from cosmic string sources have to date not yet been detected. There is a wide range of constraints from  gravitational waves \cite{CaldwellAllen92,DamourVilenkin00,DamourVilenkin01,Siemens06,Hogan06,SiemensMandicCreighton07,OlmezMandicSiemens10,BattyeMoss10} (classical gravitational wave emission from loops and infinite strings has been computed in \cite{VachaspatiVilenkin85,Burden85} and \cite{Sakellariadou90,Hindmarsh90,SiemensOlum01,Kawasaki:2010iy} respectively and from strings with junctions in \cite{BrandenbergerFirouzjahiKaroubiKhosravi09}), strong and weak lensing from strings without \cite{ChernoffTye07,KuijkenSiemensVachaspati07,DydaBrandenberger07,GaspariniMarshallTreuMorgansonDubath07,MorgansonMarshallTreuSchrabbackBlandford09,ThomasContaldiMagueijo09} (but see also \cite{ShlaerWyman05}) and with \cite{ShlaerWyman05,BrandenbergerFirouzjahiKaroubi08,Suyama08} junctions, and the CMB \cite{FraisseRingevalSpergelBouchet08,PogosianTyeWassermanWyman09,HindmarshRingevalSuyama09,WymanPogosianWasserman05,PogosianWyman08,BattyeGarbrechtMoss06,BevisHindmarshKunzUrrestilla08,BattyeMoss10,Bevis:2010gj}. Future missions searching for a polarization B-mode in the CMB will provide even stronger constraints \cite{PogosianTyeWassermanWyman03,SeljakSlosar06,Urrestilla:2008jv,GarciaBellido:2010if,Kawasaki:2010iy}. Signals from cosmic strings may also show up in ultrahigh-energy cosmic rays \cite{BhattacharjeeSigl00,BerezinskyHnatykVilenkin01}, radio wave bursts \cite{Vachaspati08}, and also diffuse X- and $\gamma$-ray backgrounds \cite{BerezinskyHnatykVilenkin01}. There is also the potential to obtain constraints on the underlying compactifications \cite{BeanChenPeirisXu08}. Even though cosmic strings can only account for a small contribution to the CMB power spectrum, they could instead be a significant source of its non-Gaussianities and are expected to dominate over inflationary perturbations at small angular scales, see \cite{Ringeval10} and references therein.

\subsection{Vertex Operators as Cosmic Strings}

Given the inherently quantum-mechanical nature of fundamental cosmic strings, the only available handle on such macroscopic objects at present that is capable of accounting for the evolution on the smallest as well as largest scales is given in terms of \emph{vertex operators} \cite{Polchinski_v1,Polchinski_v2} which completely characterize the string under consideration. For example, a vertex operator description would be required for cosmic string configurations involving a string theory analogue of cusps (i.e. points on the string that reach the speed of light at discrete instants during the loop's motion) and kinks, as presumably the effective field theory or classical description would break down close to these points.

With a vertex operator construction of cosmic strings one can address various questions, such as what is the decay rate of a given cosmic string configuration, the intercommutation and reconnection probabilities, junction decay rates, emission of massless and massive radiation and so on. The already existing quantum decay rate computations carried out in \cite{MitchellTurokWilkinsonJetzer89,MitchellSunborgTurok90,IengoRusso02,IengoRusso03,ChialvaIengoRusso03,ChialvaIengo04,Iengo06,GutperleKrym06,Chialva09} for instance make use of mass eigenstate vertex operators (with only first harmonics excited) and it is not known at this point whether these are appropriate for the description of cosmic strings. In \cite{Iengo06} for instance it was concluded that the spectrum of a particular mass eigenstate does not reproduce the classical gravitational wave spectrum, and one might expect this to be the case also for general mass eigenstates.

It is likely that cosmic strings being macroscopic and massive should have a classical interpretation. If this is the case, the appropriate vertex operators are expected to have coherent state-like properties  (from our experience with standard harmonic oscillator coherent states), and so we should be searching for coherent state vertex operators, which would be expected to have a classical interpretation. The analogous computations as the ones described above with coherent states instead of mass eigenstates would be more desirable and would probably represent a much more realistic description of cosmic strings.

A quantum-mechanical approach to computing the decay process for macroscopic and realistic cosmic string loops is highly desirable as we must also check the usual assumption that the process is classical. Furthermore, the classical computation is not well understood, as calculations based on field theory and the Nambu-Goto approximation differ, and gravitational back-reaction is not taken into account.  Back-reaction can be included very naturally in perturbative string theory.

Finally let us mention that it is very important to find tests which distinguish fundamental strings from solitonic strings;  a major difference is the quantum nature of F-strings (which for instance leads to a reduced probability for the reconnection of intersecting strings  \cite{JacksonJonesPolchinski05}, see also \cite{HananyHashimoto05} for an alternative approach).  Thus it seems that string theory computations will certainly be required in order to distinguish fundamental from solitonic strings in experiments.

\subsection{Classicality of Cosmic Strings}
Let us now say a few words concerning the classicality of quantum-mechanical string vertex operators. Consider first mass eigenstates. These are specified by certain quantum numbers, the relevant one here being the level number $N$, and a necessary (but not sufficient) condition for classicality is that these take \emph{large} values. This dates back to Niels Bohr who used this argument when he postulated that any quantum-mechanical system should satisfy the correspondence principle.  Typically the quantum numbers of interest in a given quantum system appear in the combination $(N\hbar)$ thus showing that the classical limit $\hbar\rightarrow 0$ is related to the large quantum number limit $N\rightarrow\infty$ with the combination $N\hbar$ held fixed. For example, this can be seen in the energy spectrum of the hydrogen atom, $E_N\sim{\rm const.}/(N\hbar)^2$, the harmonic oscillator, $E_N\sim{\rm const.} (N\hbar)$, and also the string spectrum \footnote{$\hbar$ is usually set equal to 1 but can be re-introduced by examining the path integral $\int\mathcal{D}X e^{\frac{i}{\hbar}S}$, with $S=\frac{1}{2\pi\alpha'}\int d^2z\,\partial X\cdot \bar{\partial}X$. Taking $z$ to be dimensionless, $\ell_s$ the string length and $[X]=L$ we see that $$2\pi \alpha'= \ell_s^2/\hbar.$$}, $E_N\sim{\rm const.} \sqrt{(N\hbar)}$. Vertex operators present in the large quantum number limit may in some sense therefore be referred to as being quasi-classical. The extent to which mass eigenstates can have a classical interpretation is not well understood. In \cite{Blanco-PilladoIglesiasSiegel07} for example, it was shown that mass eigenstates are not truly classical in the sense that they are not expected to have classical expectation values with small uncertainties, and  \cite{Iengo06} one does not expect the spectrum of gravitational radiation to match the classical computation -- whether mass eigenstates can have a classical interpretation or not is a very important issue and deserves further attention.

Coherent states on the other hand can (as we show below) easily be made macroscopic and are expected to possess classical expectation values with small uncertainties, e.g.~$\langle J^{\mu\nu}\rangle = J^{\mu\nu}_{\rm cl}$, $\langle X^{\mu}\rangle = X^{\mu}_{\rm cl}$, (with $J^{\mu\nu}$  the spacetime angular momentum and $X^{\mu}$ the target space map of the worldsheet into spacetime). It is likely that coherent states therefore should be identified with fundamental cosmic strings. There are subtleties however concerning the naive classicality requirement $\langle X^{\mu}\rangle = X^{\mu}_{\rm cl}$ (with $X_{\rm cl}^{\mu}$ non-trivially obeying the classical equations of motion, $\partial\bar{\partial}X_{\rm cl}^{\mu}=0$) and it turns out \cite{Blanco-PilladoIglesiasSiegel07} that this requirement (in the closed string case) is not compatible with the Virasoro constraints (when states are invariant under spacelike worldsheet rigid translations). Suffice it to say here that this is a gauge problem and says nothing about the classicality of the underlying states. We elaborate on this in detail later where we also propose a solution: an alternative to the $\langle X^{\mu}\rangle = X^{\mu}_{\rm cl}$ classicality condition which is compatible with the string symmetries. We will also see that it is possible for closed string (coherent) states to satisfy $\langle X^{\mu}\rangle = X^{\mu}_{\rm cl}$ in lightcone gauge when the underlying spacetime manifold is compactified in a \emph{lightlike} direction, $X^-\sim X^-+2\pi R^-$, with $X^+$ non-compact, because this compactification breaks the invariance under spacelike worldsheet shifts.

\subsection{Vertex Operator Constructions}

Various prescriptions have been given for the construction of covariant vertex operators, e.g.~the construction due to Del Giudice, Di Vecchia and Fubini (DDF) \cite{DelGiudiceDiVecchiaFubini72,AdemolloDelGuidiceDiVecchiaFubini74,Brower72,GoddardThorn72} but see also \cite{D'HokerGiddings87}, the path integral construction based on symmetry \cite{Weinberg85,Sato88,DHokerPhong87} and factorization \cite{AldazabalBoniniIengoNunez87,AldazabelBoniniIengoNunez88,Polchinski88} and operator constructions \cite{ManesVozmediano89,Nergiz94} among others. A powerful method which applies in general backgrounds is given in \cite{CallanGan86}, (although explicit results for high mass states are seemingly rather difficult to obtain in more general backgrounds, see also \cite{Polyakov02,Tseytlin03}).  To carry out the map from classical solutions to covariant vertex operators we shall make use of the DDF construction. The power of the DDF construction lies in the following: it generates the entire physical Fock space, and it can be used to translate light-cone gauge states into the corresponding covariant
vertex operators, where the standard technology for amplitude computations \cite{Polchinski_v1,DHokerPhong} can be used. This is clearly very useful indeed given that in the construction of vertex operators for cosmic strings we would like to know what the corresponding classical state is, but explicit general  classical solutions are best understood in lightcone (not covariant) gauge -- the DDF construction provides the appropriate bridge between classical lightcone gauge string solutions and covariant vertex operators.

Using these tools, in the current article we construct of a complete set of covariant vertex operators, i.e.~vertices for arbitrarily massive (closed and open) strings, for both mass eigenstates and open and closed string coherent states. We also discuss the corresponding lightcone gauge realization and provide an explicit map from these to general classical (lightcone gauge) solutions. We restrict our attention to bosonic string theory and it is likely that all results generalize to the superstring.

\subsection{Outline}

Sec.~\ref{sec:NSW} is mainly a review and is intended to provide the link between vertex operators and observable quantities, by making precise the link between the string path integral and $S$-matrix elements. We discuss in particular the normalization of vertex operators that is appropriate for scattering amplitude computations, in the sense that a string path integral with such vertex operator insertions can be directly interpreted as a \emph{dimensionless} $S$-matrix element. We also discuss tree level $S$-matrix unitarity and present some useful lightcone coordinate results that are used throughout the rest of the article.

In Sec.~\ref{DDF} we discuss the construction of a complete set of normal ordered mass eigenstate vertex operators using the DDF formalism, which can be used to translate light-cone gauge states into fully covariant vertex operators. The Virasoro constraints are solved completely and the resulting covariant vertex operators are physical for arbitrary polarization tensors that correspond to irreducible representations of SO(25). In the process we show that all covariant vertex operators can naturally be written in terms of elementary Schur polynomials.

In Sec.~\ref{SCS} we show that the construction of physical covariant coherent states becomes clear in the DDF formalism. We construct both open and closed coherent states. These fundamental string states are macroscopic and have a classical interpretation, in the sense that expectation values are non-trivially consistent with the classical equations of motion and constraints. We present an explicit map which relates  three classically equivalent descriptions: arbitrary solutions to the equations of motion, the corresponding lightcone gauge coherent states, the corresponding covariant coherent states. We gain further evidence supporting this equivalence by showing that all spacetime components of the angular momenta in all three descriptions are identical. We suggest that these quantum states should be identified with fundamental cosmic strings.

The work considered in this article has immediate applications to cosmic strings but the considerations are more general and may serve to capture a wide range of phenomena where massive string vertex operators are relevant.

A short summary of the closed string coherent state section can be found in the companion article \cite{HindmarshSkliros10}.

\section{String $S$-Matrix, Unitarity and Normalization: A Brief Review}\label{sec:NSW}
Before moving on the discuss the general DDF construction of vertex operators it will be useful to elaborate on the precise connection of vertex operators to the string $S$-matrix, as this will in turn enable us to normalize vertex operators correctly, i.e.~in such a way that the resulting $S$-matrix elements are unitary. We will follow the general approach of \cite{Weinberg85,Polchinski88,Polchinski_v1,DiVecchia:1996uq} although the reasoning here will be largely independent of these references and self-contained. We will concentrate on mass eigenstates, although these results will go through essentially untouched in the case of coherent states (Sec.~\ref{SCS}) as well.

\subsection{$S$-Matrix $=$ Path Integral}
Our objective is to use a normalization for vertex operators that is appropriate for scattering amplitude computations, and so we first discuss the precise relation between the string path integral and the $S$-matrix. 

The proper way of constructing a scattering experiment is to first construct vertex operator wave packets for the external string states of interest and then normalize each one of them to ``one string in the universe", in direct analogy to the corresponding field theory prescription. Rather than use wavepackets, we may also use momentum eigenstates instead, in which case (due to the uncertainty principle, the infinite spacetime spread of momentum eigenstates) we need to truncate the volume of spacetime at, say, $V_{d-1}$, the case of interest for the bosonic string being $d=26$ and for the superstring $d=10$. According to standard practice \cite{Weinberg_v1}, we hence identify momentum delta-functions with volume elements and energy delta functions with the time, $T$, during which the interaction is ``turned on",
\begin{equation}\label{eq:VT}
\begin{aligned}
&(2\pi)^{d-1}\delta^{d-1}({\bf p}'-{\bf p})\equiv V_{d-1},\phantom{\Big|}\\
&(2\pi)\delta(E'-E)\equiv T.
\end{aligned}
\end{equation}
By putting the system in a box of size $V_{d-1}$, the vertex operator normalization condition is changed from  ``one string in the universe" to ``one string in volume $V_{d-1}$" \cite{LandauLifshitzRQF}. Of course, physical observables (cross sections, decay rates, etc\dots) should not depend on $V_{d-1}$, although we formally think of taking $V_{d-1}\rightarrow \infty$ at the end of the computation. 

The ``one string in volume $V_{d-1}$" normalization prescription leads to an $S$-matrix such that if an initial state of a system is denoted by $|i\rangle$,  the final state will be a superposition, $\sum_f|f\rangle\langle f|S|i\rangle$. Therefore, $|S_{fi}|^2$, with 
$
S_{fi}\equiv \langle f|S|i\rangle,
$ 
 is interpreted as a transition probability associated to going from $|i\rangle$ to $|f\rangle$,
\begin{equation}\label{eq:Prob|Sfi|^2}
\begin{aligned}
&{\rm Prob}(f\leftarrow i)=|S_{fi}|^2.
\end{aligned}
\end{equation}
Conservation of probability, equivalently $S$-matrix unitarity, requires that,
$$
S^{\dagger}S=\mathds{1}.
$$
In particular, in terms of $S_{fi}$, unitarity corresponds to the statement:
\begin{equation}\label{eq:SSdagger=1}
\sum_nS^{\dagger}_{nf}S_{ni}=\delta_{fi},\qquad {\rm or}\qquad \sum_nS_{fn}S^{\dagger}_{in}=\delta_{fi},
\end{equation}
with $\delta_{fi}$ a Kronecker delta; working in the Heisenberg picture, $\delta_{fi}\equiv \langle f|i\rangle$. Setting $f=i$ it is seen that unitarity enforces conservation of probability, 
$
\sum_{f}|S_{fi}|^2=1.
$

To make the connection with the string path integral, it is conventional and convenient to define a $T$-matrix which contains the non-trivial contribution to the $S$-matrix, $S=\mathds{1}+iT$. Taking matrix elements of both sides and extracting the momentum and energy conserving delta functions leads to,
\begin{equation}\label{eq:Smatrix in terms of Tfi}
S_{fi} = \delta_{fi}+i(2\pi)^{d}\delta^{d}(P_f-P_i)T_{fi}.
\end{equation}
In terms of $T_{fi}$ the unitarity constraint (\ref{eq:SSdagger=1}) reads,
\begin{equation}\label{eq:T-T*}
T_{fi}-T_{if}^{\dagger}=i\sum_n(2\pi)^{d}\delta^{d}(P_n-P_i)T_{nf}^{\dagger}T_{ni}
\end{equation}
with $P_{i}$ or $P_f$ the total momentum associated to the in or out states respectively.  With these conventions, the $S$-matrix is given directly by the string path integral,
\begin{equation}\label{eq:S_fi in string theory}
\begin{aligned}
\langle f|(S-\mathds{1})|i\rangle&=\sum_{h=0}^{\infty}\int_{\mathcal{E}\times\mathcal{M}_h}\mathcal{D}g\mathcal{D}Xe^{-S[g,X]}V^{(1)}\dots V^{(N)}\\
& =  i(2\pi)^{d}\delta^{d}(P_f-P_i)T_{fi}.
\end{aligned}
\end{equation}
where we sum over the genus $h$ of Riemann surfaces. It is to be understood that the integrals are over a single gauge slice, i.e.~over all worldsheet embeddings, $\mathcal{E}$, into spacetime and over all worldsheet metrics (or moduli space $\mathcal{M}_h$), such that no two configurations in the integration domain are related by a symmetry. Appropriate integrations over worldsheet insertions are also implicitly included, as are the corresponding Fadeev-Popov determinants. 

To interpret the sum over final states in (\ref{eq:SSdagger=1}) or (\ref{eq:T-T*}), note that the number of ``one  string in volume $V_{d-1}$" states in a momentum space volume element, $d^{d-1}{\bf p}$, is:
\begin{equation}\label{eq:Vdp}
V_{d-1}\frac{d^{d-1}{\bf p}}{(2\pi)^{d-1}},
\end{equation}
because this is the number of sets $\{n_1,n_2,\dots,n_{d-1}\}$ (with $n_j\in \mathbb{Z}$) for which the momentum 
\begin{equation*}
{\bf p}=\frac{2\pi}{L}(n_1,n_2,\dots,n_{d-1}), \qquad {\rm with}\qquad V_{d-1}\equiv L^{d-1},
\end{equation*}
lies in the momentum space volume $d^{d-1}{\bf p}$ around ${\bf p}$. If there are additional discrete/continuous quantum numbers that label the states under consideration, we would have to sum/integrate over these. For example, in the case of coherent states, as we shall see, we would have to include a (dimensionless) integral over polarization tensors \footnote{In the case of coherent states it is simplest to use lightcone coordinates, see (\ref{eq:Vdp lightcone}), because coherent states are (as we will see) eigenstates of $p^+$, $p^i$ but not of $p^-$.}.  In particular, there will in general be a number of kinematically allowed channels and so we should also include a sum over a complete set of states -- we use the compact notation, $\suminntt$, to denote a sum over states and the associated quantum numbers, so that the sum over one-particle states in the final state will be denoted by:
\begin{equation}\label{eq:sumVdp}
\sum_f=\suminnt \,V_{d-1}\!\int \frac{d^{d-1}{\bf p}}{(2\pi)^{d-1}}.
\end{equation}
Both sides of this equation are dimensionless. In relativistic scattering experiments there is also the possibility that the number of strings in the initial and final states is different. Thus, we require the corresponding phase space of multi-particle free states, which will be a sum over products of the single string phase space,
\begin{equation}\label{eq:sumVdp multi}
\sum_f=\sum_{N_f=0}^{\infty}\prod_{a=1}^{N_f}\bigg(\suminnt\limits_a \,V_{d-1}\!\int \frac{d^{d-1}{\bf p}_a}{(2\pi)^{d-1}}\bigg),
\end{equation}
with $a$ labeling the string whose phase space we are summing/integrating over, and $d$ is the dimensionality of spacetime in which the strings are allowed to propagate in ($d\leq 26$ or 10 for the bosonic or superstring theory). The phase space sums (\ref{eq:sumVdp}) or (\ref{eq:sumVdp multi}) are not Lorentz invariant, but of course Lorentz invariance will be restored in physically observable quantities. This is the price of wanting to construct  \emph{dimensionless} $S$-matrix elements, $S_{fi}$, that can be directly interpreted as probabilities.

\subsection{Vertex Operator Normalization}
The normalization of the path integral (or $S$-matrix) and the normalization of vertex operators is completely determined in terms of the normalization of a single vertex operator by the unitarity constraint (\ref{eq:T-T*}) and the identification (\ref{eq:S_fi in string theory}). The normalization of this single vertex operator can in turn be fixed by the ``one string in the universe" normalization condition, by making contact with the corresponding field theory, and we describe this next.

A useful quantity to consider in bosonic string theory is the tachyon vertex operator, because it is the basic building block of higher mass vertex operators. Working in the flat Minkowski background,
$$
G_{\mu\nu}(X)=\eta_{\mu\nu},\quad B_{\mu\nu}(X)=0,\quad {\rm and}\quad \Phi(X)=\langle\Phi\rangle,
$$
with $\langle\Phi\rangle$ a constant, the tachyon vertex operator reads,
\begin{equation}\label{eq:Vtachyon}
V(z,\bar{z})=\mathcal{N}e^{ip\cdot X(z,\bar{z})}.
\end{equation}
We shall eventually relate the normalization of the tachyon to the normalization of all other vertex operators.

To compute the normalization constant $\mathcal{N}$, we notice that $V$ satisfies the equation of motion,
$$
\Big(\nabla^2+\frac{4}{\alpha'}\Big)V=0,
$$
with the derivative taken with respect to the zero mode $x^{\mu}$. The low energy field theory corresponding to the tachyon will therefore be that of a scalar field with mass $m^2=-4/\alpha'$ \cite{CallanGan86},
\begin{equation}\label{eq:S[V]}
S[V] = -\frac{1}{(\alpha')^{\frac{d-2}{2}}}\int d^dx\,e^{-2\langle \Phi\rangle}\Big(\frac{1}{2}(\nabla V)^2+\frac{1}{2}m^2V^2+\dots\Big),
\end{equation}
where we have taken into account the fact that the dilaton (even if it is constant in this case) couples universally as shown \cite{CallanGan86}, and we ignore all interaction terms because we are interested in the case when the string under consideration is asymptotically free and onshell, as required by conformal invariance \cite{Weinberg85}. We have found it convenient to include an appropriate power of $\alpha'$ (with $[\alpha']=L^2$) such that $V$ is dimensionless, $[V]=1$. (This will ensure that the $S$-matrix is dimensionless independently of the number of vertex operators.) Furthermore, an overall \emph{dimensionless} constant in $S[V]$ is immaterial because it can be absorbed into a shift in $\langle\Phi\rangle$. 

As discussed above, the overall normalization of the $S$-matrix and of all vertex operators \emph{other} than, say, the tachyon are fixed by unitarity. Unitarity will thus relate the normalization of all vertex operators to that of the tachyon. It is convenient to define:
\begin{equation}\label{Gd Newton}
g_c\equiv e^{\langle \Phi\rangle}(\alpha')^{\frac{d-2}{4}},\qquad {\rm and }\qquad g_s\equiv e^{\langle \Phi\rangle}.
\end{equation}

Now, the ``one string in $V_{d-1}$" constraint can be solved by requiring that the total energy, $H$, in volume $V_{d-1}$ is that of a single string, $p^0=\sqrt{{\bf p}^2+m^2}$ (with $m^2=-4/\alpha'$). We plug the plane wave solution,
$
V(x)=\mathcal{N}e^{ip\cdot x}+\mathcal{N}^*e^{-ip\cdot x},
$
into the Hamiltonian associated to (\ref{eq:S[V]}), which is given by $H(t)=\int_{V_{d-1}} d^{d-1}x [(\partial_0V)\frac{\partial \mathscr{L}}{\partial(\partial_0V)}-\mathscr{L}]$ (with $S[V]=\int dt\mathscr{L}$), and make the link with the string theory vertex operator by identifying $\mathcal{N}$ here with the $\mathcal{N}$ in (\ref{eq:Vtachyon}). It follows that, 
$
H(t) = |\mathcal{N}|^22(p^0)^2V_{d-1}g_c^{-2},
$ 
implying that there will be one string in volume $V_{d-1}$ if:
\begin{equation}\label{eq:norm mathcal{N}}
\frac{H(t)}{p^0}=1,\qquad{\rm  or,\,\, equivalently,} \qquad
\mathcal{N}=\frac{g_c}{\sqrt{2p^0V_{d-1}}}.
\end{equation}
That is, the ``one string in volume $V_{d-1}$"-normalized  tachyon vertex operator is,
\begin{equation}\label{eq:tachyon original norm}
V(z,\bar{z})=\frac{g_c}{\sqrt{2p^0V_{d-1}}}\,e^{ip\cdot X(z,\bar{z})},
\end{equation}
with $E=\sqrt{{\bf p}^2+m^2}$ (and $m^2=-4/\alpha'$). Although we will not prove this here, it is not too hard to show that this is precisely the normalization required by: (i) Lorentz invariance of the unitarity constraint of the $S$-matrix; (ii)  Lorentz invariance of the scattering cross section; (iii) the requirement that $S$-matrix elements, $S_{fi}$, be dimensionless, so as to interpret $|S_{fi}|^2$ as a probability, as in (\ref{eq:Prob|Sfi|^2}). 

Notice now that the normalization of the tachyon vertex is such that the most singular term in the operator product expansion is,
\begin{equation}\label{eq:VVope normalization*}
V(z,\bar{z})\cdot V(0,0)\cong\bigg(\frac{g_c^2}{2EV_{d-1}}\bigg)\frac{1}{|z|^4}+\dots.
\end{equation}
This suggests that we may be able to normalize \emph{arbitrarily} massive bosonic string vertex operators by requiring that (\ref{eq:VVope normalization*}) is satisfied. This is indeed the case, and it can be shown (although we shall not do so here) that this statement is compatible with unitarity (\ref{eq:T-T*}). Notice that the normalization condition (\ref{eq:VVope normalization*}) ensures that vertex operators are dimensionless.

\subsection{$S$-Matrix Unitarity and Factorization}
It is often more convenient to work with vertex operators normalized according to \footnote{This is in agreement with the conventions of Polchinski \cite{Polchinski_v1}, $g^{\rm here}_c\equiv g_c^{\rm Polchinski}$, where it is shown that the relation to the gravitational coupling is $\kappa=2\pi g_c$ with $\kappa^2=8\pi G_{(d)}$ and $G_{(d)}$ the $d$-dimensional Newton's constant.},
\begin{equation}\label{eq:VVope normalization* new}
V(z,\bar{z})\cdot V(0,0)\cong \frac{g_c^2}{|z|^4}+\dots,
\end{equation}
instead of (\ref{eq:VVope normalization*}). Starting from the original normalization (\ref{eq:VVope normalization*}), we extract the factors of $1/\sqrt{2EV_{d-1}}$ out of every vertex operator and, for $N$ asymptotic states in total, define $\mathcal{M}(1,\dots,N)$ according to,
\begin{equation}\label{eq:T-matrix in terms of M-matrix}
T_{fi}\equiv T(1,\dots,N)\equiv \frac{\mathcal{M}(1,\dots,N)}{\sqrt{2E_1V_{d-1}}\dots \sqrt{2E_NV_{d-1}}},
\end{equation}
with $T_{fi}$ defined in (\ref{eq:Smatrix in terms of Tfi}). When vertex operators are normalized according to (\ref{eq:VVope normalization* new}), the path integral yields instead,
\begin{equation}\label{eq:M(1,dots,N)}
\begin{aligned}
i(2\pi)^{d}&\delta^{d}(P_f-P_i)\mathcal{M}(1,\dots,N)\\
&=\sum_{h=0}^{\infty}\int_{\mathcal{E}\times\mathcal{M}_h}\mathcal{D}g\mathcal{D}Xe^{-S[g,X]}V^{(1)}\dots V^{(N)},
\end{aligned}
\end{equation}
and so according to (\ref{eq:S_fi in string theory}) and (\ref{eq:T-matrix in terms of M-matrix}) we need to divide (\ref{eq:M(1,dots,N)}) by the factors $\sqrt{2E_1V_{d-1}}\dots $ to get an $S$-matrix element \footnote{Note that the factors of $1/\sqrt{2EV_{d-1}}$ are absent in the $S$-matrix elements defined in \cite{Polchinski_v1}.},
\begin{equation}\label{eq:S_{fi} in terms of M}
\begin{aligned}
S_{fi}=\delta_{fi}&+ i(2\pi)^{d}\delta^{d}(P_f-P_i)\\
&\times\frac{\mathcal{M}(1,\dots,N)}{\sqrt{2E_1V_{d-1}}\dots \sqrt{2E_NV_{d-1}}}.
\end{aligned}
\end{equation}
In terms of $\mathcal{M}(1,\dots,N)$, the unitarity constraint (\ref{eq:T-T*}) in the case where the intermediate strings in the sum over states are \emph{single} string states then reads:
\begin{widetext}
\begin{equation}\label{eq:M-M*}
\begin{aligned}
\mathcal{M}(1,&\dots,N)-\mathcal{M}^*(1,\dots,N)=i\suminnt\limits_a\int \frac{d^{d-1}{\bf p}_a}{(2\pi)^{d-1}}\frac{1}{2E_a}(2\pi)^{d}\delta^{d}(p_a-P_i)
\mathcal{M}(1,\dots,a)\mathcal{M}^*(-a,\dots,N),
\end{aligned}
\end{equation}
\end{widetext}
with the sum/integral being over a complete set of states, written symbolically as $a$, and their associated quantum numbers. There is an obvious generalization for multi-string intermediate states. (Because of worldsheet duality it is also necessary to sum over both (say) $s$- and $t$-channel contributions in the case of $N=4$, and their natural generalizations for $N>4$.) It is thus clear that the volume factors have cancelled out and the factors of $\sqrt{2E_i}$ have combined to make the unitarity constraint (\ref{eq:M-M*}) Lorentz invariant. Thus, the factors $\sqrt{2E_i}$ in the vertex operator normalizations are required for Lorentz invariance when the corresponding quantities $\mathcal{M}(1,\dots,N)$ are Lorentz invariant, which is indeed the case in string theory; recall that $\frac{d^{d-1}{\bf p}}{(2\pi)^{d-1}}\frac{1}{2E_{\bf p}}$ is the Lorentz invariant phase space, with $E_{\bf p}=\sqrt{{\bf p}^2+m^2}$. Using $\int \frac{d^dp}{(2\pi)^d}\,(2\pi)\delta(p^2+m^2)\theta(p^0)=\frac{d^{d-1}{\bf p}}{(2\pi)^{d-1}}\frac{1}{2E}$ and $2\pi i\, \delta(x)=\frac{1}{x-i0}-\frac{1}{x+i0}$, it is an elementary exercise to show that tree level unitarity (\ref{eq:M-M*}) is guaranteed if the following factorization formula holds true,
\begin{equation}\label{eq:M factorization}
\begin{aligned}
&i\mathcal{M}(1,\dots,N)=\\
&=\suminnt\limits_a\,i\mathcal{M}(1,\dots,a)\cdot\frac{-i\theta(k_a^0)}{k_a^2+m_a^2-i0}\cdot i\mathcal{M}^*(-a,\dots,N),
\end{aligned}
\end{equation}
and
\begin{equation}
\begin{aligned}
\mathcal{M}(&1,\dots,a)\mathcal{M}^*(-a,\dots,N)\\
&=\big[\mathcal{M}(1,\dots,a)\mathcal{M}^*(-a,\dots,N)\big]^*.
\end{aligned}
\end{equation}
This last condition is true for $N=4$, given that $\mathcal{M}(1,2,a)\mathcal{M}^*(-a,3,4)$ is indeed real for string amplitudes, but is not in general true for $N>4$. 
Notice that  
$$
\frac{-i\theta(k^0)}{k^2+m^2-i0},
$$ 
is the propagator (in the $(-++\dots)$ signature) for a scalar particle of mass $m^2$ with the correct analytic continuation for a Minkowski process. Given the normalization of the tachyon, the formula (\ref{eq:M factorization}) can be used to derive the normalization of the tree level $S$-matrix and of all other vertex operators. (Some examples can be found in Polchinski \cite{Polchinski_v1}, where $i(2\pi)^{26}\delta^{26}(P_f-P_i)\mathcal{M}(1,\dots,N)_{\rm here}=S(1,\dots,N)_{\rm there}$ and $g_{c}^{\rm here}=g_c^{\rm there}$ when the dilaton expectation value in (\ref{Gd Newton}) has been shifted appropriately.)

\subsection{Lightcone Coordinates}
It is sometimes more convenient (especially in the case of coherent states) to use lightcone coordinates, $\{p^{\pm},p^i\}$ with $i=1,\dots,d-2$ and $p^{\pm}=\frac{1}{\sqrt{2}}(p^0\pm p^{d-1})$. In lightcone coordinates, (\ref{eq:VT}) is replaced by:
\begin{equation}\label{eq:VT lc}
\begin{aligned}
&(2\pi)\delta({p^{\pm}}'-p^{\pm})\equiv V_{\mp},\phantom{\Big|}\\
&(2\pi)^{d-2}\delta^{d-2}({\bf p}'-{\bf p})\equiv V_{d-2}.
\end{aligned}
\end{equation}
The momentum phase space analogous to (\ref{eq:Vdp}) is:
\begin{equation}\label{eq:Vdp lightcone}
\mathcal{V}_{d-1}\frac{d^{d-2}{\bf p}}{(2\pi)^{d-1}}\frac{dp^+}{2\pi},\qquad{\rm with}\qquad \mathcal{V}_{d-1}\equiv V_{d-2}V_-.
\end{equation}
For the sum over single string states (\ref{eq:sumVdp}) we thus have,
\begin{equation}\label{eq:sumVdp lc}
\sum_f=\suminnt \,\mathcal{V}_{d-1}\int_{\mathbb{R}^{d-2}} \frac{d^{d-2}{\bf p}}{(2\pi)^{d-1}}\int_0^{\infty}\frac{dp^+}{2\pi},
\end{equation}
and similarly for the multi-string case (\ref{eq:sumVdp multi}). We next need the statements analogous to (\ref{eq:tachyon original norm}) and more generally (\ref{eq:VVope normalization*}) in the case of lightcone gauge coordinates. 

In direct analogy with the procedure described in the paragraph containing (\ref{eq:tachyon original norm}), we compute the \emph{lightcone coordinate} Hamiltonian associated to the action (\ref{eq:S[V]}), which is given by $H(x^+)=\int d^{d-2}xdx^- [(\partial_+V)\frac{\partial \mathscr{L}}{\partial(\partial_+V)}-\mathscr{L}]$ (with $S[V]=\int dx^+\mathscr{L}$), and enforce the ``one string in volume $\mathcal{V}_{d-1}$" constraint by truncating the region of integration in $H(x^+)$ to $\mathcal{V}_{d-1}$ and requiring that $H(x^+)/p^-=1$.  Here $p^-=\frac{1}{2p^+}({\bf p}^2+m^2)$, is the tachyon onshell condition which yields the lightcone energy associated to a single tachyon (here $m^2=-4/\alpha'$). Plugging the plane wave solution,
$
V(x)=\mathcal{N}e^{ip\cdot x}+\mathcal{N}^*e^{-ip\cdot x},
$
into the Hamiltonian $H(x^+)$ and requiring that there is one string in volume $\mathcal{V}_{d-1}$, i.e.~$H(x^+)/p^-=1$, thus determines $\mathcal{N}$,
\begin{equation}\label{eq:norm mathcal{N} lc}
\mathcal{N}=\frac{g_c}{\sqrt{2p^+\mathcal{V}_{d-1}}}.
\end{equation}
We make the link with the string theory vertex operator by identifying this $\mathcal{N}$ with that found in  (\ref{eq:Vtachyon}), so that the ``one string in volume $\mathcal{V}_{d-1}$"-normalized  tachyon vertex operator in lightcone coordinates is,
\begin{equation}\label{eq:tachyon original norm lc}
V(z,\bar{z})=\frac{g_c}{\sqrt{2p^+\mathcal{V}_{d-1}}}\,e^{ip\cdot X(z,\bar{z})}.
\end{equation}
This normalization is such that the most singular term in the operator product expansion is,
\begin{equation}\label{eq:VVope normalization}
V(z,\bar{z})\cdot V(0,0)\cong\bigg(\frac{g_c^2}{2p^+\mathcal{V}_{d-1}}\bigg)\frac{1}{|z|^4}+\dots,
\end{equation}
and, in direct analogy to the above, this normalization can be used for arbitrarily massive bosonic vertex operators \footnote{The reason as to why lightcone coordinates are useful in the case of coherent states (as mentioned above) is that they are eigenstates of $\hat{p}^+$ and $\hat{\bf p}$, but not of $\hat{p}^-$, and so it is not possible to factor out $1/\sqrt{2p^0}$, but it is possible to factor out $1/\sqrt{2p^+}$.}.

Again, as discussed above, see (\ref{eq:VVope normalization* new}), it is sometimes more convenient to work with vertex operators normalized according to,
\begin{equation}\label{eq:VVope normalization** new}
V(z,\bar{z})\cdot V(0,0)\cong \frac{g_c^2}{|z|^4}+\dots,
\end{equation}
instead of (\ref{eq:VVope normalization}). From (\ref{eq:VVope normalization}), this implies that we should extract the factors of $1/\sqrt{2p^+\mathcal{V}_{d-1}}$ out of every vertex operator and, as in (\ref{eq:T-matrix in terms of M-matrix}), for $N$ asymptotic states in total define:
\begin{equation}\label{eq:T-matrix in terms of M-matrix lc}
T_{fi}\equiv T(1,\dots,N)\equiv \frac{\mathcal{M}(1,\dots,N)}{\sqrt{2p^+_1\mathcal{V}_{d-1}}\dots \sqrt{2p^+_N\mathcal{V}_{d-1}}}.
\end{equation}
As in (\ref{eq:M(1,dots,N)}), when vertex operators are normalized according to (\ref{eq:VVope normalization** new}), the path integral yields,
\begin{equation}\label{eq:M(1,dots,N) lc}
\begin{aligned}
i(2\pi)^{d}&\delta^{d}(P_f-P_i)\mathcal{M}(1,\dots,N)=\\
&=\sum_{h=0}^{\infty}\int_{\mathcal{E}\times\mathcal{M}_h}\mathcal{D}g\mathcal{D}Xe^{-S[g,X]}V^{(1)}\dots V^{(N)},
\end{aligned}
\end{equation}
but now we need to divide (\ref{eq:M(1,dots,N)}) by the factors $\sqrt{2p^+_1\mathcal{V}_{d-1}}\dots \sqrt{2p^+_N\mathcal{V}_{d-1}}$ to get an $S$-matrix element, and in particular,
\begin{equation}\label{eq:S_{fi} in terms of M lc}
S_{fi} = \delta_{fi}+ i(2\pi)^{d}\delta^{d}(P_f-P_i)\frac{\mathcal{M}(1,\dots,N)}{\sqrt{2p^+_1\mathcal{V}_{d-1}}\dots \sqrt{2p^+_N\mathcal{V}_{d-1}}}.
\end{equation}

The unitarity statement analogous to (\ref{eq:M-M*}) in lightcone coordinates can be derived directly from (\ref{eq:M-M*}) since (\ref{eq:M-M*}) is Lorentz invariant, or it can be derived from (\ref{eq:T-T*}) and (\ref{eq:T-matrix in terms of M-matrix lc}). It reads,
\begin{widetext}
\begin{equation}\label{eq:M-M* lc}
\begin{aligned}
\mathcal{M}(1,&\dots,N)-\mathcal{M}^*(1,\dots,N)=i\suminnt\limits_a\int_{\mathbb{R}^{d-2}}\frac{d^{d-2}{\bf p}_a}{(2\pi)^{d-1}}\int_0^{\infty}\frac{dp^+}{2\pi}\frac{1}{2p^+_a}(2\pi)^{d}\delta^{d}(p_a-P_i)
\mathcal{M}(1,\dots,a)\mathcal{M}^*(-a,\dots,N),
\end{aligned}
\end{equation}
\end{widetext}
and the result is (as above) independent of the volume $\mathcal{V}_{d-1}$. To see this let us consider the relativistic phase space integral,
$
\int \frac{d^dk}{(2\pi)^d}\,(2\pi)\delta(k^2+m^2)\theta(k^0)
$ (which as mentioned above is equivalent to $\int\frac{d^{d-1}{\bf k}}{(2\pi)^{d-1}}\frac{1}{2E_{\bf k}}$) with \footnote{It is implied here that $\alpha'=2$ in the case of closed strings or $\alpha'=1/2$ in the case of open strings.} $m^2=2N-2$. In lightcone coordinates  (where $dk^-\wedge dk^+=dk^{0}\wedge dk^{d-1}$), let us redefine the integration variable (with $i=1,\dots,24$):
\begin{equation}\label{eq:k to p}
\begin{aligned}
&k^-=p^-+\frac{N}{p^+},\qquad k^+ = p^+,\qquad k^i=p^i.
\end{aligned}
\end{equation}
This removes the $N$-dependence from the $\delta$-function, $\delta(k^2+2N-2)=\delta (p^2-2)$, and $dk^-\wedge dk^+=dp^-\wedge dp^+$. Ignoring the tachyon, so that $\theta(k^0)=\theta(p^+)$, the Lorentz invariant phase space now reads,
\begin{equation}\label{eq:phase space}
\begin{aligned}
\int \frac{d^dk}{(2\pi)^d}&\,(2\pi)\delta(k^2+2N-2)\theta(k^0)=\\
&=\int \frac{d^dp}{(2\pi)^d}\,(2\pi)\delta(p^2-2)\theta(p^+)\\
&= \int_{\mathbb{R}^{d-2}} \frac{d^{d-2}{\bf p}}{(2\pi)^{d-2}}\int_0^{\infty}\frac{dp^+}{2\pi} \frac{1}{2p^+},
\end{aligned}
\end{equation}
where we have integrated out $p^-$, so that $p^-=\frac{1}{2p^+}({\bf p}^2-2)$, the tachyon onshell condition. Therefore,
$$
\int_{\mathbb{R}^{d-1}}\frac{d^{d-1}{\bf k}}{(2\pi)^{d-1}}\frac{1}{2E_{\bf k}}=\int_{\mathbb{R}^{d-2}} \frac{d^{d-2}{\bf p}}{(2\pi)^{d-2}}\int_0^{\infty}\frac{dp^+}{2\pi} \frac{1}{2p^+},
$$
where it is understood that the integrands are taken onshell;  the aforementioned unitarity statement (\ref{eq:M-M* lc}) is proven.

\subsection{Tree Level Operator Statements}
It is sometimes desirable to compute expectation values of various operators, such as the angular momentum $J^{\mu\nu}$,
\begin{equation}\label{eq:<J>=<VJV>}
\langle J^{\mu\nu}\rangle \equiv \langle V|J^{\mu\nu}|V\rangle\equiv J_{\rm cl}^{\mu\nu},
\end{equation}
as this enables one to associate classically computed quantities, such as $J_{\rm cl}^{\mu\nu}$ that is in one-to-one correspondence with solutions of $\partial_z\partial_{\bar{z}}X^{\mu}=0$, to quantum-mechanical vertex operators that exhibit these classical characteristics (in the expectation value sense). It is convenient to work in the operator formalism here\footnote{The usual path integral definition is not useful here because the path integral associated to two vertex operator insertions vanishes (unless the state under consideration is unstable), because the volume of the CKG is infinite and two vertices are not sufficient to saturate this infinity. This is because the path integral yields only the \emph{non}-trivial contribution to the $S$-matrix, whereas in (\ref{eq:<J>=<VJV>}) it is the trivial or non-interacting part that is relevant.} and absorb the $\alpha'$ and $e^{\langle\Phi\rangle}$ dependence of $V(z,\bar{z})$ into $|0,0;p\rangle$, recall that $g_c=e^{\langle\Phi\rangle}{\alpha'}^{\frac{d-2}{4}}$, and in particular,
\begin{equation}\label{eq:|p> = g_c e^ipX}
|0,0;p\rangle \simeq g_c\,e^{ip\cdot X(z,\bar{z})}.
\end{equation}
At tree level, the factors of $e^{\langle\Phi\rangle}$ (in $g_c$ in each of the two vertex operators in e.g.~$\langle V|J^{\mu\nu}|V\rangle$ and the Euler characteristic $e^{-\chi(\Sigma)\langle\Phi\rangle}=e^{-2\langle\Phi\rangle}$) cancel. If we then normalize the state and expectation values in a relativistically invariant manner,
\begin{equation}\label{eq:tach and vac norm}
\begin{aligned}
&|V\rangle = \frac{1}{\sqrt{2E_{\bf p}V_{d-1}}}|0,0;p\rangle,\\
&\langle 0,0;p'|0,0;p\rangle = 2E_{\bf p}(2\pi)^{d-1}\delta^{d-1}({\bf p}'-{\bf p}),
\end{aligned}
\end{equation}
then, according to (\ref{eq:VT}), such states have unit norm,
$$
\langle V|V\rangle=1.
$$
The dimensionality of $g_c$ is precisely that required to make the relativistic normalization shown possible. In lightcone coordinates we have similarly the following relativistic normalization,
\begin{equation}\label{eq:tach and vac norm lcc}
\begin{aligned}
&|V\rangle = \frac{1}{\sqrt{2p^+\mathcal{V}_{d-1}}}|0,0;p\rangle,\\
&\langle 0,0;p'|0,0;p\rangle=2p^+(2\pi)\delta({p^+}'-p^+) \\
&\hspace{1.5cm}\times(2\pi)^{d-2}\delta^{d-2}({\bf p}'-{\bf p}).
\end{aligned}
\end{equation}

As we shall elaborate on extensively in the following section,  higher mass (mass eigen-)states with unit norm can be constructed by acting on the tachyon vertex with DDF operators \cite{DelGiudiceDiVecchiaFubini72,AdemolloDelGuidiceDiVecchiaFubini74}, $A_n^i$ and $\bar{A}^i_n$, which satisfy $[A_n^i,A_m^j]=n\delta^{ij}\delta_{n+m,0}$:
$$
|V\rangle = \frac{1}{\sqrt{2E_{\bf p}V_{d-1}}}\,C\xi_{i\dots,j\dots}A^{i}_{-n} \dots\bar{A}_{-\bar{n}}^j\dots |0,0;p\rangle,
$$
The combinatorial constant $C$, defined in (\ref {norm const C}), is chosen such that 
$$
\langle V|V\rangle=1,
$$
 remains true for arbitrarily massive states. There is a similar result in lightcone coordinates with $2p^+\mathcal{V}_{d-1}$ replacing $2EV_{d-1}$, with the corresponding normalization of the tachyonic lightcone vacuum implied as shown in (\ref{eq:tach and vac norm lcc}). Furthermore, the corresponding lightcone \emph{gauge} quantities can be obtained by replacing $A^i_n$ and $\bar{A}^i_n$ by $\alpha^i_n$ and $\tilde{\alpha}^i_n$ respectively. Similarly, we will see that the closed string covariant coherent states are of the form,
 \begin{widetext}
\begin{equation}\label{V_0A*}
\begin{aligned}
|V\rangle= \frac{1}{\sqrt{2p^+\mathcal{V}_{d-1}}}\,\mathcal{C}_{\lambda\bar{\lambda}}\int_0^{2\pi}\!\!\!&\dslash s\exp\Big\{\sum_{n=1}^{\infty}\frac{1}{n}e^{ins}\lambda_n\cdot A_{-n}\Big\}\exp\Big\{\sum_{m=1}^{\infty}\frac{1}{m}e^{-ims}\bar{\lambda}_m\cdot \bar{A}_{-m}\Big\}\,|0,0;p\rangle,
\end{aligned}
\end{equation}
\end{widetext}
see (\ref{V_0A}), which again has unit norm, 
$$
\langle V|V\rangle=1, 
$$
 as do the mass eigenstates. Notice that, as mentioned above, for coherent states lightcone coordinates are more convenient. 

Similar results hold for open strings, with $g_o$ and $|0;p\rangle$ replacing $g_c$ and $|0,0;p\rangle$, both vacua being normalized in the same manner, as in (\ref{eq:tach and vac norm}) or (\ref{eq:tach and vac norm lcc}) depending on the choice of coordinates. In addition, in the case of open strings left- and right-movers are related and hence one can construct states using only, say, the holomorphic quantities $A_n^i$ or $\alpha_n^i$. The closed and open string couplings, $g_c$ and $g_o$, are related by unitarity \cite{Polchinski_v1}, e.g.~by factorizing the annulus diagram on a closed string pole; in $d=26$, $g_o^2=2^{18}\pi^{25/2}{\alpha'}^6g_c$, and in our conventions, see (\ref{Gd Newton}), where $g_c=e^{\langle\Phi\rangle}{\alpha'}^6$,
\begin{equation}\label{eq:g_o in terms of alpha' phi}
g_o=8\pi^{\frac{1}{4}}(2\pi\alpha')^6e^{\langle\Phi\rangle/2}.
\end{equation}
Note that the dimensionality of both $g_c$ and $g_o$ is the same. Below we will consider both open and closed string vertex operators in detail.

\section{Arbitrarily Massive Vertex Operators}\label{DDF}

Before we can hope to understand the covariant vertex operator description of cosmic strings we must first understand how to construct arbitrarily massive covariant vertex operators, and this is the question we address in the present section. We base our exposition on the general (yet practical) approach of Del Giudice, Di Vecchia and Fubini \cite{DelGiudiceDiVecchiaFubini72,AdemolloDelGuidiceDiVecchiaFubini74} (henceforth referred to as DDF), see also \cite{Brower72,GoddardThorn72,GSW1,D'HokerGiddings87}, although we will adopt a somewhat more modern viewpoint.

\begin{figure*}
\includegraphics[width=.8\textwidth]{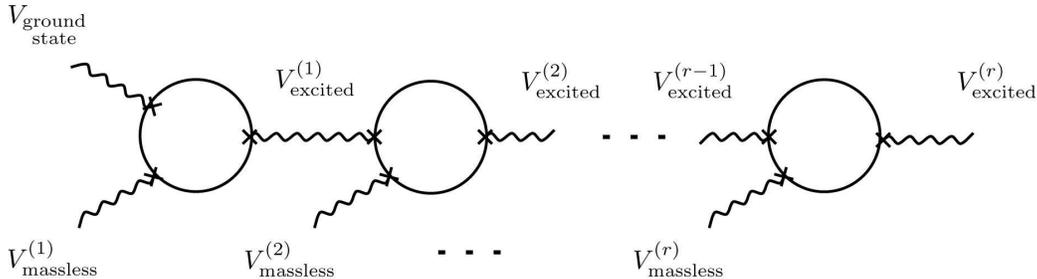}\caption{The DDF construction of a complete set of open string physical covariant vertex operators. \label{fig:DDFop}}
\end{figure*}

The geometrical string picture underlying the DDF vertex operator construction is as follows. Arbitrary vertex operators can be extracted from a certain factorization of an $n$-point scattering amplitude. The setup we have in mind is the following: a string in its vacuum state absorbs some number of massless excitations, resulting in an excited state -- the resulting excited vertex operator is what we wish to extract. The first non-trivial statement is that a complete set of vertex operators can be obtained from the factorization of a diagram with an arbitrary number of massless open string vertex operator insertions and a vacuum insertion. When the vertex operator we wish to extract is an open string state the appropriate factorization is shown in Fig.~\ref{fig:DDFop}.

It turns out that (as we shall show) a complete set of states can be obtained if the $i^{\rm th}$ massless photon vertex operator has momentum $k_{(i)}^{\mu}=-n_iq^{\mu}$ and polarization tensor $\xi_{(i)}^j$ with $q^2=0$ and $n_i$ a positive integer. All photons therefore approach the vacuum string state from the same angle of incidence with momenta that are only allowed to differ by some integer multiple of a so far arbitrary null vector $q^{\mu}$. Conformal invariance enforces the vector $q^{\mu}$ to be transverse to all photon polarization tensors, $\xi_{(i)}^j$, and this leads to spacetime gauge invariance \cite{GSW1}. The vacuum vertex operator, $e^{ip\cdot X}$, which absorbs these photons has momentum $p^{\mu}$ and is tachyonic in the bosonic string, $p^2=1/\alpha'$. 

Let us now become more explicit. This procedure is to be thought of in a step-wize sense: first consider a single photon absorbed by an open string vacuum state. Vertices produced in this process are then given by the residue of the OPE (operator product expansion) as these two initial states approach on the boundary of the worldsheet, $$V_{\rm excited}^{(1)}(w)\cong \oint_{w} \dslash z_1\,V^{(1)}_{\rm massless}(z_1)\cdot V_{\substack{\rm ground\\ \rm state}}(w).$$ The resulting state, $V_{\rm excited}^{(1)}(w)$ has momentum $(p-n_1q)^{\mu}$ with $n_1$ a positive integer of our choice. $V_{\rm excited}^{(1)}(w)$ is then brought close to an additional photon, $V^{(2)}_{\rm massless}(z)$, the residue of this OPE now giving rise to a new state, $$V_{\rm excited}^{(2)}(w)\cong \oint_{w} \dslash z_2\,V^{(2)}_{\rm massless}(z_2)\cdot V_{\rm excited}^{(1)}(w),$$ with momentum $(p-n_1q-n_2q)^{\mu}$ and so on. Carrying this out $r$ times gives rise to a general vertex operator,
\begin{widetext}
$$V_{\rm excited}^{(r)}(w)\cong \oint_{w} \dslash z_r\,V^{(r)}_{\rm massless}(z_r)\dots\oint_{w} \dslash z_2\,V^{(2)}_{\rm massless}(z_2)\cdot  \oint_{w} \dslash z_1\,V^{(1)}_{\rm massless}(z_1)\cdot V_{\substack{\rm ground\\ \rm state}}(w),$$ 
\end{widetext}
where it is to be understood that the rightmost integrals are carried out first so as to respect the order with which the photons are absorbed by the vacuum. To ensure that the internal strings (see Fig.~\ref{fig:DDFop}) are onshell we must require that $(p-Nq)^2=(1-N)/\alpha'$ for $N=\sum_in_i$, which will be satisfied provided: $$p\cdot q=1/(2\alpha').$$ The choice of integers $\{n_1,n_2,\dots,n_r\}$ determines the mass level $N$ of the vertex operator of the final state and $(p-Nq)^{\mu}$ is the corresponding momentum of this excited state. Defining $A_n^i=\sqrt{\frac{2}{\alpha'}}\oint \dslash z\,\partial_z X^i(z)e^{inq\cdot X(z)}$, the above state can be equivalently written as,
\begin{equation}\label{eq:V(w)DDF}
V_{\rm excited}^{(r)}(w)\cong \frac{g_o}{\sqrt{2p^+\mathcal{V}_{d-1}}}\,C\xi_{i\dots j}A_{-n_1}^i\dots A_{-n_r}^j\cdot e^{ip\cdot X(w)},
\end{equation}
with $C$ a to-be-determined normalization constant and formally write $\xi^{ij\dots }=\xi_{(1)}^i\xi_{(2)}^{j}\dots$. We have included the factor of $\frac{g_o}{\sqrt{2p^+\mathcal{V}_{d-1}}}$ that we computed (by the `one string in volume $\mathcal{V}_{d-1}$' condition) in Sec.~\ref{sec:NSW} that ensures that $S$-matrix elements transform correctly under Lorentz transformations. Recall from above, see (\ref{eq:g_o in terms of alpha' phi}), that we denote the open string coupling by $g_o$. 

The  $A_n^i$ are the so called DDF operators \cite{DelGiudiceDiVecchiaFubini72,AdemolloDelGuidiceDiVecchiaFubini74}. After carrying out the contour integrals the resulting vertex operator, $V(w)\equiv V_{\rm excited}^{(r)}(w)$, will be composed of a linear superposition of normal ordered terms of the form $\zeta^{\mu\nu\dots}\partial^{\#}X\partial^{\#}X\dots$ with an overall factor of $e^{i(p-Nq)\cdot X(z)}$ (we shall compute these explicitly). The polarization tensors $\zeta^{\mu\nu\dots}$ will be composed of the quantities, $\xi^{ij\dots}$, $p^{\mu}$, and $q^{\mu}$. There is clearly a one-to-one correspondence between vertex operators $V(w)$ and lightcone gauge states, 
$$
|V\rangle_{\rm lc}=\frac{1}{\sqrt{2p^+\mathcal{V}_{d-1}}}\,C'\xi_{i\dots j}\alpha_{-n_1}^i\dots \alpha_{-n_r}^j|0;p^+,p^i\rangle,
$$ 
with $C'$ an \emph{a priori} different normalization constant to $C$. It is determined by the condition 
$\langle V|V\rangle_{\rm lc}=1$ and, writing $|0;p\rangle=|0;p^+,p^i\rangle$,
$$
\langle 0;p'|0;p\rangle=2p^+(2\pi)\delta({p^+}'-p^+)(2\pi)^{d-2}\delta^{d-2}({\bf p}'-{\bf p}).
$$ 
Therefore, we reach the important conclusion that covariant vertex operators extracted via factorization of a scattering amplitude with photons and a ground state tachyon form a complete set. A rather non-trivial statement is that $V(w)$ has the same mass and angular momenta as $|V\rangle_{\rm lc}$ (we prove this later for arbitrary coherent states), and we take this correspondence further and \emph{conjecture} that $V(w)$ and $|V\rangle_{\rm lc}$ also share identical interactions.

Note that the above construction \emph{is} covariant \cite{D'HokerGiddings87}, although not manifestly so: even though the $\xi^j_{(i)}$ do not contain any timelike directions (as is also the case for the lightcone gauge states) the resulting polarization tensors $\zeta^{\mu\nu\dots}$ potentially have all components non-vanishing, thus restoring manifest covariance -- we shall prove this with some examples below. We have not enforced any constraint, e.g.~$X^+\propto\tau$, on the target space coordinates in the vertex operator $V(w)$, and so the path integral with vertex insertions $V(w)$ includes a measure $\int_{\mathcal{E}} \mathcal{D}X^0\dots\mathcal{D}X^{25}e^{\frac{i}{\hbar}\,S[X]}$. Making covariance manifest is of course not required in order to plug such vertices into covariant path integrals. The correspondence with the lightcone gauge states suggests also the following: the quantity $\xi^{ij\dots}$ that appears in the covariant vertex operators are to be identified with tensors corresponding to irreducible representations of SO(25), the little group of SO(25,1) for massive states: that is, $\xi^{ij\dots}$ have the symmetries of Young tableaux \cite{Hamermesh}.

A good consistency check is the following. Given that the DDF operators are integrals of photon vertex operators, i.e.~integrals of (1,0) conformal primary operators, they must be gauge invariant: $[L_n,A_m^i]=0$. Therefore, $V(w)$ must satisfy the Virasoro constraints: the operator $L_{n>0}$ will commute through to hit the vacuum, $e^{ip\cdot X}$, which will be annihilated if it is physical, i.e.~if $p^2=1/\alpha'$. The $L_0$ operator similarly commutes through to hit the vacuum and given that $L_0\cdot e^{ip\cdot X}\cong e^{ip\cdot X}$, the full vertex operator $V(w)$ satisfies the Virasoro constraints automatically: 
$$L_0\cdot V(w)\cong V(w),\qquad {\rm and}\qquad L_{n>0}\cdot V(w)\cong 0.$$
In direct analogy to the lightcone gauge states, the vertices $V(w)$ are transverse to null states \cite{Brower72} as one would expect given the underlying geometrical string picture on which the construction is based.

For the construction of closed string vertex operators it turns out that the naive expression, namely,
\begin{equation}\label{eq:DDF state intro}
\begin{aligned}
V(z,&\bar{z})=\frac{g_c}{\sqrt{2p^+\mathcal{V}_{d-1}}}\,C\xi_{ij\dots,kl\dots}\\
&\times A^{i}_{-n_1} A^j_{-n_2}\dots\bar{A}_{-\bar{n}_1}^k\bar{A}_{-\bar{n}_2}^l\dots\,e^{ip\cdot X(z,\bar{z})},
\end{aligned}
\end{equation}
with the DDF operators $A_n^i$ and $\bar{A}_n^i$ defined in (\ref{eq: DDF As}), is also the correct expression. The lightcone gauge realization of this state is the expression  \cite{Brower72,GoddardThorn72}, 
\begin{equation}\label{eq:lc state intro}
\begin{aligned}
|V&\rangle_{\rm lc}=\,\frac{g_c}{\sqrt{2p^+\mathcal{V}_{d-1}}}\,C\xi_{ij\dots,kl\dots}\\
&\times\alpha^{i}_{-n_1}\alpha_{-n_2}^{j}\dots\tilde{\alpha}_{-\bar{n}_1}^{k}\tilde{\alpha}_{-\bar{n}_2}^{l}\dots |0,0;p^+,p^i\rangle.
\end{aligned}
\end{equation}
We as usual need to introduce the constraint, $N=\bar{N}$ by hand \footnote{Vertex operators (\ref{eq:DDF state intro}) or (\ref{eq:lc state intro}) that do not satisfy the constraint $N=\bar{N}$ still satisfy the Virasoro constraints, $L_0=\bar{L}_0$, but require the presence of a lightlike compactified background. We will discuss vertex operators in lightlike compactified backgrounds in detail when we construct closed string coherent states.}. The closed string constraints analogous to the open string case are $p^2=4/\alpha'$, $p\cdot q=2/\alpha'$, $q^2=0$ and $q\cdot \xi=0$. The DDF operators commute with the Virasoro generators and so (\ref{eq:DDF state intro}) again satisfies the Virasoro constraints. The normalization of the vacuum is again:
\begin{equation}\label{eq:lcc vac norm op}
\begin{aligned}
 \langle 0,&0;p'|0,0;p\rangle =\\
 & =2p^+(2\pi)\delta({p^+}'-p^+)(2\pi)^{d-2}\delta^{d-2}({\bf p}'-{\bf p}),
 \end{aligned}
\end{equation}
and determine $C$ by the condition $\langle V|V\rangle_{\rm lc}=1$, see Sec.~\ref{sec:NSW}.

Caution however is needed in interpreting this expression as a vertex arising in a scattering experiment of massless states and a vacuum (as we did above for the open string). If for example, the vacuum and the corresponding massless states in a string scattering experiment are all bulk vertex operators then a complete set of states would not be generated: e.g., vertices with an asymmetry corresponding to the lightcone gauge states $\alpha_{-1}^i\alpha_{-1}^j\tilde{\alpha}_{-2}^k|p^i,p^+;0,0\rangle$ could not be generated in a closed string scattering experiment of massless vertices and a tachyon. It is likely that rather vertex operators (\ref{eq:DDF state intro}) can instead be created in an open string scattering experiment: factorization of a (one loop open string) scattering amplitude involving photons and a closed string tachyon should give rise to an arbitrary closed string vertex operator of the form (\ref{eq:DDF state intro}). It might be worth mentioning that a closed string scattering experiment in a lightlike compactified spacetime, $X^-\sim X^-+2\pi R^-$ with $R^-=\frac{\alpha'}{2}q^-$, of massless vertex operators (with lightlike winding) and a tachyon (without lightlike winding) would generate a complete set of vertex operators of the form (\ref{eq:DDF state intro}), without the need of introducing open string interactions. 

Crucially, the above prescription for extracting vertex operators results in explicit polarization tensors for which there are no additional constraints to be solved, which is a common serious drawback of many other approaches to vertex operator constructions, see e.g.~\cite{Weinberg85,FriedanMartinecShenker86,Polchinski87,DHokerPhong,Sato88,SasakiYamanaka85,Nergiz94} among others.

\subsection{Momentum Phase Space}
We now examine a subtlety related to the fact that the operators $A_n^i$ depend on the momenta $q^{\mu}$. The question we want to address here is: when we compute expectation values, can different vertex operators be labelled by different null vectors $q^{\mu}$? DDF operators satisfy an oscillator algebra, $[A_n^i,A_m^j]=n\delta^{ij}\delta_{n+m,0}$, which is identical to the algebra associated to the $\alpha_n^i$ operators, $[\alpha_n^i,\alpha_m^j]=n\delta^{ij}\delta_{n+m,0}$. 
In general, one might expect however that different vertex operators should be constructed out of DDF operators which in turn are defined with different $q^{\mu}$ -- different choices of $q^{\mu}$ for different vertices corresponds to different choices of momentum, $k^{\mu}=p^{\mu}-Nq^{\mu}$. It would then seem that the relevant commutator is $[A_n^i,{A_m^j}']$ rather than $[A_n^i,A_m^j]$ with ${A_n^i}'$ a DDF operator constructed out of $q'$. To examine this possibility, let us analyze the constraints and momentum phase space. 

Consider the case of open strings with both ends attached to a single D$p$-brane, and take $p=25$. In this case, we can write down results that hold for both open strings and closed strings when the choice $\alpha'=1/2$ and $\alpha'=2$ is made respectively. As discussed above, in the DDF formalism, the momentum of a level $N$ mass eigenstate is: 
$$
k^{\mu}=p^{\mu}-Nq^{\mu}.
$$
Two 26-dimensional vectors $p^{\mu}$, $q^{\mu}$ are therefore needed to specify the momentum of the state, but there are only 3 constraint equations: $p^2=2$, $p\cdot q=1$, and  $q^2=0$, so that there remain, $2\times 26-3=49$ free parameters. Given that $k^{\mu}$ has only 26 parameters, one of them being eliminated by making use of the mass shell condition, it follows that only 25 of the 49 free parameters are needed in order to completely specify the momentum of a state. 
Therefore, we can fix $49-25=24$ of the $2\times 26$ parameters in $p^{\mu}$, $q^{\mu}$ while still spanning the \emph{full} the phase space. Use this freedom to set 
$$
q^i=0,\qquad{\rm for}\qquad i=1,\dots,24,
$$
for \emph{all} states constructed by DDF operators. Substituting this into the constraint equations (\ref{eq:pq cond_open}), leads to the positive energy solution \footnote{Here for notational simplicity $\alpha'=1/2$, or $\alpha'=2$ for the open or closed string case respectively. Also, ${\bf p}=(p^1,\dots,p^{24})$ and as usual $p^{\pm}=\frac{1}{\sqrt{2}}(p^0\pm p^{25})$, or in the case of open strings attached to a D$p$-brane, $p^{\pm}=\frac{1}{\sqrt{2}}(p^0\pm p^{p})$.},
\begin{equation}\label{eq:p,q;Q}
\begin{aligned}
p^{\mu}&=\Big(\frac{c}{2}\big({\bf p}^2-2\big)+\frac{1}{2c},{\bf p},-\frac{c}{2}\big({\bf p}^2-2\big)+\frac{1}{2c}\Big),\\
q^{\mu}&=\big(-c,0,\dots,0,c\big).\phantom{\Big)}
\end{aligned}
\end{equation}
As required, this choice satisfies $-(p-Nq)^2\equiv m^2=2N-2$ for any $p^i, c$. In terms of $p^+$ we have $c=1/(\sqrt{2}p^+)$, and $k^-=\frac{1}{2p^+}({\bf p}^2+2N-2)$. The positive energy condition requires $c>0$ (for non-tachyonic states, $N\geq 1$), and the full phase space (neglecting the tachyon) is:
 $$
 -\infty\leq {\bf p}\leq \infty\qquad {\rm and}  \qquad p^+>0,
 $$
with $p^+=-1/q^-$ \footnote{As an example, if we boost to the rest frame where the $k^i=0$ and $k^0=\sqrt{2N-2}$, the vectors $p^{\mu}$ and $q^{\mu}$ are determined completely, and $c^{-1}=\sqrt{2 N-2}$.}. We reach the important conclusion that different vertex operators may indeed be labelled by different $q^{\mu}$ when their momenta differ, but that \emph{all} vertices may be taken to have $q^i=q^+=0$ while spanning the \emph{full} phase space. For instance, when we compute the inner product of two covariant vertex operators of the form (\ref{eq:V(w)DDF}), we may take one vertex to be constructed out of DDF operators with $q^{'-}\neq0$, $q^{'i}=q^{'+}=0$ and a vacuum with momentum $p^{'\mu}$ and the other to be constructed from DDF operators with $q^{-}\neq0$, $q^{i}=q^{+}=0$ and a vacuum with momentum $p^{\mu}$. The important point is now that 
$$
q\cdot q'=0,
$$
and it is due to this fact that $[A_n^{'i},A^j_m]=n\delta^{ij}\delta_{n+m,0}$, with $A_n^{'i}$ and $A_n^{i}$ the DDF operators constructed out of $q'$ and $q$ respectively. Therefore, different vertex operators can be constructed out of different $q^{\mu}$ provided $q\cdot q'=0$, which in the coordinate system shown above is equivalent to saying that different vertex operators can be labelled by $\{{\bf p},p^+\}$, which can be taken to be independent for every vertex operator, as required.

In the next two sections we summarize what we have learnt and fill in the details on some of the finer points. We first discuss the closed string and then the modifications required for the open string.

\subsection{Closed String Mass Eigenstates}
As discussed above, the DDF formalism provides a dictionary which relates every light-cone gauge state to the corresponding covariant gauge vertex operator. Writing $N=\sum_jn_j$ and $\bar{N}=\sum_j\bar{n}_j$ with $N=\bar{N}$, a general light-cone gauge mass eigenstate state is of the form
\begin{equation}\label{eq:lc state}
\begin{aligned}
|V&\rangle_{\rm lc}=\,\frac{1}{\sqrt{2p^+\mathcal{V}_{d-1}}}\,C\xi_{ij\dots,kl\dots}\\
&\times\alpha^{i}_{-n_1}\alpha_{-n_2}^{j}\dots\tilde{\alpha}_{-\bar{n}_1}^{k}\tilde{\alpha}_{-\bar{n}_2}^{l}\dots |0,0;p^+,p^i\rangle,
\end{aligned}
\end{equation}
with $|0,0;p^+,p^i\rangle$ an eigenstate of $p^+,p^i$ and annihilated by the (dimensionless) lowering operators $\alpha^i_{n>0}$, $\tilde{\alpha}^i_{n>0}$, normalized according to (\ref{eq:tach and vac norm lcc}). If the polarization tensor $\xi_{ij\dots\,,kl\dots}$ is normalized to unity,
$$
\xi_{ij\dots\,,kl\dots}\xi^{ij\dots\,,kl\dots}=1,
$$ 
then the combinatorial normalization constant, $C$, contains  \cite{Polchinski_v1} a factor of $\frac{1}{\sqrt{n}}$ for every $\alpha_{-n}^i$ that appears and factors of $\frac{1}{\sqrt{\mu_{n,i}!}}$, with $\mu_{n,i}$ the multiplicity of $\alpha^i_n$ in the above product. Similar factors are required for the anti-holomorphic sector; in total \footnote{The constant $C$ should not be confused with that obtained in the previous sections. Throughout the rest of the section $C$ will be defined according to (\ref{norm const C}). For coherent states (in later sections) $C$ will again be different.},
\begin{equation}\label{norm const C}
C\equiv \frac{1}{\sqrt{\prod_rn_r\prod_{n,i}\mu_{n,i}!}}\times \frac{1}{\sqrt{\prod_s\bar{n}_s\prod_{\bar{n},i}\bar{\mu}_{\bar{n},i}!}}.
\end{equation}
To every light-cone gauge state (\ref{eq:lc state}) there corresponds \cite{D'HokerGiddings87} the correctly normalized covariant vertex operator of momentum $k$,
\begin{equation}\label{eq:DDF state}
\begin{aligned}
V(z,&\bar{z})=\,\frac{g_c}{\sqrt{2p^+\mathcal{V}_{d-1}}}\,C\xi_{ij\dots,kl\dots}\\
&\times A^{i}_{-n_1} A^j_{-n_2}\dots\bar{A}_{-\bar{n}_1}^k\bar{A}_{-\bar{n}_2}^l\dots\,e^{ip\cdot X(z,\bar{z})},
\end{aligned}
\end{equation}
with the (dimensionless) DDF operators, $A_n^i$, $\bar{A}_n^i$, defined by,
\begin{equation}\label{eq: DDF As}
\begin{aligned}
&A_n^i\equiv \sqrt{\frac{2}{\alpha'}}\oint \dslash z\,\partial_z X^i(z)e^{inq\cdot X(z)},\\
&\bar{A}_n^i\equiv \sqrt{\frac{2}{\alpha'}}\oint  \dslash \bar{z}\,\partial_{\bar{z}} X^i(\bar{z})e^{inq\cdot X(\bar{z})}.
\end{aligned}
\end{equation}
The indices $i$ are understood to be transverse to $q^{\mu}$. In accordance with the above considerations the null spacetime vector $q^{\mu}$ and the (tachyonic) vacuum momentum $p^{\mu}$ are such that,
\begin{equation}\label{eq:pq cond_closed}
p^2=\frac{4}{\alpha'},\qquad p\cdot q=\frac{2}{\alpha'},\qquad {\rm and}\qquad q^2=0.
\end{equation}
The quantity $k^{\mu}\equiv p^{\mu}-Nq^{\mu}$, as discussed above is identified with the momentum of the vertex operator (\ref{eq:DDF state}): from the definitions of $p$ and $q$ we can confirm that the mass shell condition is automatically satisfied if $N$ is identified with the level number, $N=\sum_in_i$,
\begin{equation}\label{eq:mass-shell_DDF}
k^{\mu} = p^{\mu}-Nq^{\mu},\qquad {\rm and}\qquad k^2 =\frac{4}{\alpha'}(1 -N).
\end{equation}

As an example, it is also useful to note that one can always Lorentz boost to a frame where (for simplicity here $\alpha'=2$),
\begin{equation}\label{eq:p,q;Q}
\begin{aligned}
&p=\big(c-1/(2c),0,\dots,0,c+1/(2c)\big),\\
&q=\big(c,0,\dots,0,c\big),
\end{aligned}
\end{equation}
given that these satisfy $p^2=2$, $p\cdot q=1$ and $q^2=0$ as required for any $c$, see Sec.~\ref{DDF}. As an example, let us boost to the rest frame where the $k^i=0$ and $k^0=\sqrt{2N-2}$. $p$ and $q$ are determined completely, with $c^{-1}=-\sqrt{2 N-2}$.

The vertex (\ref{eq:DDF state}) is not yet normal ordered and can be brought into a manifestly normal ordered form by bringing the operators in the integrands close to the vacuum, summing over all Wick contractions using the standard sphere two-point function for scalars,
\begin{equation}\label{eq:<XX>_cl}
\big\langle X^{\mu}(z,\bar{z})X^{\nu}(w,\bar{w})\big\rangle = -\frac{\alpha'}{2}\eta^{\mu\nu}\ln|z-w|^2,
\end{equation}
and evaluating the resulting contour integrals so as to extract the residues which correspond to the physical states. The contour integrals in (\ref{eq:DDF state}) are to contain the ground-state vacuum.  We are to bring the rightmost operators close to the vacuum first so as to respect the order with which these hit the vacuum. When the right-most DDF operator is brought close to the vacuum we evaluate the associated contour integral (with all other insertions placed outside the contour). We then bring the next DDF operator close to the resulting object, evaluate the operator products and the associated contour integral and so on, see Fig.~\ref{fig:DDFop}. The procedure is analogous to the usual procedure of extracting vertex operators from Fock space states \cite{SasakiYamanaka85}.

Using the operator product interpretation of the commutators\label{eq:commutatots} (see Appendix \ref{C}) it is seen that the DDF operators satisfy an oscillator algebra and annihilate the vacuum when $n>0$ in direct analogy with the corresponding oscillators $\alpha_{n}$ and $\tilde{\alpha}_n$,
\begin{equation}\label{eq: [A_nA_m]}
\begin{aligned}
&\big[A_n^i,A_m^j\big]\cong n\delta^{ij}\delta_{n+m,0},\quad{\rm and} \quad A_{n>0}^i\cdot e^{ip\cdot X(z,\bar{z})}\cong0,
\end{aligned}
\end{equation}
In addition, they commute with the Virasoro generators \footnote{\label{VirasoroGen}Recall that the Virasoro generators read, $L_n=\oint \frac{dz}{2\pi i} z^{n+1}\big(-\frac{1}{\alpha'}\,\partial X\cdot \partial X\big)$,  and $\bar{L}_n=\oint \frac{d\bar{z}}{2\pi i} \bar{z}^{n+1}\big(-\frac{1}{\alpha'}\,\bar{\partial} X\cdot \bar{\partial} X\big).$}, $L_m\cdot A_n\cong \bar{L}_m\cdot \bar{A}_n\cong \bar{L}_m\cdot A_n\cong L_m\cdot \bar{A}_n\cong 0,$ for all $m,n\in \mathbb{Z}$ and the (tachyonic) vacuum on which the DDF operators act has conformal dimension $(1,1)$ and is therefore an $L_0$, $\bar{L}_0$ eigenstate, $L_0\cdot e^{ip\cdot X(z,\bar{z})}\cong\bar{L}_0\cdot e^{ip\cdot X(z,\bar{z})}\cong e^{ip\cdot X(z,\bar{z})}$. It follows that $V(z,\bar{z})$ is a physical vertex operator given that, $(L_0-1)\cdot V(z,\bar{z})\cong0$, $L_{m>0}\cdot V(z,\bar{z})\cong 0,$
 and $(\bar{L}_0-1)\cdot V(z,\bar{z})\cong0$, $\bar{L}_{m>0}\cdot V(z,\bar{z})\cong0$.

An important point that can be mentioned here is that level matching, $(L_0-\bar{L}_0)\cdot V(z,\bar{z})\cong0$, is satisfied even for states with asymmetrically excited left- and right-movers, one such state being e.g.~$V(z,\bar{z})=\xi_{i,j} A^i_{-n} \bar{A}^j_{-m} e^{ip\cdot X(z,\bar{z})}$ with $n\neq m$ and positive. In fact, when we normal order this expression it will be seen that the presence of such states requires a lightlike compactification of spacetime -- we will have more to say about this later on when we discuss covariant coherent states for closed strings.

We suggest that the states (\ref{eq:lc state}) and (\ref{eq:DDF state}) are different descriptions of the same state. This is supported from various points of view: (a) there is a one-to-one correspondence between (\ref{eq:lc state}) and (\ref{eq:DDF state}), and the lightcone gauge states (\ref{eq:lc state}) describe a complete set of states for the bosonic string; (b) the lightcone and covariant expressions have the same mass and angular momenta; (c) the first mass level states are identical. We \emph{conjecture} and work on the assumption that the lightcone and covariant states share identical correlation functions (provided these are gauge invariant).

As discussed above, that (\ref{eq:DDF state}) is covariant is not manifest due to the explicit presence of transverse indices. However, when the operator products and contour integrals are carried out the resulting object can be given a manifestly covariant form \cite{D'HokerGiddings87} -- we will show this explicitly with a couple of examples (and in particular, (\ref{eq:cov_graviton}) and (\ref{eq:ManiCovVert})).

In the next section we fill in the details for the open string covariant vertex operator construction before discussing the normal ordered expression of the closed string vertex operators.

\subsection{Open String Mass Eigenstates}\label{DDFop}
The open string vertex operator construction proceeds in a similar manner, but there are certain differences that we mention here. First of all note that our open string conventions are presented in Appendix \ref{O}. We restrict our attention to open strings with both ends attached to a single D$p$-brane (with $p\geq1$ \cite{Giddings96}), although such vertex operators are also relevant in scattering amplitude computations involving open  string vertices stretched between parallel D$p$-branes, the so called $p$-$p$ strings. The construction may be generalized to $p$-$p'$ string vertex operators that stretch between a D$p$- and a D$p'$-brane along the lines of \cite{Hashimoto96} by making use of the notion of a twist field. 

Consider the case of $p$-$p$ vertex operators where a string worldsheet is attached to two parallel D$p$-branes. In a direction transverse to the brane the string satisfies Dirichlet boundary conditions \cite{Polchinski96},
$$X|_{\partial \Sigma} = x(s),$$
with $x(s)$ parametrizing the boundary of the worldsheet, $\Sigma$, which is fixed to the brane. For a worldsheet conformally transformed to the upper half plane with the boundary on the real axis, an example would be a vertex inserted on the real axis at ${\rm Im}\,z=0$ and ${\rm Re}\,z=y$, in which case the Dirichlet boundary conditions become,
\begin{equation*}
\begin{aligned}
X = 0\qquad {\rm for}&\qquad {\rm Im}\,z=0\,\,\,\,{\rm Re}\,z <y,\\
X = L\qquad {\rm for}&\qquad {\rm Im}\,z=0\,\,\,\,{\rm Re}\,z >y,
\end{aligned}
\end{equation*}
for the two parallel branes separated by a distance $L$. A useful formula has been given in \cite{Giddings96} for the functional integral,
\begin{widetext}
\begin{equation}\label{eq:DirichletPathIntegrals}
\begin{aligned}
\int_{X|_{\partial\Sigma}=x(z)}\mathcal{D}&Xe^{-S[X]}\dots \\
&= \int_{X|_{\partial\Sigma}=0}\mathcal{D}Xe^{-S[X]}\exp\bigg\{\frac{1}{(2\pi\alpha')^2}\oint_{\partial \Sigma} ds\oint_{\partial \Sigma} ds'x(s)x(s')\partial_{\perp}\partial_{\perp}'G_{\rm D}(z,z')\bigg\}\dots,
\end{aligned}
\end{equation}
\end{widetext}
with $S$ the Polyakov action, the normal derivatives $\partial_{\perp}$  acting on the Green's function with Dirichlet boundary conditions, $G_{\rm D}(z,z')=\langle X(z,\bar{z})X(z',\bar{z}')\rangle$ with the normalization convention $\partial_z\partial_{\bar{z}}G(z,w) = -\pi\alpha^{\prime}\delta^2(z-w)+\frac{\pi\alpha^{\prime}g_{z\bar{z}}}{\int_{\Sigma}d^2z\sqrt{g}}$ and $G_{\rm D}(z,z')|_{z\in\partial \Sigma}=0$, and the dots ``\dots'' denoting vertex operator insertions. This expression shows \cite{Giddings96} that we may restrict our attention to the construction of vertex operators with both ends attached to a single brane, say at $X^i|_{\partial\Sigma}=0$, keeping in mind that one is to include the above exponential factor as appropriate for $p$-$p$ strings stretching between parallel branes in the various scattering amplitude computations.

Spacetime directions tangent to the D$p$-brane are labelled by lower case latin letters from the beginning of the alphabet, $X^a$, with $a=0,\dots,p$, and directions transverse to the brane by upper case latin letters from the middle of the alphabet, $X^I$, with $I=p+1,\dots 25$.  It is sometimes useful to work in lightcone coordinates in both covariant and lightcone gauge as this enables us to make the correspondence between the two gauges explicit. Assuming the associated lightcone directions satisfy Neumann boundary conditions we may define, $$X^{\pm}=\tfrac{1}{\sqrt{2}}\big(X^0\pm X^{p}\big).$$ Note that it is necessary \cite{Giddings96} for the $X^{\pm}$ directions to lie in the Neumann directions in order to make the correspondence with lightcone gauge for which $X^+= (2\alpha') p^+\tau_{\rm M}$, with $\tau=i\tau_{\rm M}$, as this is not compatible with Dirichlet boundary conditions, see (\ref{eq:ND BC's}). To place the lightcone directions in the Dirichlet directions one needs to instead reformulate lightcone gauge quantization with $X^+= (2\alpha') p^+\sigma$. A general spacetime direction is as always labelled by Greek lower case letters, $X^{\mu}$. To summarize,
\begin{equation}\label{eq:indices open}
\begin{aligned}
&X^{a}=\{X^{\pm},X^A\}, \qquad{\rm  with}\qquad A=1,\dots,p-1,\\
&X^{i}\,=\{X^A,X^I\}, \,\qquad{\rm  with} \qquad I=p+1,\dots,25,\\
&X^{\mu}=\{X^{\pm},X^i\}.
\end{aligned}
\end{equation}
and so the directions $X^A$ satisfy Neumann boundary conditions, whereas directions $X^I$ satisfy Dirichlet boundary conditions. In the Euclidean worldsheet coordinates, $z=e^{-i(\sigma+i\tau)}$, $\bar{z}=e^{i(\sigma-i\tau)}$ with $\sigma\in [0,\pi]$ and $\tau\in(-\infty,\infty)$, (considering only the case of NN and DD strings) Neumann and Dirichlet boundary conditions read respectively, 
\begin{equation}\label{eq:ND BC's}
N:\quad\partial_{\sigma}X^a|_{\partial \Sigma_{1,2}}=0\quad {\rm and}\quad D:\quad\partial_{\tau}X^I|_{\partial \Sigma_{1,2}}=0.
\end{equation}
Note that, $\partial_{\sigma}=i(\bar{z}\bar{\partial}-z\partial)$ and $\partial_{\tau}=\bar{z}\bar{\partial}+z\partial$. In the $(z,\bar{z})$ coordinates the open string physical worldsheet, $\Sigma$, is conformally mapped to the upper half plane with the identification, $z\sim \bar{z}$. The associated fixed point, the real line $z=\bar{z}$, defines the open string boundaries.

Using the doubling trick we can as usual write the various expressions needed in terms of holomorphic quantities only \cite{Polchinski_v1}: one identifies antiholomorphic quantities in the upper half plane with holomorphic quantities in the lower half plane and therefore one may just as well work with holomorphic quantities only provided one works in the full complex plane. The open string vertex operators are inserted on the real axis. We assume that both ends of the string satisfy the same boundary conditions for any given direction, we thus consider the cases of NN and DD directions only and do not consider mixed boundary conditions ND and DN. 

The relevant DDF operators now read,
\begin{equation}\label{eq: DDF A op}
\begin{aligned}
&A_n^A=\sqrt{\frac{2}{\alpha'}}\oint \dslash z\,\partial X^A(z)e^{inq\cdot X(z)},\\
&A_n^I=\sqrt{\frac{2}{\alpha'}}\oint \dslash z\,\partial X^I(z)e^{inq\cdot X(z)},
\end{aligned}
\end{equation}
for oscillators parallel or transverse to the brane respectively and the closed contour integrals are to contain the operators they act on, which are on the real axis. In a Minkowski signature worldsheet the integrals are along the boundary of the worldsheet which is coincident with the D$p$-brane. The null vectors $q^{\mu}$ are restricted to lie within the D-brane worldvolume and are transverse to the DDF operators: $$q^A=q^I=0.$$  In direct analogy to the closed string case we create open string vertex operators with fluctuations in the $X^A$ or $X^I$ directions by acting on the vacuum with DDF operators see also Appendix \ref{O}),
\begin{equation}\label{eq:DDF state op}
V(z,\bar{z})=\frac{g_o}{\sqrt{2p^+\mathcal{V}_{d-1}}}\,C\xi_{ij\dots}A^i_{-n_1} A^j_{-n_2}\dots\,e^{ip\cdot X(z)},
\end{equation}
the vacuum, $e^{ip\cdot X(z)}$ being restricted to the worldsheet boundary (e.g.~the real axis in the complex $z$-plane) and the combinatorial normalization constant $C$,
\begin{equation}\label{norm const C op}
C\equiv \frac{1}{\sqrt{\prod_rn_r\prod_{n,i}\mu_{n,i}!}}.
\end{equation}
The vertex operators (\ref{eq:DDF state op}) are mass level $N=\sum_in_i$ states with momenta $k^{\mu}=p^{\mu}-Nq^{\mu}$, the onshell constraints now reading,
\begin{equation}\label{eq:pq cond_open}
p^2=\frac{1}{\alpha'},\qquad p\cdot q=\frac{1}{2\alpha'},\qquad {\rm and}\qquad q^2=0,
\end{equation}
so as to ensure that $m^2=-(p-Nq)^2=(N-1)/\alpha'$ as appropriate for open strings. The contractions appearing in (\ref{eq:pq cond_open}) are with respect to all spacetime indices $\mu$. The boundary conditions require in addition, $p^I=0$, see Appendix \ref{O}.

Normal ordered vertex operators are obtained from (\ref{eq:DDF state op}) by bringing the operators in the integrands close to the vacuum, summing over all Wick contractions using e.g.~the upper half plane two-point function for scalars (given in (\ref{eq:<XX>_op apx}) in Appendix \ref{O} for completeness) for Neumann (N) or Dirichlet (D) directions, and evaluating the resulting contour integrals so as to extract the residues which correspond to the physical states. In evaluating the operator products we are to restrict the integrands of the DDF operators to the real axis
. Only after the operator products have been computed are we to analytically continue in the variable of integration so as to circle the tachyonic vacuum in order to extract the residue. This is best understood by realizing that the vertex operator (\ref{eq:DDF state op}) can be thought of as being created in a sequence of open string scattering events as explained in the introduction and depicted in Fig.~\ref{fig:DDFop}. 

The massless states, $V_{\rm massless}^{(i)}$, that are absorbed by the ground state string, $V_{\rm ground\,\,state}=e^{ip\cdot X(z)}$, are the integrands of the DDF operators polarized in some direction, $\xi^i$, of our choice, and the final excited state $V_{\rm excited}^{(r)}$ is given by the vertex operator (\ref{eq:DDF state op}) after normal ordering, when a sequence of $r$ DDF operators have acted on the vacuum. In what follows we compute this normal ordered expression for a complete set of such open string covariant vertex operators. We give explicit results for the closed string and consider the open string explicitly when we construct coherent states. Open string vertices constructed from the $A^A_n$ operators are related by T-duality to vertices constructed out the $A^I_n$ \cite{NairShapereStromingerWilczek87,DaiLeighPolchinski89,Polchinski96}. The latter are interpreted as ripples in the D-brane worldvolume. The remaining possibility is vertex operators with excitations associated to both transverse and tangent directions to the D-brane, and these may be interpreted as the usual Neumann boundary condition vertices with excitations within the D-brane worldvolume which in addition generate ripples of the D-brane.  In the open string coherent state section we shall consider vertices constructed from the $A^A_n$. 

As in the closed string case there is a one-to-one correspondence with the lightcone gauge states,
\begin{equation}\label{eq:lc state op}
|V\rangle_{\rm lc}=\frac{1}{\sqrt{2p^+\mathcal{V}_{d-1}}}\,C\xi_{ij\dots}\,\alpha^i_{-n_1}\alpha_{-n_2}^j\dots |0;p^+,p^i\rangle,
\end{equation}
with $|0;p^+,p^i\rangle$ an eigenstate of $p^+,p^i$ and annihilated by the (dimensionless) lowering operators, $\alpha^i_{n>0}$, where $$\alpha^{\mu}_{n}=\sqrt{\frac{2}{\alpha'}}\oint \dslash z\,\partial X^{\mu}(z)\,z^n.$$ 
 and defined so that (\ref{eq:tach and vac norm lcc}) holds true.

The fact that the covariant gauge vertex operators (\ref{eq:DDF state op}) are in one- to one correspondence with the lightcone gauge states (\ref{eq:lc state op}) proves that the former comprise a complete set. We conjecture and work on the assumption that the states $|V\rangle_{\rm lc}$ and $V(z,\bar{z})$ are identical states in the sense that they share identical masses, angular momenta and interactions. We shall obtain evidence supporting this conjecture as we go along. 

We next discuss the correspondence between lightcone gauge states and covariant gauge vertex operators, and consider the issue of normal ordering in detail.  We start from the graviton and subsequently move on to arbitrarily excited vertex operators.

\subsection{The covariant equivalent of $\alpha_{-1}^{i}\tilde{\alpha}_{-1}^{j}|0,0;p^+,p^i\rangle$}
We wish to obtain the covariant equivalent of the lightcone gauge graviton (or other massless) state,
$$
|V\rangle_{\rm lc} =\frac{1}{\sqrt{2p^+\mathcal{V}_{d-1}}}\, \xi_{i,j}\,\alpha_{-1}^{i}\tilde{\alpha}_{-1}^{j}|0,0;p^+,p^i\rangle.
$$
Here $m^2=0$, and so from (\ref{eq:mass-shell_DDF}) $k^{\mu}= p^{\mu}-q^{\mu}$. We see from (\ref{eq:lc state}) and (\ref{eq:DDF state}) that the light-cone to covariant vertex map is realized by:
\begin{equation}\label{eq:grav_lc->ddf}
\xi_{i,j}\,\alpha_{-1}^{i}\tilde{\alpha}_{-1}^{j}|0,0;p^+,p^i\rangle\,\,\rightarrow\,\, \xi_{i,j}\,A_{-1}^{i}\bar{A}_{-1}^{j}e^{ip\cdot X(z,\bar{z})},
\end{equation}
with $\xi\cdot q\equiv 0$. To bring this into a manifestly covariant form we substitute into the right hand side  the definitions (\ref{eq:DDF state}). Using the operator products we bring the integrands close to the vacuum and evaluate the resulting contour integrals as explained below (\ref{eq: DDF As}). For the graviton this procedure can be seen to lead to \footnote{We use the convention $X(z,\bar{z})=X(z)+X(\bar{z})$ which can be used inside correlation functions in the absence of sources \cite{Polchinski87}.}:
\begin{widetext}
\begin{equation}
\begin{aligned}
\xi_{i,j}\,A_{-1}^{i}\bar{A}_{-1}^{j}e^{ip\cdot X(z,\bar{z})}&=\frac{2}{\alpha'}\, \xi_{i,j}\oint_{z}\dslash w\,\partial_w X^i(w)e^{-iq\cdot X(w)}\,\oint_{\bar{z}} \dslash \bar{w}\,\,\partial_{\bar{w}} X^j(\bar{w})e^{-iq\cdot X(\bar{w})}\,\,e^{ip\cdot X(z,\bar{z})}\\
&\cong \frac{2}{\alpha'}\,\xi_{i,j}\Big(\delta^i_{\mu}-\frac{\alpha'}{2}\,p^iq_{\mu}\Big)\Big(\delta^j_{\nu}-\frac{\alpha'}{2}\,p^jq_{\nu}\Big)\,\partial X^{\mu}(z)\bar{\partial}X^{\nu}(\bar{z})e^{i(p-q)\cdot X(z,\bar{z})}.\phantom{\int}
\end{aligned}
\end{equation}
\end{widetext}
With the identification $\zeta_{\mu,\nu}=\xi_{i,j}(\delta^i_{\mu}-\frac{\alpha'}{2}p^iq_{\mu})(\delta^j_{\nu}-\frac{\alpha'}{2}p^jq_{\nu})$, we find the manifestly covariant and normal-ordered expression for the graviton vertex  \cite{D'HokerGiddings87},
\begin{equation}\label{eq:cov_graviton}
V(z,\bar{z})=\frac{g_c}{\sqrt{2p^+\mathcal{V}_{d-1}}}\,\frac{2}{\alpha'}\,\zeta_{\mu,\nu}\,\partial X^{\mu}(z)\bar{\partial}X^{\nu}(\bar{z})e^{ik\cdot X(z,\bar{z})},
\end{equation}
which has been derived from the corresponding light-cone gauge graviton via the DDF formalism. Note that we could just as well have written $\frac{g_c}{\sqrt{2E_{\bf k}V_{d-1}}}$ (with $E_{\bf k}=|{\bf k}|$) instead of $\frac{g_c}{\sqrt{2p^+\mathcal{V}_{d-1}}}$, provided the momentum phase space in $S$-matrix elements is taken to be (\ref{eq:Vdp}) instead of $(\ref{eq:Vdp lightcone})$, as discussed in Sec.~\ref{sec:NSW}. This remark applies also to the other mass eigenstate vertex operators given below as well, but does not apply in the case of coherent states (see later).

The polarization tensor $\zeta_{\mu,\nu}$ is transverse to the graviton momentum $k^{\mu}$ as can be explicitly verified \footnote{Recall that $\xi_{i,j}$ is transverse to $q^{\mu}$.}.  Notice that depending on our choice of $\xi$, $p$ and $q$ all entries of the covariant polarization tensor, $\zeta_{\mu,\nu}$, may be non-vanishing in general. Whether or not the corresponding polarization tensor is traceless depends on our choice of $\xi_{i,j}$.

The above procedure generalizes to arbitrarily massive vertices and given that the DDF operators generate the complete set of physical states \cite{Brower72,GoddardThorn72} it is clear that all arbitrarily massive vertices in covariant gauge may be extracted via this method. The fact that the physical content of the light-cone gauge states (where there are no ghost excitations) is clearer than covariant gauge vertex operators has been one of the great virtues of the light-cone gauge approach -- it is seen that this virtue is also present in the covariant gauge if one makes use of the DDF formalism.

\subsection{The covariant equivalent of $\alpha^i_{-N}\tilde{\alpha}_{-N}^j|0,0;p^+,p^i\rangle$}
Consider now a not so obvious example which in fact, as will become apparent in the next subsection, is the basic building block of all vertex operators whose polarization tensors are traceless. In this subsection we derive the normal ordered covariant vertex operator corresponding to the lightcone state 
\begin{equation}\label{eq:lc |V> with A_-NAbarA_-N}
|V\rangle_{\rm lc} =\frac{1}{\sqrt{2p^+\mathcal{V}_{d-1}}}\, \frac{1}{N}\,\xi_{i,j}\,\alpha^i_{-N}\tilde{\alpha}_{-N}^j|0,0;p^+,p^i\rangle,
\end{equation}
with the normalization $C=1/N$, see (\ref{norm const C}). Here the mass, $m^2=4(N-1)/\alpha'$, and so from (\ref{eq:mass-shell_DDF}), $k^{\mu}= p^{\mu}-Nq^{\mu}$. Following the DDF prescription, we consider the state
\begin{equation}\label{eq:nn DDF map}
V(z,\bar{z})=\frac{g_c}{\sqrt{2p^+\mathcal{V}_{d-1}}}\,\frac{1}{N}\,\xi_{i,j}\,A_{-N}^{i}\bar{A}_{-N}^{j}e^{ip\cdot X(z,\bar{z})}.
\end{equation}
As in the graviton example, we use the definitions of the DDF operators and carry out the relevant operator products. Let us consider the holomorphic sector and shift the vertex to $z=0$. This leads us to consider,
\begin{widetext}
\begin{equation}\label{eq:A_-N DDF}
\begin{aligned}
A_{-N}^{i}\cdot e^{ip\cdot X(0)}&=\sqrt{\frac{2}{\alpha'}} \oint_{0}\dslash w\,\partial X^i(w)e^{-iNq\cdot X(w)}\cdot e^{ip\cdot X(0)}\\
&\cong \sqrt{\frac{2}{\alpha'}}\oint_0 \frac{\dslash w}{iw}\Big{(}p^i\,w^{-N}+\sum_{r=1}^{\infty}\frac{i}{(r-1)!}\,\partial^rX^i(0)\,w^{r-N}\Big{)}\sum_{m=0}^{\infty}w^{m}S_m(Nq;0)\,e^{i(p-Nq)\cdot X(0)}\\
&= \sqrt{\frac{2}{\alpha'}}\Big(\frac{\alpha'}{2}p^iS_N(Nq;0)+\sum_{m=1}^{N}\frac{i}{(m-1)!}\,\partial^mX^i(0)S_{N-m}(Nq;0)\Big)e^{i(p-Nq)\cdot X(0)},
\end{aligned}
\end{equation}
\end{widetext}
with the definition, 
\begin{equation}\label{eq:Sm(nqz)}
S_m(nq;z)\equiv S_m(a_1,\dots,a_m),
\end{equation}
when the following identification is made,
$$
a_s= -\frac{inq\cdot \partial_z^sX}{s!}.
$$ 
The elementary Schur (or complete Bell) polynomials, $S_m(a_1,\dots,a_m)$, are in turn defined in general by:
\begin{subequations}\label{eq:S_m(a)mt}
\begin{align}
S_m(a_1,\dots,a_m)&\equiv-i\oint_0\dslash u\,u^{-m-1} \exp \sum_{s=1}^{m}a_su^s\label{eq:S_m(a)b}\\
&=\!\!\sum_{k_1+2k_2+\dots+mk_m=m}\!\!\frac{a_1^{k_1}}{k_1!}\dots\frac{a_m^{k_m}}{k_m!}.\label{eq:S_m(a)a}
\end{align}
\end{subequations}
Similar remarks hold for the anti-holomorphic sector, see Appendix \ref{SP}. Note that $\oint_0 \frac{dw}{2\pi iw}=-\oint_0 \frac{d\bar{w}}{2\pi i\bar{w}}=1$, 
and we have made use of the standard correlator on the complex plane (\ref{eq:<XX>_cl}), as well as the onshell constraints (\ref{eq:mass-shell_DDF}). The elementary Schur polynomials arise from the Taylor expansion (inside the normal ordering) of $e^{-iNq\cdot X(z)}=\sum_{m=0}^{\infty}z^mS_m{(Nq;0)}e^{-iNq\cdot X(0)}$ which can be derived from Fa\`a di Bruno's formula  \cite{Riordan58} for the $m^{\rm th}$ derivative of the exponential, $(e^{iNq\cdot X(z)}\partial^m e^{-iNq\cdot X(z)})_{z=0}$. As a preliminary consistency check note that the subscript $N$ on $S_N(Nq)$ denotes the total number of derivatives and so the level number on both sides of the equation is the same. We have noted also the corresponding expression, $\bar{S}_m(nq;\bar{z})$, for the anti-holomorphic sector. Shifting the insertion back to $z,\bar{z}$ we conclude that the level $N$  lightcone state $\frac{1}{N}\xi_{i,j}\,\alpha^i_{-N}\tilde{\alpha}_{-N}^j|0,0;p^+,p^i\rangle$  has the covariant manifestation:
\begin{equation}\label{eq:A_-NA_-N DDF}
\begin{aligned}
V(z,\bar{z})=&\,\frac{g_c}{\sqrt{2p^+\mathcal{V}_{d-1}}}\,:\frac{1}{N}\,\xi_{i,j}H_N^i(z)\bar{H}_N^j(\bar{z})e^{i(p-Nq)\cdot X(z,\bar{z})}:
\end{aligned}
\end{equation}
We have found it convenient to define the polynomials $H_N^i(z)$, $\bar{H}_N^i(\bar{z})$, in $\partial^{\#}X$ and $\bar{\partial}^{\#}X$ respectively,
\begin{subequations}\label{eq:H_n}
\begin{align}
H_N^i(z)&\equiv \sqrt{\frac{\alpha'}{2}}p^i S_{N}(Nq;z)+P_N^i(z),\phantom{\bigg|}\\ 
\bar{H}_N^i(\bar{z})&\equiv \sqrt{\frac{\alpha'}{2}}p^i \bar{S}_{N}(Nq;\bar{z})+\bar{P}_N^i(\bar{z}),
\end{align}
\end{subequations}
with $P_N^i(z)$, $\bar{P}_N^i(\bar{z})$ in turn defined by,
\begin{subequations}\label{eq:PPbardfn}
\begin{align}
P_N^i(z)&=\sqrt{\frac{2}{\alpha'}}\sum_{m=1}^{N}\frac{i}{(m-1)!}\, \partial^mX^i(z)S_{N-m}(Nq;z),\label{eq:Pn}\\
\bar{P}_N^i(\bar{z}) &=\sqrt{\frac{2}{\alpha'}}\sum_{m=1}^{N}\frac{i}{(m-1)!}\,\bar{\partial}^mX^i(\bar{z})\bar{S}_{N-m}(Nq;\bar{z}).\label{eq:Pnbar}
\end{align}
\end{subequations}
These polynomials are the fundamental building blocks of normal ordered covariant vertex operators when these correspond in lightcone gauge to a traceless state as we shall see \footnote{For vertices that correspond to lightcone states whose trace is non-vanishing there is an additional polynomial, $\mathbb{S}_{n,m}(z)$, see below. All these polynomials however are ultimately composed of elementary Schur polynomials, $S_m(nq;z)$.}. In the rest frame we are to replace, $H_N^i(z)$, $\bar{H}_N^i(\bar{z})$ with, $P_N^i(z)$, $\bar{P}_N^i(\bar{z})$, respectively as in this case the momenta, $k^{\mu}=p^{\mu}-Nq^{\mu}$, are transverse to the polarization tensors and consequently $\xi_{\dots i\dots}p^i=0$. Some examples for $N=0,1$ and 2 have been given in Appendix \ref{C}. We next give an explicit example for $m^2=4/\alpha'$, mass levels, where $N=2$, to illustrate that the vertices generated in this manner are the standard covariant vertex operators \cite{FriedanMartinecShenker86}, see also \cite{Weinberg85,SasakiYamanaka85, IchinoseSakita86,Nergiz94}, with polarization tensors that range over the entire range of spacetime indices. The difference to the traditional approach (taken in the above cited papers) is that here physical polarization tensors are automatically generated -- there are no additional constraints to be solved. First of all note that for $N=1$ we recover the graviton (or in general the massless) vertex operator(s) \footnote{Recall that in the CFT language there is no Ricci scalar in the dilaton vertex, see Polchinski \cite{Polchinski87}.}. For $N=2$, we have $k^{\mu}=p^{\mu}-2q^{\mu}$. The covariant vertex operator which is equivalent to the lightcone state $\frac{1}{\sqrt{2p^+\mathcal{V}_{d-1}}}\,\frac{1}{2}\xi_{i,j}\,\alpha^i_{-2}\tilde{\alpha}_{-2}^j|0,0;p^+,p^i\rangle$ follows as a corollary of (\ref{eq:A_-NA_-N DDF}),
\begin{equation}\label{eq:ManiCovVert}
\begin{aligned}
|V\rangle=&\,\frac{1}{\sqrt{2E_{\bf k}V_{d-1}}}\,\frac{1}{2}\big(\chi_{\mu\nu}\alpha_{-1}^{\mu}\alpha_{-1}^{\nu}+\zeta_{\mu}\alpha_{-2}^{\mu}\big)\\
&\quad\times\big(\bar{\chi}_{\rho\sigma}\tilde{\alpha}_{-1}^{\rho} \tilde{\alpha}_{-1}^{\sigma}+\bar{\zeta}_{\rho}\tilde{\alpha}_{-2}^{\rho}\big)|0,0;k^{\mu}\rangle,
\end{aligned}
\end{equation}
where we  have made use of the operator-state correspondence, $\alpha_{-n}^{\mu}\simeq \sqrt{\frac{2}{\alpha'}}\frac{i}{(n-1)!}\partial^nX^{\mu}(z)$, $|0,0;k^{\mu}\rangle\simeq g_c\,e^{ik\cdot X(z,\bar{z})}$, and have written $|V\rangle\simeq V(z,\bar{z})$, in order to make manifest the differences to the equivalent lightcone gauge state. We have chosen to write (\ref{eq:ManiCovVert}) in the more conventional coordinates used in covariant gauge, where the vacuum is normalized according to (\ref{eq:tach and vac norm}) and $E_{\bf k}=\sqrt{{\bf k}^2+m^2}$. 
From (\ref{eq:A_-NA_-N DDF}) one can derive (by expanding out the various polynomials for $N=2$) the manifestly covariant polarization tensors, 

\begin{equation}\label{eq:polzetachi}
\begin{aligned}
&\zeta_{\mu}=\xi_i\big(\delta^i_{\phantom{i}\mu}-\tfrac{\alpha'}{2}p^iq_{\mu} \big)\\ &\chi_{\mu\nu}=\sqrt{\tfrac{\alpha'}{2}} \xi_i\big(\alpha' p^iq_{\mu}q_{\nu}-\delta^i_{\phantom{i}\mu}q_{\nu}-\delta^i_{\phantom{i}\nu}q_{\mu}\big),
\end{aligned}
\end{equation}
with the properties $|\zeta|^2=|\xi|^2$ (with $|\xi|^2=1$ so that the lightcone state is correctly normalized), $\chi_{\mu\nu}=\chi_{\nu\mu}$, $|\chi|^2=\zeta\cdot k=\chi^{\mu}_{\phantom{i}\mu}=\chi_{\mu\nu}k^{\mu}k^{\nu}=0$. As a consistency check note that these polarization tensors solve the physical state conditions, $2\zeta_{\mu}+k^{\nu}\chi_{\mu\nu}=0$, $2k_{\mu}\zeta^{\mu}+\eta^{\mu\nu}\chi_{\mu\nu}=0$, which were derived by completely different methods in  \cite{Nergiz94}. There are similar expressions for $\bar{\zeta}_{\mu}$, $\bar{\chi}_{\mu\nu}$ with $\bar{\xi}_i$ replacing $\xi_i$. One thing to notice is that all components of these polarization tensors may be non-vanishing in general so that the resulting states really are covariant in the usual sense even though the state (\ref{eq:nn DDF map}) from which (\ref{eq:ManiCovVert}) was derived seems to break spacetime covariance by the explicit choice of transverse indices.

There has been some confusion concerning a state of the form (\ref{eq:ManiCovVert}) in the literature \cite{SasakiYamanaka85,Nergiz94} where it is concluded that such a state may satisfy the Virasoro constraints but has zero norm. We disagree in that we find that the state $|V\rangle$ has positive norm \footnote{Here we have included the `one string in volume $V_{d-1}$' normalizing factor $\frac{1}{\sqrt{2E_{\bf k}V_{d-1}}}$ and use the relativistic normalization $\langle 0,0;k'|0,0;k\rangle = 2E_{\bf k}(2\pi)^{d-1}\delta^{d-1}({\bf k}'-{\bf k})$.}, $\langle V|V\rangle=1$, while satisfying all the Virasoro constraints, $L_{n>0}|V\rangle=0$, $L_0|V\rangle=|V\rangle$ and is hence physical. In fact, all covariant states generated by the DDF formalism are positive norm physical states. The reason as to why there is disagreement with \cite{SasakiYamanaka85,Nergiz94} is because the constraints on the polarization tensors $\zeta_{\mu}$, $\chi_{\mu\nu}$ obtained there do not have a unique solution; the solution identified there corresponds to a zero norm state but there is the additional solution, namely (\ref{eq:polzetachi}), which gives rise to the positive norm state (\ref{eq:ManiCovVert}).

What we learn from the above exercises is that the DDF vertex operators (\ref{eq:DDF state}) are fully covariant, they all have a lightcone gauge equivalent which can be identified explicitly, and last but not least they generate a complete set of physical states (given that they are in one-to-one correspondence with the light-cone gauge states).

\subsection{The covariant equivalent of $\alpha^{i}_{-n_1}\alpha_{-n_2}^{j}\dots\tilde{\alpha}_{-\bar{n}_1}^{k}\tilde{\alpha}_{-\bar{n}_2}^{l}\dots |0,0;p^+,p^i\rangle$}\label{arb_vert}
We next generalize the result of the previous subsection and discuss the covariant manifestation of a general lightcone gauge state, 
\begin{equation}\label{eq:lc |V> general}
\begin{aligned}
|V\rangle_{\rm lc}\,&=\frac{1}{\sqrt{2p^+\mathcal{V}_{d-1}}}\,C\xi_{ij\dots,kl\dots}\\
&\times\alpha^{i}_{-n_1}\alpha_{-n_2}^{j}\dots\tilde{\alpha}_{-\bar{n}_1}^{k}\tilde{\alpha}_{-\bar{n}_2}^{l}\dots |0,0;p^+,p^i\rangle,
\end{aligned}
\end{equation}
which according the DDF prescription is given by, 
\begin{equation}\label{eq:general DDF mass eigenst}
\begin{aligned}
V(z,\bar{z})&\,=\frac{g_c}{\sqrt{2p^+\mathcal{V}_{d-1}}}\,C \xi_{ij\dots,kl\dots}\\
&\times A^{i}_{-n_1}A_{-n_2}^{j}\dots\tilde{A}_{-\bar{n}_1}^{k}\tilde{A}_{-\bar{n}_2}^{l}\dots e^{ip\cdot X(z,\bar{z})}.
\end{aligned}
\end{equation}
Here the relevant level numbers associated to left- and right-moving modes are $N=\sum_{\ell}n_{\ell}$ and $\bar{N}=\sum_r\bar{n}_r$, and for non-compact spacetimes we are to enforce \footnote{The $L_0-\bar{L}_0$ Virasoro constraint is satisfied without the requirement $N=\bar{N}$ but as we discuss later this is only possible in a spacetime with lightlike compactification given that for $N\neq \bar{N}$ we have $k_{\rm L}-k_{\rm R}=-(N-\bar{N})q$ with $q^2=0$.} $N=\bar{N}$. The associated momentum is then, $k^{\mu}=p^{\mu}-Nq^{\mu}$, and the mass shell constraint, $k^2=4(1-N)/\alpha'$.

Writing formally $\xi_{ij\dots,kl\dots}=\xi_{ij\dots}\bar{\xi}_{kl\dots}$ we first consider the case when the polarization tensors $\xi$ and $\bar{\xi}$ are \emph{traceless}, $$\xi_{\dots i\dots j \dots }\eta^{ij}=\bar{\xi}_{\dots i\dots j \dots }\eta^{ij}=0,$$ but with $\xi_{\dots j\dots}k^j$, $\bar{\xi}_{\dots j\dots}k^j$ non-vanishing in general. The normal ordered vertex operator corresponds to a straightforward generalization of (\ref{eq:A_-NA_-N DDF}), $\prod_{r}A^{i_r}_{-n_r}e^{ip\cdot X(z)}\cong\prod_{r}H_{n_r}^{i_r}e^{i(p-Nq)\cdot X(z)}$ for the holomorphic sector. Therefore, the covariant normal ordered vertex operator associated to a general traceless lightcone state $C\xi_{i_1i_2\dots,j_1j_2\dots}\,\alpha^{i_1}_{-n_1}\alpha_{-n_2}^{i_2}\dots\tilde{\alpha}_{-\bar{n}_1}^{j_1}\tilde{\alpha}_{-\bar{n}_2}^{j_2}\dots |0,0;p^+,p^i\rangle$ is,
\begin{equation}\label{eq: many A_-N DDF}
\begin{aligned}
V&(z,\bar{z})\cong \frac{g_c}{\sqrt{2p^+\mathcal{V}_{d-1}}}\,\,:C\xi_{ij\dots,kl\dots}\\
&\times H^{i}_{n_1}(z) H^j_{n_2}(z)\dots\bar{H}_{\bar{n}_1}^k(\bar{z})\bar{H}_{\bar{n}_2}^l(\bar{z})\dots\,e^{i(p-Nq)\cdot X(z,\bar{z})}:
\end{aligned}
\end{equation}
with $C$ as given in (\ref{norm const C}). Without referring explicitly to the lightcone state we see that $C$ contains a factor of $\frac{1}{\sqrt{n}}$ for every $H_{n}^i$ that appears and factors of $\frac{1}{\sqrt{\mu_{n,i}!}}$, with $\mu_{n,i}$ the multiplicity of $H^i_n$. 

We can always boost to a frame where $\xi_{\dots i\dots} k^i=0$ (e.g.~the rest frame) given that there are no timelike directions in the lightcone gauge polarization tensor, $\xi$, in which case the above vertex simplifies to,
\begin{equation*}
\begin{aligned}
V(&z,\bar{z})\cong \,\,\frac{g_c}{\sqrt{2p^+\mathcal{V}_{d-1}}}:C\xi_{ij\dots,kl\dots}\\
&\times P^{i}_{n_1}(z) P^j_{n_2}(z)\dots\bar{P}_{\bar{n}_1}^k(\bar{z})\bar{P}_{\bar{n}_2}^l(\bar{z})\dots\,e^{i(p-Nq)\cdot X(z,\bar{z})}:
\end{aligned}
\end{equation*}

We therefore learn that when the polarization tensor of a given light-cone state is traceless we can build the corresponding normal ordered covariant vertex operator by making the following replacements,
\begin{equation}\label{eq: alpha map}
\begin{matrix} 
      \phantom{\Big(}\alpha_{-n}^i \\
      \phantom{\Big(}\tilde{\alpha}_{-\bar{n}}^i  \\
      \phantom{\Big(}|0,0;p^+,p^i\rangle  \\
\end{matrix}
\hspace{0.4cm}
\begin{matrix} 
      \phantom{\Big(}\rightarrow \\
      \phantom{\Big(}\rightarrow  \\
      \phantom{\Big(}\rightarrow  \\
\end{matrix} 
\hspace{0.4cm}
\begin{matrix} 
      \phantom{\Big(}H_n^i(z) \\
      \phantom{\Big(}\bar{H}_{\bar{n}}^i(\bar{z})  \\
      \phantom{\Big(}g_c\,e^{i(p-Nq)^{\mu} X_{\mu}(z,\bar{z})}  \\
\end{matrix} 
\end{equation}
with an overall combinatorial normalization constant $C$ given in (\ref{norm const C}), and the lightcone operator vacuum normalized as in (\ref{eq:lcc vac norm op}). If the lightcone states in addition to $\xi_{\dots i\dots j \dots }\eta^{ij}=0$ satisfy $\xi_{\dots j\dots}k^j=0$ (and similarly for the anti-holomorphic sector), the above identification simplifies to, $\alpha_{-n}^i\,\sim \, P_n^i(z)$ and $\tilde{\alpha}_{-n}^i\,\sim\, \bar{P}_n^i(\bar{z})$. The resulting covariant vertex operator formed in this way is normal ordered. Note that the normalization of the lightcone state carries over to the covariant vertex unaltered because the normalization for the DDF states is set by the DDF commutation relations (\ref{eq: [A_nA_m]}) which are identical to those of the usual creation and annihilation operators.

We next construct covariant normal ordered vertex operators in the case when the polarization tensors of the corresponding lightcone gauge states are arbitrary, for which in general, $$\xi_{\dots i\dots j \dots }\eta^{ij},\quad\bar{\xi}_{\dots i\dots j \dots }\eta^{ij},\quad \xi_{\dots i\dots }k^{i},\quad \bar{\xi}_{\dots i \dots }k^{i},$$ need not vanish. We start from the simplest non-trivial case and then move on to more general cases. Proceeding by induction we then obtain the general result.

For this purpose we'll be needing the following local dimensionless polynomial functionals of $q\cdot \partial^{\#}X(z)$, and $q\cdot\bar{\partial}^{\#}X(\bar{z})$ respectively,
\begin{subequations}\label{eq:HS}
\begin{align}
&\mathbb{S}_{m,n}(z) \equiv \sum_{r=1}^nrS_{m+r}(mq;z)S_{n-r}(nq;z),\phantom{\bigg|}\label{eq:HSa}\\
&\bar{\mathbb{S}}_{m,n}(\bar{z}) \equiv \sum_{r=1}^nr\bar{S}_{m+r}(mq;\bar{z})\bar{S}_{n-r}(nq;\bar{z}),\label{eq:HSb}
\end{align}
\end{subequations}
with the elementary Schur polynomials, $S_m(nq;z)$, $\bar{S}_m(nq;\bar{z})$, defined in Appendix \ref{C}. In (\ref{eq:A_-N DDF}) we showed that normal ordering of $A_{-n}^i\cdot \,e^{ip\cdot X(z)}$ leads to,
\begin{equation}
\phantom{\Big{(}}A_{-n}^k\cdot \,e^{ip\cdot X(z)}\cong H_n^k(z)\,e^{i(p-nq)\cdot X(z)}.
\end{equation}
Let us apply an additional DDF operator from the left to this expression and normal order the resulting object. We find,
\begin{equation}
\begin{aligned}
&A^j_{-m}A_{-n}^k\cdot \,e^{ip\cdot X(z)}\\
&\cong \Big[H_m^jH_n^k +\delta^{jk}\,\mathbb{S}_{m,n}\Big](z)\,e^{i[p-(m+n)q]\cdot X(z)}.
\end{aligned}
\end{equation}
Proceeding in a similar manner we apply another DDF operator to the resulting expression and normal order the right-hand-side. An important point to note now is that $\mathbb{S}_{m,n}(z)$ commutes with the DDF operators, $A_{\ell}^i$, because $\mathbb{S}_{m,n}(z)$ is a functional of $q\cdot \partial^{\#}X$ and $[A^i_n,q\cdot \partial^{\#}X]=0$. We find,
\begin{equation}
\begin{aligned}
&A^i_{-\ell}A^j_{-m}A_{-n}^k\cdot \,e^{ip\cdot X(z)}\cong  \Big[H^i_{\ell}H_m^jH_n^k+\delta^{ij}\,\mathbb{S}_{\ell,m}H^k_n\\
& +\delta^{ik}\,\mathbb{S}_{\ell,n}H^j_m+\delta^{jk}\,\mathbb{S}_{m,n}H^i_{\ell}\Big](z)\,e^{i[p-(\ell+m+n)q]\cdot X(z)}\\ 
\end{aligned}
\end{equation}
By induction it follows from the above that the general normal ordered expression reads,
\begin{widetext}
\begin{equation}\label{eq:xiA^g e^ipX1}
\begin{aligned}
&A_{-n_1}^{i_1}\dots A_{-n_g}^{i_g}\cdot e^{ip\cdot X(z)}\cong\\
&\hspace{2cm} \cong 
\sum_{a=0}^{\lfloor g/2\rfloor}\sum_{\pi\in S_g/\!\sim}\prod_{\ell=1}^a\delta^{i_{\pi(2\ell-1)}i_{\pi(2\ell)}}\,\mathbb{S}_{n_{\pi(2\ell-1)},n_{\pi(2\ell)}}(z)\prod_{q=2a+1}^gH_{n_{\pi(q)}}^{i_{\pi(q)}}(z)\,e^{i(p-\sum_rn_rq)\cdot X(z)},
\end{aligned}
\end{equation}
\end{widetext}
with $S_g$ the permutation group of $g$ elements and the equivalence relation $\sim$ being such that $\pi_i\sim \pi_j$ with $\pi_i,\pi_j\in S_g$ when they define indistinguishable terms in (\ref{eq:xiA^g e^ipX1}). In all terms where $\mathbb{S}_{n_i,n_j}$ appears we are to only include permutations which preserve the inequality $i\leq j$.  Furthermore, the notation $\lfloor \cdot\rfloor$ in the summation indicates that the upper limit saturates the inequality $a\leq g/2$. The number of terms in the sum over permutations at fixed $a$ is 
\begin{equation}\label{eq:number of terms in sum_pi}
\frac{2^{-a}g!}{a!(g-2a)!}.
\end{equation}

For every lightcone gauge state $|V\rangle =C\xi_{i_1i_2\dots}\bar{\xi}_{j_1j_2\dots}\,\alpha^{i_1}_{-n_1}\alpha_{-n_2}^{i_2}\dots\tilde{\alpha}_{-\bar{n}_1}^{j_1}\tilde{\alpha}_{-\bar{n}_2}^{j_2}\dots |0,0;p^+,p^i\rangle$, with $C$ is as given in (\ref{norm const C}), there exists a covariant normal ordered vertex operator 
\begin{equation}\label{eq:Vgeneral}
V(z,\bar{z})=\frac{g_c}{\sqrt{2p^+\mathcal{V}_{d-1}}}\,CU(z)\bar{U}(\bar{z}).
\end{equation}
The normal ordered chiral half $U(z)$ is equal to the right hand side of (\ref{eq:xiA^g e^ipX1}) when contracted with the lightcone gauge polarization tensor, $\xi_{i_1\dots i_g}$, which corresponds to an arbitrary irreducible representation of SO(25) (or SO(24) for massless states). 
\begin{equation}\label{eq:U(z) mass eigenstates}
\begin{aligned}
U(z)&\,=\sum_{a=0}^{\lfloor g/2\rfloor}\sum_{\pi\in S_g/\!\sim}\\
&\times \xi_{i_1\dots i_g}\prod_{\ell=1}^a\delta^{i_{\pi(2\ell-1)}i_{\pi(2\ell)}}\,\mathbb{S}_{n_{\pi(2\ell-1)},n_{\pi(2\ell)}}(z)\\
&\times\prod_{q=2a+1}^gH_{n_{\pi(q)}}^{i_{\pi(q)}}(z)\,e^{-i(\sum_{r=1}^gn_r)q\cdot X(z)}.
\end{aligned}
\end{equation}
There is a similar expression for $\bar{U}(\bar{z})$ with $\bar{\xi}_{ij\dots}$, $\bar{\mathbb{S}}_{n,m}(\bar{z})$, $\bar{H}_{\bar{n}}^i(\bar{z})$ and $e^{i(p-\sum_r{\bar{n}_r}q)\cdot X(\bar{z})}$ replacing $\xi_{ij\dots}$, $\mathbb{S}_{n,m}(z)$, $H_{n}^i(z)$ and $e^{i(p-\sum_rn_rq)\cdot X(z)}$ respectively. If the underlying spacetime manifold is not compactified in a lightlike direction we are to enforce \emph{in addition}: 
$$
\sum_rn_r=\sum_r{\bar{n}_r};
$$
we elaborate on this in the closed string coherent state section in detail.

 When the polarization tensor is traceless, $\xi_{\dots i\dots j\dots}\delta^{ij}=0$, $U(z)$ reduces to the result obtained in (\ref{eq: many A_-N DDF}), the chiral half of which reads, $ \xi_{i_1\dots i_s}H_{n_1}^{i_1}\dots H_{n_s}^{i_s}e^{i(p-\sum_rn_rq)\cdot X(z)}$. In the rest frame, $ \xi_{\dots i\dots }p^{i}=0$, all the $H_n^i(z)$ in $U(z)$ in turn reduce to $P^i_n(z)$.

There are specific and very interesting examples where the sum over permutations may be carried out explicitly. In fact, this is precisely possible in the case of coherent states. In particular, we construct coherent states below, we will be interested in expressions of the form: $\frac{1}{g!}\big(\sum_{n=1}^{\infty}\lambda_n\cdot A_{-n}\big)^{g}\,e^{ip\cdot X(z)}$. Clearly, in this case the indices on the $A_{-n}$ are dummy variables, and hence from (\ref{eq:xiA^g e^ipX1}) and (\ref{eq:number of terms in sum_pi}) we deduce that:
\begin{widetext}
\begin{equation}\label{eq:A^g e^ipX}
\begin{aligned}
\frac{1}{g!}\Big(&\sum_{n>0}\frac{1}{n}\lambda_n\cdot A_{-n}\Big)^{g}\,e^{ip\cdot X(z)}\cong\\
&\cong \sum_{a=0}^{\lfloor g/2\rfloor}\frac{1}{a!(g-2a)!}\Big(\frac{1}{2nm}\sum_{n,m>0}\lambda_{n}\cdot \lambda_m\,\mathbb{S}_{n,m}\,e^{-i(n+m)q\cdot X(z)}\Big)^a\Big(\sum_{n>0}\frac{1}{n}\lambda_n\cdot H_n\,e^{-inq\cdot X(z)}\Big)^{g-2a}\,e^{ip\cdot X(z)}.
\end{aligned}
\end{equation}
\end{widetext}
When we sum over $g$ (from 0 to $\infty$) such a object has an interpretation of the chiral half of a closed string coherent state or 
an open string coherent state as we shall demonstrate in Sec.~\ref{SCS}, where we discuss string coherent states in great detail. The corresponding lightcone gauge state is $\exp(\sum_{n>0}\frac{1}{n}\lambda_n\cdot \alpha_{-n})|0,p^+,p^i\rangle$, which is an eigenstate of $\alpha^i_{n>0}$ with eigenvalue $\lambda_n^i$ and $\lambda_n^*=\lambda_{-n}$.  The covariant gauge expression is not an eigenstate of $\alpha^{\mu}_{n>0}$ but nevertheless satisfies the definition of a coherent state (which is given in the opening lines of Sec.~\ref{OS} or Sec.~\ref{CS}).

Note finally that all vertex operators in this section have been normalized to `one string in volume $\mathcal{V}_{d-1}$' as discussed in Sec.~\ref{sec:NSW}. Therefore, for instance, the normalization of the general lightcone state (\ref{eq:lc |V> general}) is such that:
$$
\langle V(p')|V(p)\rangle_{\rm lc} = \delta_{p',p},
$$
with $\delta_{p',p}$ a \emph{Kronecker} delta which reduces to unity when ${p^+}'=p^+$ and ${\bf p}'={\bf p}$ and vanishes otherwise. The associated covariant vertex operator (\ref{eq:general DDF mass eigenst}) or (\ref{eq:Vgeneral}), is normalized by the most singular term in the operator product expansion (\ref{eq:VVope normalization}):
$$
V^{\dagger}(z,\bar{z})\cdot V(0,0)\cong \Big(\frac{g_c^2}{2p^+\mathcal{V}_{d-1}}\Big)\frac{1}{|z|^4}+\dots,
$$
the dimensionless coefficient having been fixed by Lorentz covariance and unitarity of the $S$-matrix. We have made use of the relation between operator product expansions and commutators. (Recall that for arbitrary operators of the form,
$$
A=\oint dz\,a(z), \qquad B=\oint dw\, b(w),
$$ 
there exists the interpretation, see e.g.~\cite{DiFrancescoMatheuSenechal97},
\begin{equation}\label{eq:[A,B]}
\begin{aligned}
&[A,B]\cong A\cdot B= \oint_0 dw\oint_w dz \,a(z) \cdot b(w),\\
&[A,b(w)]\cong A\cdot b(w)=\oint_w dz\, a(z)\cdot b(w),\phantom{\Bigg(}
\end{aligned}
\end{equation}
with `$\cdot $' denoting operator product expansion.) With this normalization the string path integral yields the $S$-matrix directly, see (\ref{eq:S_fi in string theory}). 

\section{String Coherent States}\label{SCS}
It is possible that cosmic strings being macroscopic and massive should have a classical interpretation. If this is the case, one may suspect that the appropriate vertex operators for the description of cosmic superstrings (from our experience with standard harmonic oscillator coherent states) would have coherent state-like properties. With this motivation in mind we will be searching for coherent state vertex operators, which from the standard coherent state properties would be expected to have a classical interpretation.

The states we have considered in the previous sections are mass eigenstates. The dictionary described above, which identifies the states (\ref{eq:lc state}) and (\ref{eq:DDF state}), is tailor-made for light-cone to covariant mass eigenstate maps. Coherent states however are not mass eigenstates in general \footnote{The coherent states constructed here are eigenstates of momentum however in the spacetime directions transverse to $q^a$ as we shall see.}. In the construction of string coherent states one normally proceeds in direct analogy with the construction of coherent states in the harmonic oscillator, whereby coherent states are constructed by exponentiation of the creation operator, $e^{-|\lambda|^2/2}e^{\lambda a^{\dagger}}|0\rangle$, with $a|0\rangle=0$ and $[a,a^{\dagger}]=1$. In the string case there is an infinite number of creation operators and the vacuum depends on the center of mass momentum. The usual approach is to proceed in lightcone gauge where the constraints are solved automatically and the open string construction is trivial, see e.g.~\cite{Calucci89}. Rather than drop spacetime covariance we shall make use of the spectrum generating DDF operators which can be used to generate covariant physical states.

In the current section we construct covariant and lightcone gauge open and closed coherent states and show that these states have a classical interpretation by associating them to general classical solutions. Let us primarily define what we mean by a quantum state with a classical interpretation:
\begin{itemize}
\item[-] \emph{String states with a classical interpretation} should possess classical expectation values (with small uncertainties modulo zero mode contributions) provided these are compatible with the symmetries of string theory. These classical expectation values should be non-trivially consistent with the classical equations of motion and constraints.
\end{itemize}

Starting with the \emph{open string} we begin by defining a string coherent state and using DDF operators proceed by analogy to the harmonic oscillator. The definition of a coherent state that we adopt is very general but standard  \cite{KlauderSkagerstam85} which we minimally extend to include the string theory requirements \footnote{The naive definition, that a coherent state should be an eigenstate of the annihilation operators is not in general compatible with the string theory symmetries.} (see the opening lines of Sec.~\ref{OS} below). After establishing that the coherent state properties are satisfied for the states under consideration we go on to show that the covariant and lightcone gauge states share identical angular momenta and present the explicit map to general classical solutions. We show that these coherent states indeed possess classical expectation values, thus proving that the above definition of classicality is satisfied.

We then go on to discuss the construction of \emph{closed string} coherent states. Here the naive construction leads to the requirement of a lightlike compactification of spacetime, $X^-\sim X^-+2\pi R^-$. We show that all states considered are indeed physical and single-valued under translations around the compact direction, $X^-$.

We are then, according to the above definition of classicality, led to search for classical expectation values. In the closed string case the string symmetries  forbid \cite{Blanco-PilladoIglesiasSiegel07} the naive expectation that $\langle X^{\mu}(z,\bar{z})\rangle=X^{\mu}_{\rm cl}(z,\bar{z})$ \footnote{Here $X^{\mu}_{\rm cl}(z,\bar{z})$ is an arbitrary non-trivial solution of the wave equation, $\partial \bar{\partial}X^{\mu}_{\rm cl}(z,\bar{z})=0$.} should be satisfied by a state with a classical interpretation. We elaborate on this and discuss various definitions of classicality and their range of applicability. Here we provide a new classicality requirement (in accordance with the above definition) that applies in all the usual gauges of interest (e.g.~lightcone and covariant, but not in static gauge for instance) where the vertices are invariant under spacelike worldsheet shifts where the naive definition $\langle X^{\mu}\rangle=X_{\rm cl}^{\mu}$ does not apply.

Finally, we construct coherent closed string states in fully non-compact spacetimes by projecting out the lightlike winding states in the underlying Hilbert space and go on to show that all the coherent state properties are satisfied by the projected states as well, and therefore that the projected states have a classical interpretation. We also compute the angular momenta of the projected states in both lightcone and covariant gauge and show that they are both identical to the angular momentum associated to the corresponding classical solutions which we identify explicitly.

For a good overview of coherent states (but not explicitly in the context of string theory) see Klauder and Skagerstam's book \cite{KlauderSkagerstam85} and the excellent review article by Zhang, Feng and Gilmore \cite{ZhangFengGilmore90}.

\subsection{Open String Coherent States}\label{OS}
We define an \emph{open string coherent state}, $V(\lambda)\cong |V(\lambda)\rangle$, to be a state that: 
\begin{itemize}
\item[(a)] is specified by a set of \emph{continuous} labels $\lambda=\{\lambda_n^i\}$; 
\item[(b)] produces a resolution of unity,\begin{equation}\label{eq:unity open}
\mathds{1} =\suminnt\int d\lambda  \big|V(\lambda,\dots)\big\rangle \big\langle V(\lambda,\dots)\big|;
\end{equation}
\item[(c)] transforms correctly under all symmetries of bosonic (or super-) string theory.
\end{itemize}
We also allow for the possibility that the state depends on other discrete or continuous quantum numbers (such as momentum), denoted by ``$\dots$", which are to be summed or integrated over respectively -- this is what is meant by the symbol $\ssumint$ \footnote{We will normally not exhibit these additional labels explicitly, and hence write $V(\lambda)$ instead of $V(\lambda,\dots)$, or even $V(z)$ when there is no possibility for confusion with the mass eigenstates of the previous section.}.
The measure associated to the continuous labels explicitly reads $d\lambda =\frac{1}{N}\prod_{n,i}d^2\lambda_n^i$ with $N$ an appropriate normalization (to be determined) and as usual $d^2\lambda_n^i=id\lambda_n^i\wedge d\lambda_n^{*i}$ (no sum over $i$). The labels $n$ and $i$ will be related to the distribution of harmonics present and spacetime directions respectively. The requirements (a,b) are the minimal requirements for a state to be termed coherent \cite{KlauderSkagerstam85} and to these we add the minimal string theory requirement (c).

As we discussed in Sec.~\ref{DDFop}, we may construct open string vertex operators using the $A^A_n$ and $A^I_n$ DDF operators for excitations in spatial directions tangent and transverse to the D$p$-brane respectively with $A=\{1,\dots,p-1\}$ and $I=\{p+1,\dots ,25\}$. (Note that $p\geq1$, see Sec.~\ref{DDFop}, and our open string conventions are given in Appendix \ref{O}). We shall here consider the construction of coherent state vertex operators with excitations in the directions tangent to the brane.  Let us then consider the normalized open string DDF vertex operator,
\begin{equation}\label{eq:DDF_coherent}
\begin{aligned}
V(\lambda)&=\frac{g_{o,p}}{\sqrt{2p^+\mathcal{V}_{\paral}}}\,C_{\lambda}\exp\Big(\sum_{n=1}^{\infty}\frac{1}{n}\lambda_n^A A^A_{-n}\Big)\,e^{ip_aX^a(z)},
\end{aligned}
\end{equation}
with $a=\{0,1,\dots,p\}$. with $a=\{0,1,\dots,p\}$. We have found it convenient to define \footnote{The index $p$ on $g_{o,p}$ denotes the dimensionality of the D$p$-brane in which the string is propagating and should not be confused with the momentum of the vacuum $p^{a}$.}:
$$
g_{o,p}\equiv \frac{g_o}{\sqrt{V_{\perp}}}, \qquad{\rm with}\qquad \mathcal{V}_{d-1}\equiv V_{\perp}\mathcal{V}_{\paral},
$$
with $V_{\perp}$ the volume of spacetime transverse to the D$p$-brane, and $\mathcal{V}_{\paral}$ the volume tangent to the brane (so that $V_{\perp}\mathcal{V}_{\paral}$ is the total volume of spacetime transverse to $x^+$). \footnote{The dimensionalities are such that $[g_{o,p}]=L^{\frac{d-2}{2}}L^{-\frac{d-1-p}{2}}=L^{\frac{p-1}{2}}$, so that $[g_{o,p}/\sqrt{2p^+\mathcal{V}_{\paral}}]=1$ as required by unitarity.} In parallel to (\ref{eq:VT lc}) in particular, we thus define:
\begin{equation}
\begin{aligned}
&\mathcal{V}_{\paral}\equiv \lim_{p'\rightarrow p}(2\pi)\delta(p^{'+}-p^+)(2\pi)^{p-1}\delta^{p-1}(
{\bf p}'-{\bf p}),\\
&V_{\perp}\equiv \lim_{p'\rightarrow p}(2\pi)^{d-1-p}\delta^{d-1-p}({\bf p}'-{\bf p}).
\end{aligned}
\end{equation}
The total volume of spacetime is $V_d=V_+\mathcal{V}_{d-1}$. The kinematic pre-factor and the normalization $C_{\lambda}$ is chosen such that the vertex operator is normalized to `one string in volume $\mathcal{V}_{d-1}$' as shown in (\ref{eq:VVope normalization}) for the case of closed strings. 

$p^a$ is the (tachyonic) vacuum momentum of the string, the DDF operators, $A_n^A$, defined in (\ref{eq: DDF A op}) and the normalization constant, 
$$
C_{\lambda}\equiv \exp\Big(-\sum_{n=1}^{\infty}\frac{1}{2n}|\lambda_n|^2\Big),
$$
chosen such that the operator product expansion has the leading singularity $V^{\dagger}(\lambda;z)\cdot V(\lambda;0)\simeq \big(\frac{g_{o,p}^2}{2p^+\mathcal{V}_{\paral}}\big)\frac{1}{|z|^2}+\dots$, corresponding to `one string in volume $\mathcal{V}_{\paral}$' as required by unitarity of the $S$-matrix. 

The vertex operators associated to ripples of the brane are related by T-duality \cite{NairShapereStromingerWilczek87,Polchinski96} to the vertices (\ref{eq:DDF_coherent}). The onshell constraints are given by (\ref{eq:pq cond_open}), repeated here for convenience: $p\cdot q=1/(2\alpha')$, $q^2=0$, and $p^2=1/\alpha'$. The polarization complex vectors $\{\lambda_n^A\}$ are defined such that $\lambda_n\cdot q=0$, $\lambda_n^*=\lambda_{-n}$, and require \cite{Calucci89} that $\sum_n|\lambda_n|^2<\infty$ to ensure that the vertex is well behaved.

First of all we show that the vertex operator (\ref{eq:DDF_coherent}) is a coherent state. To prove this recall that a coherent state must by definition satisfy three properties: (a) it must be labelled by a set of continuous parameters, these here being $\{\lambda_n^A\}$, (b) there must exist a completeness relation of the form (\ref{eq:unity open}), and (c) it must transform correctly under the symmetries of string theory. (a) is trivially satisfied and the state remains correctly normalized for arbitrary values of the $\lambda_n^A$ when $\sum_n|\lambda_n|^2<\infty$. To prove that a completeness relation exists it is convenient to write (\ref{eq:DDF_coherent}) in operator form,
\begin{equation}\label{eq:DDF_coherent op}
\begin{aligned}
|V(\lambda,p)\rangle&=\frac{1}{\sqrt{2p^+\mathcal{V}_{\paral}}}\,C_{\lambda}\exp\Big(\sum_{n=1}^{\infty}\frac{1}{n}\lambda_n^A A^A_{-n}\Big)\,|0;p^a\rangle,
\end{aligned}
\end{equation}
with the correspondence $|0;p^a\rangle \simeq g_{o,p}\,e^{ip_aX^a}$ and we use the relativistic normalization:
$$
\langle 0;{p^a}'|0;p^a\rangle = 2p^+(2\pi)\delta(p^{'+}-p^+)(2\pi)^{p-1}\delta^{p-1}(
{\bf p}'-{\bf p}).
$$ 
Note primarily that from  the DDF operator commutation relations, $V(\lambda)$ is an eigenstate of the annihilation operators, $A_{n>0}^A\cdot V(\lambda)\cong \lambda_{n>0}^AV(\lambda)$, from which on account of (\ref{eq:DDF_coherent})  it follows that states are not orthogonal, the inner product of two states being given by,
$$
\big\langle V(\lambda,p')|V(\zeta,p)\big\rangle =\delta_{p',p} C_{\lambda} C_{\zeta}\exp\Big(\sum_{n>0}\frac{1}{n}\lambda^*_{n}\cdot\zeta_{n}\Big).
$$
The factor $C_{\lambda} C_{\zeta}\exp\big(\sum_{n>0}\frac{1}{n}\lambda^*_{n}\cdot\zeta_{n}\big)$ reduces to unity when $\lambda_n^A=\zeta_n^A$, for all $n,A$, so that,
$$
\big\langle V(\lambda,p)|V(\lambda,p)\big\rangle=1.
$$ 
Recall that coherent states are (when we choose $q^i=q^+=0$) eigenstates of momentum in the $k^+$ and ${\bf k}$ directions (but not in the $k^-$ direction). So, as one would expect, these coherent states are over-complete, the overlap between any two being non-zero for a wide range of $\lambda_n^i,\zeta_n^i$. From this expression we then deduce (by forming appropriate inner products and integrating) that there exists the completeness relation,
\begin{equation}
\begin{aligned}
\mathds{1} = \mathcal{V}_{\paral}\int_0^{\infty} &\frac{dp^+}{2\pi}\int_{\mathbb{R}^{p-1}}\frac{d^{p-1}{\bf p}}{(2\pi)^{p-1}}\int \bigg(\prod_{n,A}\frac{d^2\lambda_n^A}{2\pi n}\bigg) \\
&\,\times \big|V(\lambda,p)\big\rangle\big\langle V(\lambda,p)\big|,
\end{aligned}
\end{equation}
with  $d^2\lambda_n^A=id\lambda_n^A\wedge d\lambda_n^{*A}$. Finally, that the vertex operator (\ref{eq:DDF_coherent}) is physical (requirement c) follows from the fact that $L_{n\in \mathbb{Z}}$ commutes with all the DDF operators, $L_{n>0}$ annihilates the vacuum $e^{ip\cdot X(z)}$ and $L_0\cdot e^{ip\cdot X(z)}\cong e^{ip\cdot X(z)}$. Therefore, $V(\lambda)$ satisfies the Virasoro constraints, $(L_0-1)\cdot V(\lambda)\cong0$, $ L_{n>0}\cdot V(\lambda)\cong0$ and is hence physical. Recall from Sec.~\ref{DDF} that in addition all states formed from DDF operators are transverse to null states. We conclude that the string coherent state defining properties (a-c) are satisfied.
\vspace{0.5cm}

\subsubsection{Functional Representation}
Let us now consider the corresponding local normal ordered representation of $V(\lambda)$, which in practice means subtracting all self contractions from the vertex (\ref{eq:DDF_coherent}). The vacuum $e^{ip\cdot X(z)}$ is already normal ordered and so the remaining self-contractions that need to be subtracted are those associated to contractions with one leg in the DDF operators and one leg in the vacuum. In Sec.~\ref{DDF} we computed the normal ordered representation of arbitrary covariant states. For the above coherent state this is obtained by using the integral representation of the DDF operators (\ref{eq: DDF As}) in (\ref{eq:DDF_coherent}) and carrying out the operator products on account of the onshell constraints (given below (\ref{eq:DDF_coherent})) and the property $\lambda_n\cdot q=0$. The integrands of the DDF operators are to lie on the real axis as they are brought close to the vacuum which is also on the real axis, $z=\bar{z}$, and so the relevant propagator takes the form, 
\begin{equation}\label{eq:<XX>hol_op}
\langle X^a(z)X^b(w)\rangle=-(2\alpha')\eta^{ab}\ln (z-w).
\end{equation}
From Fig.~\ref{fig:DDFop} where the open string DDF construction is exhibited it can be seen that this is the correct procedure -- in the figure we have conformally mapped to the disc with boundary $z\bar{z}=1$ (instead of the upper half plane) where the propagator is again of the form (\ref{eq:<XX>hol_op}) on the boundary (up to terms that drop out of correlation functions). We then compute all Wick contractions and subsequently analytically continue in the variable of integration and choose an integration contour that circles the vacuum. The same procedure can then be repeated, with additional DDF operators which may be brought close to the resulting state in the same manner as above and so on. The resulting normal ordered vertex assumes a particularly simple form when we assume in addition, $\lambda_{n>0}\cdot \lambda_{m>0}=0$, see (\ref{eq:xiA^g e^ipX1}). In this case the normal ordered open string coherent states are given by a linear combination of the traceless mass eigenstates (\ref{eq: many A_-N DDF}),
\begin{equation}\label{eq:DDF_coherent_cov_tr}
\begin{aligned}
V(&\lambda)=\frac{g_{o,p}}{\sqrt{2p^+\mathcal{V}_{\paral}}}\,C_{\lambda}\\
&\times \exp\Big(\sum_{n=1}^{\infty}\frac{1}{n}\lambda_n\cdot H_n(z)\,e^{-inq\cdot X(z)}\Big)\,e^{ip\cdot X(z)},
\end{aligned}
\end{equation}
the difference being that for open strings the dimensionless quantity $H_n(z)$ reads,
\begin{subequations}\label{eq:H_n open}
\begin{align}
H_N^A(z)&\equiv \sqrt{2\alpha'}p^A S_{N}(Nq;z)+P_N^A(z),\\
P_N^A(z)&=\sqrt{\frac{2}{\alpha'}}\sum_{m=1}^{N}\frac{i}{(m-1)!}\, \partial^mX^A(z)S_{N-m}(Nq;z).
\end{align}
\end{subequations}
The general result for arbitrary (but of course transverse) $\lambda_n^i$ follows directly from (\ref{eq:A^g e^ipX}), 
\begin{widetext}
\begin{equation}\label{eq:V(lambda)A^g e^ipX gen norm ord}
\begin{aligned}
V(\lambda)&=\frac{g_{o,p}}{\sqrt{2p^+\mathcal{V}_{\paral}}}\,C_{\lambda}\exp\Big(\sum_{n>0}\frac{1}{n}\lambda_n\cdot A_{-n}\Big)\,e^{ip\cdot X(z)}\\
&\cong \frac{g_{o,p}}{\sqrt{2p^+\mathcal{V}_{\paral}}}\,C_{\lambda}\sum_{g=0}^{\infty}\sum_{a=0}^{\lfloor g/2\rfloor}\frac{1}{a!(g-2a)!}\\
&\hspace{1.5cm}\times\Big(\frac{1}{2nm}\sum_{n,m>0}\lambda_{n}\cdot \lambda_m\,\mathbb{S}_{n,m}\,e^{-i(n+m)q\cdot X(z)}\Big)^a\Big(\sum_{n>0}\frac{1}{n}\lambda_n\cdot H_n\,e^{-inq\cdot X(z)}\Big)^{g-2a}\,e^{ip\cdot X(z)},
\end{aligned}
\end{equation}
\end{widetext}
which of course reduces to (\ref{eq:DDF_coherent_cov_tr}) when $\lambda_n\cdot \lambda_m=0$ (for $n,m>0$). The quantities $\mathbb{S}_{n,m}(z)$ are related to elementary Schur polynomials, $S_N(nq;z)$, and have been defined in (\ref{eq:HS}). For later reference, define the quantity $U(\lambda)$ in (\ref{eq:V(lambda)A^g e^ipX gen norm ord}) by the expression, 
$$
V(\lambda)\equiv\frac{g_{o,p}}{\sqrt{2p^+\mathcal{V}_{\paral}}}\,C_{\lambda}U(\lambda)e^{ip\cdot X(z)}.
$$ 

\subsubsection{Open String Coherent State Properties}
Series expanding the exponential in (\ref{eq:DDF_coherent_cov_tr}) it is seen that the mass eigenstates in the underlying Hilbert space are polynomials in $\partial^{\#}X$, multiplied by $e^{i(p-\sum_nns_nq)\cdot X(z)}$, for some sequence of positive integers, $\{s_1,s_2,\dots\}$, with $\sum_nns_n$ equal to the level number. Also, $V(\lambda)$ is an eigenstate of momentum in the directions transverse to $q^{\mu}$; given that $q^2=0$ one may take for example, $q^+=q^A=q^I=0$ and $q^{-}$ non-vanishing (see also the discussion in Sec.~\ref{sec:NSW}), in which case one learns that $\hat{p}^A\cdot V(\lambda)=p^AV(\lambda)$ and $\hat{p}^+\cdot V(\lambda)=p^+V(\lambda)$, with $\hat{p}^{\mu}=\frac{1}{\alpha'}\oint \dslash z\,\partial X^{\mu}$. The full momentum expectation value is in turn given by,
\begin{equation}\label{eq:<mass-shell_DDF>}
\langle \hat{p}^{a}\rangle =p^{a}-\langle N\rangle q^{a},\qquad \langle \hat{p}^2\rangle = -\frac{1}{\alpha'}\big(\langle N\rangle-1\big),\phantom{\Big|}
\end{equation}
where we have identified an \emph{effective} level number, 
$$
\langle N\rangle\equiv\sum_{n=1}^{\infty}|\lambda_n|^2,
$$
in direct analogy to the generic DDF state momentum (\ref{eq:mass-shell_DDF}). (These tree level operator statements are to be interpreted as $\langle A\rangle = \langle V^{\dagger}AV\rangle$ for an operator $A$ (with $\langle V^{\dagger}V\rangle=1$), and $V^{\dagger}$ is obtained from $V$ by reversing the momenta and complex-conjugating the polarization tensors. Note also that $\hat{p}^{\mu}=\hat{p}^{\mu}_{\rm open}=\frac{1}{\alpha'}\oint \dslash z\,\partial X^{\mu}$ in this section; in the rest of the paper, $\hat{p}^{\mu}=\hat{p}^{\mu}_{\rm closed}=\frac{2}{\alpha'}\oint \dslash z\partial X^{\mu}$.)

The above considerations imply that $V(\lambda)$ carries an effective mass associated to $\langle N\rangle$, which is in agreement with the usual open string mass shell constraint, $m^2=(N-1)/\alpha'$, when $N$ is identified with $\langle N\rangle$. Notice that $\langle N\rangle$ is a \emph{continuous} function of the $|\lambda_n|$ as required from the definition of a coherent state, not necessarily an integer. Therefore, coherent states can in particular have masses which are non-zero, but yet much smaller than the string scale (a common draw-back of mass eigenstates), or, in the opposite extreme, they may have large mass and represent macroscopic string states; although we have not \emph{yet} proven that the states constructed are macroscopic.

From the well known properties of coherent states \cite{KlauderSkagerstam85} we expect the limit $|\lambda_n|\gg1$ to be associated to  the macroscopic or long string limit. To show that this is indeed the case we next consider the open string coherent state (\ref{eq:DDF_coherent}) in lightcone gauge. Using the map discussed in Sec.~\ref{DDF} we immediately write down the lightcone gauge analogue of the covariant state (\ref{eq:DDF_coherent}),
\begin{equation}\label{eq:lc_coherent open}
\begin{aligned}
|V(\lambda)\rangle_{\rm lc}&=\frac{1}{\sqrt{2p^+\mathcal{V}_{\paral}}}\,C_{\lambda}\exp\Big(\sum_{n=1}^{\infty}\frac{1}{n}\lambda_n\cdot \alpha_{-n}\Big)\,\big|0;p^+,p^A\big\rangle.
\end{aligned}
\end{equation}
This is also an eigenstate of $p^+,p^A$, as was the covariant state above (when $q^+=q^A=0$). The contractions are associated to indices, $A$,  and are transverse to the longitudinal, $\pm$, directions (with $v^{\pm}=\frac{1}{\sqrt{2}}(v^0\pm v^p)$ for some generic spacetime vector $v^{\mu}$). This state is an eigenstate of the annihilation operators, $\alpha_{n>0}^A|V(\lambda)\rangle_{\rm lc}=\lambda^A_n|V(\lambda)\rangle_{\rm lc}$ and so the lightcone gauge position expectation value is given by  (\ref{eq:X open ND apx}), 
$$
\langle X^A(z,\bar{z})-\hat{x}^A\rangle_{\rm lc}=(X^A(z,\bar{z})-x^A)_{\rm cl},
$$ 
with,
\begin{equation}\label{eq:<X>_coherent_lc}
\begin{aligned}
\big(X^A(z,\bar{z}&)-x^A\big)_{\rm cl}=-i\alpha'p^A\ln |z|^2\\
&+i\Big(\frac{\alpha'}{2}\Big)^{1/2}\sum_{n\neq0}^{\infty}\frac{\lambda_n^A}{n}\big(z^n+\bar{z}^{-n}\big),
\end{aligned}
\end{equation}
where we have identified $\langle\hat{p}^A\rangle$ with $p^A$ (given that $q^A=0$). 
Equation (\ref{eq:<X>_coherent_lc}) is the general solution to the equations of motion, $\partial \bar{\partial}X_{\rm cl}^A(z,\bar{z})=0$, the constraints, $(\partial X_{\rm cl})^2=(\bar{\partial}X_{\rm cl})^2=0$ having been solved by the gauge choice:\footnote{Recall that $\tau=(\tau)_{\rm Euclidean}=i(\tau)_{\rm Minkowski}$, $z=e^{-i(\sigma+i\tau)}$, $\bar{z}=e^{i(\sigma-i\tau)}$.} $X_{\rm cl}^+(z,\bar{z})=-i\alpha'p^+\ln |z|^2$, reached by the conformal map $z=e^{2iq\cdot X(z)}$, $\bar{z}=e^{2iq\cdot X(\bar{z})}$ (recall that $q\cdot p=1/(2\alpha')$ for open strings). The corresponding longitudinal components of the position expectation value are likewise computed. On account of the operator equation, $\sqrt{2\alpha'}\alpha_n^-=\frac{1}{2p^+}\sum_{\ell\in \mathbb{Z}}:\alpha_{n-\ell}^i\alpha_{\ell}^i:$ (for $n\neq0$), and the fact that the coherent state is an eigenstate of $\alpha_{n>0}^A$ with eigenvalue $\lambda_n^A$ one learns that,
$$
\big\langle X^-(z,\bar{z})-\hat{x}^-\big\rangle_{\rm lc}=\big(X^-(z,\bar{z})-x^-\big)_{\rm cl},
$$
with
\begin{equation}\label{eq:<X^->_coherent_lc open}
\begin{aligned}
\big(X^-(z&,\bar{z})-x^-\big)_{\rm cl}= -i\frac{1}{p^+}\Big(\alpha'{\bf p}^2+\sum_{n=1}^{\infty}|\lambda_n|^2-1\Big)\ln |z|^2\\
&+i\sum_{n\neq0}\frac{1}{n}\sum_{r\in \mathbb{Z}}\frac{1}{4p^+}\,\lambda_{n-r}\cdot\lambda_{r}\,\big(z^{-n}+\bar{z}^{-n}\big),\\
\end{aligned}
\end{equation}
with the definitions $\lambda_0^A\equiv \sqrt{2\alpha'}p^A$, ${\bf p}^2=p^Ap^A$. Recall that for open strings,
\begin{equation}\label{X lc_open}
\begin{aligned}
X^-(z,\bar{z})&-x^-=-i\alpha'\hat{p}^-\ln |z|^2\\
&+i\Big(\frac{\alpha'}{2}\Big)^{1/2}\sum_{n\neq0}\frac{\alpha^-_n}{n}\,\big(z^{-n}+\bar{z}^{-n}\big).
\end{aligned}
\end{equation}

Finally, in the Dirichlet directions, on account of (\ref{eq:X open ND apx}), it follows that,
$$\big\langle X^I(z,\bar{z})-\hat{x}^I\big\rangle_{\rm lc}=\big(X^I(z,\bar{z})-x^I\big)_{\rm cl}=0,$$
with
\begin{equation*}
X^I(z,\bar{z}) = x^I-i\alpha'w^I\ln\frac{z}{\bar{z}}+i\sqrt{\frac{\alpha'}{2}}\sum_{n\neq0}\frac{\alpha_n^I}{n}\Big(\frac{1}{z^n}-\frac{1}{\bar{z}^n}\Big),
\end{equation*}
which shows that the open string coherent state vertex operators we have constructed are restricted to lie on a single D$p$-brane, and that for vertices stretched between two parallel D-branes of the same dimensionality one can still work with these vertex operators provided the exponential factor given in (\ref{eq:DirichletPathIntegrals}) is inserted into the path integral.

The position operator is not a gauge invariant quantity and so the corresponding covariant gauge position expectation value, although of the form (\ref{eq:<X>_coherent_lc}), would be a more complicated expression whose polarization tensors are not independent, being subject to the constraints $(\partial X)^2=(\bar{\partial}X)^2=0$. Therefore, the covariant position expectation value is not a particularly useful quantity in practice because the classical solutions we want to match vertex operators to are not known in covariant gauge. The angular momentum on the other hand is a gauge invariant operator, $[L_n,J^{\mu\nu}]=0$, and so a good consistency check is to show that both the covariant, $\langle J^{ab}\rangle_{\rm cov}$, and the lightcone, $\langle J^{ab}\rangle_{\rm lc}$, angular momentum expectation values are equal (in the unit norm representation) to the classical angular momentum, $J_{\rm cl}^{ab}$. Such an equivalence would support the conjecture that (\ref{eq:DDF_coherent})  and (\ref{eq:lc_coherent open}) are different manifestations of the same state and correspond classically to the lightcone gauge solution (\ref{eq:<X>_coherent_lc}). The total angular momentum operator is the integral of the current associated to Lorentz invariance over a spacelike curve, say $|z|^2=1$ in the coordinates $z=e^{-i(\sigma+i\tau)},\bar{z}=e^{i(\sigma-i\tau)}$, that cuts once across the string worldsheet \cite{GSW1}. For the open string,
\begin{equation}\label{eq:Jmunu}
\begin{aligned}
J^{\mu\nu}& =\frac{2}{\alpha'} \oint \dslash z X^{[\mu}\partial X^{\nu]},\\
S^{\mu\nu}&=-i\sum_{\ell=1}^{\infty}\frac{1}{\ell}\big(\alpha_{-\ell}^{\mu}\alpha_{\ell}^{\nu}-\alpha_{-\ell}^{\nu}\alpha_{\ell}^{\mu}\big),
\end{aligned}
\end{equation}
with $a^{[\mu\nu]}=\frac{1}{2}(a^{\mu\nu}-a^{\nu\mu})$ and $J^{\mu\nu}=L^{\mu\nu}+S^{\mu\nu}$. Due to the anti-symmetry there are no normal ordering ambiguities.  $L^{\mu\nu}$ is the zero mode contribution \footnote{\label{L_munu}In particular, in covariant gauge, $L^{\mu\nu}= x^{\mu}p^{\nu}-x^{\nu}p^{\mu}$ and in lightcone gauge, $L^{ij}= x^ip^j-x^jp^i$, $L^{-i}= x^-p^i-\frac{1}{2}\big(x^ip^--p^-x^i\big)$, $L^{-+}=\frac{1}{2}\big(x^-p^++p^+x^-\big)$ and $L^{i+}= x^ip^+$ which may be interpreted either classically or quantum-mechanically.} and we have used the doubling trick \cite{Polchinski_v1}. Notice furthermore that $S^{\mu\nu}=\sum_{\ell=1}^{\infty}\frac{2}{\ell}{\rm Im}\big(\alpha_{-\ell}^{\mu}\alpha_{\ell}^{\nu}\big)$. For simplicity focus on these non-zero mode components, $S^{\mu\nu}$, and consider first the components, $S^{AB}$. For the lightcone gauge classical computation we find, $S_{\rm cl}^{AB}=\sum_{n>0}\frac{2}{n}{\rm Im}\big(\lambda_n^{*A}\lambda_n^B\big),$ which follows from (\ref{eq:<X>_coherent_lc}) and (\ref{eq:Jmunu}). In the lightcone gauge the quantity 
$$
\langle S^{AB}\rangle_{\rm lc}\equiv\langle V(\lambda)|S^{AB} |V(\lambda)\rangle_{\rm lc},
$$
is computed using $S^{AB}=\sum_{\ell>0}\frac{2}{\ell}{\rm Im}\big(\alpha_{-\ell}^{A}\alpha_{\ell}^B\big),$ and  (\ref{eq:lc_coherent open}). Given that $|V(\lambda)\rangle_{\rm lc}$ is an eigenstate of the annihilation operators it follows immediately that $\langle S^{AB}\rangle_{\rm lc}=\sum_{n>0}\frac{2}{n}{\rm Im}\big(\lambda_n^{*A}\lambda_n^B\big)$. Finally, the covariant gauge quantity 
$$
\langle S^{AB}\rangle_{\rm cov}=\langle V(\lambda)|S^{AB} |V(\lambda)\rangle_{\rm cov},
$$ is also computed using $S^{AB}=\sum_{\ell>0}\frac{2}{\ell}{\rm Im}\big(\alpha_{-\ell}^{A}\alpha_{\ell}^B\big),$ and we are to identify $V(\lambda)$ with the covariant vertex operator (\ref{eq:DDF_coherent}), or, equivalently the operator state (\ref{eq:DDF_coherent op}). For this computation one may readily derive the following commutators  \cite{CornalbaCostaPenedonesVieira06}, 
$$
\big[\alpha_m^{A},A^B_n\big] = m\delta^{A,B}B^n_{m}, \qquad [A^A_n,B^m_{\ell}\big]=0=[B^n_m,B^{\ell}_r],
$$
with $B^{n}_{m} \equiv -i\oint \dslash z\,z^{m-1}\,e^{inq\cdot X(z)}$, see Appendix \ref{C}. Using these one can show primarily that 
\begin{equation}\label{eq:aiV}
\alpha_{m>0}^A\cdot V(\lambda)\cong\sum_{n=1}^{\infty}\frac{m}{n}\lambda_n^AB^{-n}_{m}\cdot V(\lambda).
\end{equation}
From the definition of $B^{-n}_{m}$ and $[A^A_n,B^m_{\ell}\big]=0$ follows the operator product, 
$$
B^{-n}_{m}\cdot V(\lambda)\cong \,\,:S_{n-m}(nq;z)\,e^{-inq\cdot X(z)}V(\lambda):
$$ 
From this latter expression and the properties (see Appendix \ref{SP} and \ref{C}), $S_0=1$ and $S_{n<0}=0$, we find that $B^{-n}_{\phantom{i}m}$ annihilates $V(\lambda)$ when $m>n$ and shifts the vacuum momentum, $p^a\rightarrow p^a-nq^a$, leaving the state otherwise unaltered, when $n=m$. From $(B^{-n}_m)^{\dagger}=B^{n}_{-m}$ we find that terms with $m>n$ similarly annihilate the out state, $V(\lambda)^{\dagger}$, in the expectation value $\langle S^{AB}\rangle_{\rm cov}$ where similar considerations apply. Therefore, only the term $n=m$ survives in the sum over $n$ in (\ref{eq:aiV}). We thus find the covariant gauge expectation value, $\langle S^{AB}\rangle_{\rm cov}=\sum_{n>0}\frac{2}{n}{\rm Im}\big(\lambda_n^{*A}\lambda_n^B\big)$. Collecting the classical, lightcone gauge and covariant gauge computations, we have shown that,
\begin{equation}\label{eq:Jcorr}
\langle S^{AB}\rangle_{\rm cov}=\langle S^{AB}\rangle_{\rm lc}=\sum_{n>0}\frac{2}{n}{\rm Im}\big(\lambda_n^{*A}\lambda_n^B\big)=S_{\rm cl}^{AB}.
\end{equation}

The angular momentum components in the longitudinal directions are similarly computed. For the lightcone gauge computation, 
$$
\langle S^{A-}\rangle_{\rm lc}=\langle V(\lambda)|S^{AB} |V(\lambda)\rangle_{\rm lc},
$$
one can use the commutator $[\alpha^-_{\ell},\alpha_{-n}^A]=n\alpha^A_{\ell-n}/(\sqrt{2\alpha'}p^+)$, but since $|V\rangle_{\rm lc}$ is an eigenstate of $\alpha_{n>0}^A$ with eigenvalue $\lambda_n^A$ it is advantageous to use the expression, $\sqrt{2\alpha'}\alpha_{\ell}^-=\frac{1}{2p^+}\sum_{m\in\mathbb{Z}}:\alpha_{m}^A \alpha_{\ell-m}^A:$, in $S^{A-}$. This then leads to, $\langle S^{A-}\rangle_{\rm lc}=\frac{1}{\sqrt{2\alpha'}p^+}\sum_{\ell>0}\sum_{m\in\mathbb{Z}}\frac{1}{\ell }{\rm Im}\big(\lambda_{\ell}^{*A}\,\lambda_{m}\!\cdot \!\lambda_{\ell-m}\big)$, with $\lambda^A_0\equiv \sqrt{2\alpha'}p^A$ as above. For the covariant gauge computation, 
$$
\langle S^{A-}\rangle_{\rm cov}=\langle V(\lambda)|S^{AB} |V(\lambda)\rangle_{\rm cov},
$$
to match to the lightcone gauge  we use lightcone \emph{coordinates} where, $q^{+}=q^A=0$ and $q^-=-1/(2\alpha'p^+)$ (which solve the constraints $q^2=0$ and $p\cdot q=1/(2\alpha')$). One can readily derive the commutators  \cite{CornalbaCostaPenedonesVieira06}, $$\big[\alpha_m^-,A^A_n\big] = n\sqrt{2\alpha'}q^-D_{m,n}^A,\quad\big[A^A_{\ell},D_{m,n}^B\big] = \ell\delta^{AB}E^{\ell+n}_m,$$
and $\big[A^A_{\ell},E^n_{m}\big] = 0,$ with $D^A_{m,n}$ and $E^n_m$ defined in Appendix~\ref{O}, from which follows the operator product,
\begin{equation}\label{eq:aaV}
\begin{aligned}
\alpha_{\ell}^-\cdot V(\lambda)\cong&\sqrt{2\alpha'}q^-\sum_{n=1}^{\infty}\Big(-\lambda_n\cdot D_{\ell,-n} \\
&+\sum_{m=1}^{\infty}\frac{1}{2}\lambda_n\cdot \lambda_mE^{-n-m}_{\ell}\Big)\cdot V(\lambda).
\end{aligned}
\end{equation}
Consider the second term in this expression. Given that $[A_{\ell},E^n_m]=0$ we may commute the $E^{-n-m}_{\ell}$ through to hit the vacuum, $e^{ip\cdot X(z)}$, where the following operator product is required,
\begin{equation}\label{eq:E-nell eigenvalue}
\begin{aligned}
&E^{-n-m}_{\ell}\cdot e^{ip\cdot X(z)}\cong\\
& :\sqrt{2\alpha'}q\cdot H_{n+m-\ell}\big((n+m)q;z\big)e^{i(p-n-m)\cdot X(z)}:,
\end{aligned}
\end{equation} 
with $\sqrt{2\alpha'}q\cdot H_{0}=1$ and $q\cdot H_{m<0}=0$, the polynomial $H_m$ having been defined in Appendix~\ref{SP}. (See also comments below (\ref{eq:opes BDE}).) In the expectation value, $\langle V^{\dagger}S^{A-}V\rangle$, this implies that we should only bring $E^{-n-m}_{\ell}$ to the right to hit $V(\lambda)$ if $n+m-\ell\leq 0$. Of these, the $n+m-\ell=0$ subset will shift the vacuum momentum, $p\rightarrow p-(n+m)q$, leaving the state otherwise unaltered, and the $n+m-\ell<0$ subset will annihilate it. Therefore of the terms with $n+m-\ell\leq 0$ in the sum over $m$ only the $m=\ell-n$ term will contribute. The remaining terms with, $n+m-\ell>0$, will not contribute either. These are to be commuted through to the out-state, $V^{\dagger}$, which is annihilated by them. In doing so these latter terms first encounter $\alpha_{-\ell}^A$ from $S^{A-}$. We here use the fact that $\langle V|\alpha^A_{-\ell}=\big(\alpha^A_{\ell}|V\rangle \big)^{\dagger}\cong \big(\sum_{n=1}^{\infty}\frac{\ell}{n}\lambda_n^AB^{-n}_{\phantom{a}\ell}|V\rangle\big)^{\dagger}=\sum_{n=1}^{\infty}\frac{\ell}{n}\lambda_n^{*A}\langle V| B^{n}_{-\ell}$, and $[B^n_{-\ell},E^{-m}_{r}]=0$, so that the quantities, $E^{-n-m}_{\ell}$, with $n+m-\ell>0$ commute freely through to hit and annihilate the out state, $V^{\dagger}$, and so indeed only the term $m=\ell-n$ will survive in the second term in (\ref{eq:aaV}) in the computation of $\langle S^{A-}\rangle$.

Next consider the first term in (\ref{eq:aaV}). On account of the operator product,
$$
D_{\ell,-n}^A\cdot e^{ip\cdot X(z)}\cong \,\,:H^A_{n-\ell}(nq;z)e^{i(p-nq)\cdot X(z)}:\,,
$$
and the properties, $H^A_{0}=p^A$ and $H^A_{n<0}=0$, we will commute the $D^A_{\ell,-n}$ through to hit the $e^{ip\cdot X(z)}$ vacuum when $n-\ell\leq0$. Of these the subset of $D^A_{\ell,-n}$ for which $n-\ell=0$ shifts the vacuum momentum, $p^a\rightarrow p^a-nq^a$, leaving the state otherwise unaltered, whereas the subset satisfying $n-\ell<0$ annihilates it. The $D_{\ell,-n}^a$ terms with $n-\ell>0$ are to be commuted through to the out state, $V^{\dagger}$, in the expectation value $\langle V^{\dagger}S^{A-}V\rangle$, just like we did above for the $E^{-n-m}_{\ell}$ terms with $n+m-\ell>0$. From the commutators, $\big[A^A_{\ell},D_{m,n}^B\big] = \ell\delta^{AB}E^{\ell+n}_m$ and $\big[A^A_{\ell},E^n_{m}\big] = 0$ we find that, $\left[D^A_{\ell,-n},\exp\left(\sum_{m>0}\frac{1}{m}\lambda_m\cdot A_{-m}\right)\right] = \sum_{m>0}\lambda_m^AE^{-n-m}_{\ell}$. For the terms with $n-\ell\leq0$, for which $D^A_{\ell,-n}\cdot e^{ip\cdot X}\cong\,:\delta_{n,\ell}\,\sqrt{2\alpha'}p^A\,e^{i(p-n)\cdot X(z)}$:, we find,
\begin{equation}\label{eq:lDV}
\begin{aligned}
\lambda_n\cdot &D_{\ell,-n}\cdot V(\lambda)\cong \sum_{m>0}\lambda_n\cdot\lambda_mE^{-n-m}_{\ell}\cdot V(\lambda)\\
&+:\delta_{n,\ell}\,\sqrt{2\alpha'}\lambda_n\cdot p\,e^{-inq\cdot X(z)}V(\lambda):\quad (n\leq\ell)
\end{aligned}
\end{equation}
Now, the same argument that applied to the second term in (\ref{eq:aaV}) applies to the first term in (\ref{eq:lDV}) and so again only the $m=\ell-n$ term will contribute in the sum over $m$ to the expectation value $\langle S^{A-}\rangle$. Finally, for the first term in (\ref{eq:aaV}), for which $n-\ell>0$, we  commute $\lambda_n\cdot D_{\ell,-n}$ through to the out state $V^{\dagger}$ using the fact that $[B^n_m,D^A_{\ell,-n}]=0$ and $V^{\dagger}\cdot D^A_{\ell,-n}\cong \sum_{m>0}\lambda^{A}_{-m} V^{\dagger}\cdot E^{-n+m}_{\ell}$. The same argument as above applies and only the term $m=n-\ell$ contributes in the sum over $m$ (which is consistent with $n-\ell>0$ as $m$ is positive).

Identifying $-q^-$ with $1/(2\alpha'p^+)$, the above considerations are summarized in the expression,
\begin{equation}\label{eq:S-A}
\begin{aligned}
\big\langle V^{\dagger}&S^{A-}V\big\rangle_{\rm cov}=\frac{1}{\sqrt{2\alpha'}p^+}\sum_{\ell>0}\frac{2}{\ell}{\rm Im}\Big\langle\Big(\lambda_{\ell}^{*A} B^{\ell}_{-\ell}\Big)\\
&\!\!\!\!\!\!\!\!\times\Big(\frac{1}{2}\sum_{n=1}^{\ell}\sum_{m>0}\lambda_n\cdot\lambda_{m}\delta_{n+m,\ell}E^{-\ell}_{\ell}+\lambda_{\ell}\cdot \lambda_0e^{-A\ell q\cdot X(0)}\\
&+\sum_{n=\ell+1}^{\infty}\sum_{m>0}\lambda_n\cdot \lambda_{-m}\delta_{n-m,\ell}E^{-\ell}_{\ell}\Big)\Big\rangle\\
&=\frac{1}{\sqrt{2\alpha'}p^+}\sum_{\ell>0}\sum_{m\in\mathbb{Z}}\frac{1}{\ell}{\rm Im}\big(\lambda_{\ell}^{*A} \lambda_m\cdot\lambda_{\ell-m}\big),
\end{aligned}
\end{equation}
and this is in agreement with the lightcone gauge and classical computation. In going from the first to the second equality in (\ref{eq:S-A}) there are a number of steps. Let us write $f(n,m)=\lambda_n\cdot \lambda_mE^{-\ell}_{\ell}$. Focus on the second parenthesis and recall from (\ref{eq:E-nell eigenvalue}) that one may replace $e^{-i\ell q\cdot X}$ in the second term with $E^{-\ell}_{\ell}$, which identifies the second term as $f(\ell,0)$. The delta function in the first term restricts the summations appearing, $\frac{1}{2}\sum_{n=1}^{\ell}\sum_{m>0}f(n,m)\delta_{n,\ell-m}= \frac{1}{2}\sum_{m=1}^{\ell-1}f(\ell-m,m)$, and when the resulting expression is combined with the second term, $\sum_{m=1}^{\ell-1}\rightarrow \sum_{m=0}^{\ell}$. Similarly, the delta function in the third term restricts the summations appearing according to $\sum_{n=\ell+1}^{\infty}\sum_{m>0}f(n,-m)\delta_{n,\ell+m} =\sum_{m<0}f(\ell-m,m)$. The second parenthesis in (\ref{eq:S-A}) is therefore equal to $\frac{1}{2}\sum_{m=0}^{\ell}f(\ell-m,m)+\sum_{m<0}f(\ell-m,m)$, half of the second term of which can be absorbed into the first term leading to $\frac{1}{2}\sum_{m=-\infty}^{\ell}f(\ell-m,m)+\frac{1}{2}\sum_{m<0}f(\ell-m,m)$. After a change of variables in the second term, $m'=m-\ell$ with $m'\in[\ell+1,\infty)$, these two terms can be combined into the expression $\frac{1}{2}\sum_{m\in\mathbb{Z}}f(\ell-m,m)$. On account of the fact that $\langle V^{\dagger}B^{\ell}_{-\ell}E^{-\ell}_{\ell}V\rangle=1$ it follows that the first equality in (\ref{eq:S-A}) implies the second.

Collecting the classical, lightcone gauge and covariant gauge computations, we have shown that the longitudinal components of the angular momentum for the classical, lightcone gauge and covariant gauge computations are in agreement,
\begin{equation}\label{eq:Jcorrlong}
\begin{aligned}
\langle S^{-A}&\rangle_{\rm cov}=\langle S^{-A}\rangle_{\rm lc}=S_{\rm cl}^{-A}\\
&=\frac{1}{\sqrt{2\alpha'}p^+}\sum_{\ell>0}\sum_{m\in\mathbb{Z}}\frac{1}{\ell}{\rm Im}\big(\lambda_{\ell}^{*A} \lambda_m\cdot\lambda_{\ell-m}\big)
\end{aligned}
\end{equation}
The non-zero mode contributions to the angular momentum components involving $S^{+-}$, and $S^{+A}$, are all vanishing in the chosen coordinate system where $q^+=0$ (and $q^-=-1/p^+$). Recall furthermore that $\lambda_0^i\equiv \sqrt{2\alpha'}p^i$.

We have shown that there is a one-to-one correspondence between the covariant vertex operators (\ref{eq:DDF_coherent}), lightcone gauge states (\ref{eq:lc_coherent open}) and classical macroscopic string evolution (\ref{eq:<X>_coherent_lc}) and (\ref{X lc_open}). The preceding angular momentum computations provide further support for the conjecture that the covariant and lightcone gauge descriptions are different manifestations of the same state, both of which have a classical interpretation.

\subsection{Closed String Coherent States}\label{CS}
In close analogy to the open string case above, we define a closed string \emph{coherent state}, $V(\lambda,\bar{\lambda},\dots)$, to be a state that \footnote{We are adopting the rather general definition of a coherent state as given in \cite{KlauderSkagerstam85} and minimally extend it to include the string theory requirements. For instance, under this definition, coherent states need not (in general) be eigenstates of the annihilation operators, $\alpha_{n>0}^{\mu}$, $\tilde{\alpha}_{n>0}^{\mu}$, in order for this definition to be satisfied.}:
\begin{itemize}
\item[(a)] is specified by a set of \emph{continuous} labels $(\lambda,\bar{\lambda})=\{\lambda_n^i,\bar{\lambda}_n^i\}$ (with $\lambda$ and $\bar{\lambda}$ associated to the left- and right-moving modes respectively of the string);
\item[(b)] there must exist a resolution of unity,
\begin{equation*}
\mathds{1} =\suminnt\int d\lambda  d\bar{\lambda} \big|V(\lambda,\bar{\lambda},\dots)\big\rangle \big\langle V(\lambda,\bar{\lambda},\dots)\big|,
\end{equation*}
so that the $V(\lambda,\bar{\lambda},\dots)$ span the string Hilbert space, $\mathcal{H}$;
\item[(c)] it must transform correctly under all symmetries of the bosonic (or super-) string.
\end{itemize}
The dots $``\dots"$ in $V(\lambda,\bar{\lambda},\dots)$ allow for the possibility that the vertex operator depends on additional continuous or discrete quantum numbers and these are all to be summed over in the completeness relation. (We will often not exhibit these latter labels explicitly, and hence write instead $V(\lambda,\bar{\lambda})$, or even $V_{\lambda\bar{\lambda}}$, all of which refer to the same object $V(\lambda,\bar{\lambda},\dots)$.) The unit operator on the left is defined with respect to $\mathcal{H}$ \footnote{We are being pedantic here for a subtle reason that will become clear later. Recall that the Hilbert space $\mathcal{H}$ is in general a background dependent quantity, and so the explicit realization of the unit operator, $\mathds{1}$, is also background dependent.}, $\mathds{1}\cdot \big|V(\lambda,\bar{\lambda})\big\rangle\equiv  \big|V(\lambda,\bar{\lambda})\big\rangle$. The measures for the case of interest explicitly read $d\lambda  d\bar{\lambda} =\prod_{n,i}\frac{d^2\lambda_n^id^2\bar{\lambda}_n^i}{N}$ with $N$ a to-be determined  normalization and as usual $d^2\lambda_n^i=id\lambda_n^i\wedge d\lambda_n^{*i}$ (no sum over $i$), and so on.

In the next two subsections we construct two realizations of closed string covariant coherent states that satisfy the above definition.

\subsection{DLCQ Coherent States}\label{CSLCB}
We next construct closed string coherent states that satisfy the above definition. Our first approach will be naive and we will discover that internal consistency requires the underlying spacetime manifold be lightlike-compactified: 
$$
X^-\sim X^-+2\pi R^-.
$$ 
Quantization on a lightlike compactified background is known as `discrete lightcone quantization' (DLCQ) \cite{PauliBrodsky85,Susskind97,Seiberg97,HellermanPolchinski99}. In the following section we shall make the appropriate refinements and construct coherent states in a fully non-compact Minkowski spacetime background.

The closed string coherent state candidate that we consider in this section is obtained by joining two copies of the open string state (\ref{eq:DDF_coherent}),
\begin{equation}\label{eq:DDF_coherent_closed}
\begin{aligned}
V(\lambda,\bar{\lambda}&)=\frac{g_c}{\sqrt{2p^+\mathcal{V}_{d-1}}}\,C_{\lambda\bar{\lambda}}\exp\Big(\sum_{n=1}^{\infty}\frac{1}{n}\lambda_n\cdot A_{-n}\Big)\\
&\times\exp\Big(\sum_{m=1}^{\infty}\frac{1}{m}\bar{\lambda}_m\cdot \bar{A}_{-m}\Big)\,e^{ip\cdot X(z,\bar{z})},
\end{aligned}
\end{equation}
with the normalization, 
$$
C_{\lambda\bar{\lambda}}=\exp\Big(\sum_{n=1}^{\infty}-\frac{1}{2n}|\lambda_n|^2-\frac{1}{2n}|\bar{\lambda}_n|^2\Big),
$$ 
chosen such that 
if we write $V(z,\bar{z})=V(\lambda,\bar{\lambda},p)$, the most singular term in the operator product expansion is as in (\ref{eq:VVope normalization}),
\begin{equation}\label{eq:norm V coh}
V(z,\bar{z})\cdot V(0,0)\cong \Big(\frac{g_c^2}{2p^+\mathcal{V}_{d-1}}\Big)\frac{1}{|z|^4}+\dots,
\end{equation}
corresponding to `one string in volume $\mathcal{V}_{d-1}$' as required by unitarity of the $S$-matrix, which was discussed in Sec.~\ref{sec:NSW}. In operator language, we have:
$$
\langle V(\lambda,\bar{\lambda},p)|V(\lambda,\bar{\lambda},p)\rangle =1,\qquad |0,0;p\rangle \cong g_c\,e^{ip\cdot X(z,\bar{z})}.
$$
This corresponds to a relativistic unit norm normalization with, see (\ref{eq:tach and vac norm lcc}),
$$ 
\langle 0,0;p'|0,0;p\rangle = 2p^+(2\pi)\delta({p^+}'-p^+)(2\pi)^{d-2}\delta^{d-2}({\bf p}'-{\bf p}).
$$
Furthermore, $(\lambda,\bar{\lambda})=\{\lambda_n^i,\bar{\lambda}_n^i\}$, are the polarization tensors, defined by, $\lambda_n\cdot q=0$, $\lambda^*_n=\lambda_{-n}$, and $\sum_{n=1}^{\infty}|\lambda_n|^2<\infty$, and similarly for the anti-holomorphic sector $\{\bar{\lambda}_n^i\}$. The real vectors $p^{\mu}$ and $q^{\mu}$ are as usual subject to the constraints (\ref{eq:pq cond_closed}), repeated here for convenience: $p\cdot q=2/\alpha'$, $q^2=0$, and $p^2=4/\alpha'$. 

First let us prove that the vertex operator (\ref{eq:DDF_coherent_closed})  is a coherent state by showing that the defining properties (a-c) above are satisfied. (a) is trivially satisfied, the state is specified by the set of continuous labels $(\lambda,\bar{\lambda})=\{\lambda_n^i,\bar{\lambda}_n^i\}$ and remains normalized for arbitrary values provided \cite{Calucci89} $\sum_{n=1}^{\infty}|\lambda_n|^2+|\bar{\lambda}_n|^2<\infty$. To prove that (b) is satisfied note that primarily that $V(\lambda,\bar{\lambda})$ is an eigenstate of the annihilation operators, $A_{n>0}^i\cdot V\cong \lambda_{n}^iV$ and $\bar{A}_{n>0}^i\cdot V\cong \bar{\lambda}_{n}^iV$, which follows from the DDF operator commutation relations (\ref{eq: [A_nA_m]}) and the corresponding anti-holomorphic expression with $\bar{A}_n$ replacing $A_n$. Therefore, we find the following inner product, \begin{equation}\label{eq:<VV> coh closed}
\begin{aligned}
\big\langle &V(\lambda,\bar{\lambda},p')|V(\zeta,\bar{\zeta},p)\big\rangle =\\
&= \delta_{p',p} C_{\lambda\bar{\lambda}} C_{\zeta\bar{\zeta}}\exp\Big(\sum_{n>0}\frac{1}{n}\lambda^*_{n}\cdot\zeta_{n}+\frac{1}{n}\bar{\lambda}^*_{n}\cdot\bar{\zeta}_{n}\Big),
\end{aligned}
\end{equation}
which reduces to unity when $(\lambda,\bar{\lambda})=(\zeta,\bar{\zeta})$ and $p'=p$. Note that $\delta_{p',p}$ is a \emph{Kronecker} delta which reduces to unity when ${p^+}'=p^+$ and ${\bf p}'={\bf p}$ and vanishes otherwise. By then forming appropriate inner products and integrating we find that there exists the completeness relation, \label{eq:completeness}
\begin{equation}\label{eq:completeness closed}
\begin{aligned}
\mathds{1} &=\mathcal{V}_{d-1}\int_0^{\infty} \frac{dp^+}{2\pi}\int_{\mathbb{R}^{24}}\frac{d^{24}{\bf p}}{(2\pi)^{24}}\\
&\times \int \bigg(\prod_{n,A}\frac{d^2\lambda_n^A}{2\pi n}\bigg)\bigg(\prod_{n,A}\frac{d^2\bar{\lambda}_n^A}{2\pi n}\bigg) \big|V(\lambda,\bar{\lambda},p)\big\rangle \big\langle V(\lambda,\bar{\lambda},p)\big|,
\end{aligned}
\end{equation}
with $n=\{1,2,\dots,\infty\}$ 
\footnote{We shall occasionally write $V_{\lambda\bar{\lambda}}(p)$, $V(\lambda,\bar{\lambda})$, $V(\lambda,\bar{\lambda};p)$, or even $V(z,\bar{z})$ (with $z,\bar{z}$ the worldsheet location where the vertex is inserted) to denote the same object $V(\lambda,\bar{\lambda},p)$.}. The phase space integrals are precisely as anticipated from Sec.~\ref{sec:NSW}, and in particular (\ref{eq:sumVdp lc}), for the sum over single string states. In the case of closed string coherent states therefore we see that the additional sums over quantum numbers in (\ref{eq:sumVdp lc}) correspond to integrals over the polarization tensors:
$$
\suminnt = \int \bigg(\prod_{n,A}\frac{d^2\lambda_n^A}{2\pi n}\bigg)\bigg(\prod_{n,A}\frac{d^2\bar{\lambda}_n^A}{2\pi n}\bigg).
$$

Finally, to show that (c) is satisfied we must prove that $V(\lambda,\bar{\lambda})$  satisfies the Virasoro constraints, $L_0\cdot V\cong V$, $L_{n>0}\cdot V\cong0$. These are trivially satisfied given that: the DDF operators commute with the $L_n$, $\bar{L}_n$ for all $n$, and the vacuum $e^{ip\cdot X(z,\bar{z})}$ is physical,  $L_0\cdot e^{ip\cdot X}\cong e^{ip\cdot X}$, $L_{n>0}\cdot e^{ip\cdot X}\cong0$. Similar results hold for the antiholomorphic sector with $\bar{L}_n$ replacing $L_n$. Therefore, the vertex (\ref{eq:DDF_coherent_closed}) is a coherent state and respects the string theory symmetries.

\subsubsection{Functional Representation}
We postulated that closed string covariant coherent states are described by the vertex operator (\ref{eq:DDF_coherent_closed}). These vertices however are not what we are looking for, and to see why let us normal order $V(\lambda,\bar{\lambda})$. To simplify the computation we initially assume that $\lambda_{n>0}\cdot \lambda_{m>0}=0$ and similarly for the antiholomorphic sector, and then generalize the result. The normal ordering procedure has been explained in great detail in Sec.~\ref{DDF} for arbitrary mass eigenstates, the difference here being that the coherent state $V(\lambda,\bar{\lambda})$ is instead a linear superposition of mass eigenstates. As in the open string, the normal ordered version of (\ref{eq:DDF_coherent_closed}) is obtained by using the integral representation of the DDF operators (\ref{eq: DDF As}), the integration contour being taken around the vacuum $e^{ip\cdot X(z)}$ and $e^{ip\cdot X(\bar{z})}$ for the holomorphic and antiholomorphic sectors respectively. Holomorphy then allows us to shrink the contours and hence the computation only requires knowledge of the leading behaviour of the integrand close to the vacuum, which is determined by operator product expansions using the scalar propagator, $\langle X^{\mu}(z,\bar{z})X^{\nu}(w,\bar{w})\rangle = -\frac{\alpha'}{2}\eta^{\mu\nu}\ln |z-w|^2.$ This procedure leads to,
\begin{equation}\label{eq:DDF_coherent_cov_closed_transverse}
\begin{aligned}
V(&\lambda,\bar{\lambda})=\frac{g_c}{\sqrt{2p^+\mathcal{V}_{d-1}}}\\
&\times C_{\lambda\bar{\lambda}}\exp\Big(\sum_{n=1}^{\infty}\frac{1}{n}\lambda_n\cdot H_{n}(z)e^{-inq\cdot X(z)}\Big)\\
&\times\exp\Big(\sum_{m=1}^{\infty}\frac{1}{m}\bar{\lambda}_m\cdot \bar{H}_{m}(\bar{z})e^{-inq\cdot X(\bar{z})}\Big)\,e^{ip\cdot X(z,\bar{z})}.
\end{aligned}
\end{equation}
More generally, (i.e.~had we not assumed that $\lambda_{n>0}\cdot \lambda_{m>0}=0$) we would have found instead:
$$
V(\lambda,\bar{\lambda})=\frac{g_c}{\sqrt{2p^+\mathcal{V}_{d-1}}}\,C_{\lambda\bar{\lambda}}U(\lambda)\bar{U}(\bar{\lambda}) e^{ip\cdot X(z,\bar{z})},
$$ 
with $U(\lambda)$ defined below (\ref{eq:V(lambda)A^g e^ipX gen norm ord}),
\begin{widetext}
$$
U(\lambda)=\sum_{g=0}^{\infty}
\sum_{a=0}^{\lfloor g/2\rfloor}\frac{1}{a!(g-2a)!}\Big(\sum_{n,m>0}\frac{1}{2nm}\lambda_n\cdot \lambda_m\,\mathbb{S}_{n,m}\,e^{-i(n+m)q\cdot X(z)}\Big)^a\Big(\sum_{n>0}\frac{1}{n}\lambda_n\cdot H_n\,e^{-inq\cdot X(z)}\Big)^{g-2a},
$$
\end{widetext}
and $\bar{U}(\bar{\lambda})$ given by a similar expression with $\bar{\lambda}_m^i$, $\bar{z}$, $\bar{\mathbb{S}}_{m,m}(\bar{z})$ and $\bar{H}_m^i(\bar{z})$ replacing the corresponding holomorphic quantities. Note that the positive integers $n,m$ need not be equal.

\subsubsection{DLCQ Coherent State Properties}
The underlying Hilbert space consists of the states we are superimposing in order to construct the closed string coherent states. These can be obtained by series expanding the exponentials which leads to an expression of the form,
\begin{equation}\label{eq:mom op string coh_closed}
\begin{aligned}
V(\lambda,&\bar{\lambda})\propto\sum_{\{s_1,s_2,\dots\}=0}^{\infty}
{\rm Pol}[\partial^{\#}X]e^{i(p-\sum_nns_nq)\cdot X(z)}\\
&\times\sum_{\{\bar{s}_1,\bar{s}_2,\dots\}=0}^{\infty}\overline{\rm Pol}[\bar{\partial}^{\#}X]\,e^{i(p-\sum_mm\bar{s}_mq)\cdot X(\bar{z})},
\end{aligned}
\end{equation}
with ${\rm Pol}[\partial^{\#}X]$ and $\overline{\rm Pol}[\bar{\partial}^{\#}X]$ being certain polynomials of the arguments which depend on the sets of uncorrelated positive integers $\{s_1,s_2,\dots\}$ and $\{\bar{s}_1,\bar{s}_2,\dots\}$ respectively. Let us write $N=\sum_{n=1}^{\infty}ns_n$ and $\bar{N}=\sum_{n=1}^{\infty}n\bar{s}_n$ for an arbitrary sequence of positive integers $\{s_1,s_2,\dots\}$ and $\{\bar{s}_1,\bar{s}_2,\dots\}$ respectively. We learn that the left- and right-moving momenta associated to a given mass eigenstate in (\ref{eq:mom op string coh_closed}) satisfy, $k_{\rm L}^{\mu}-k_{\rm R}^{\mu}=-\big(N-\bar{N}\big)q^{\mu}$, the associated total momentum being $k^{\mu}=\frac{1}{2}(k^{\mu}_{\rm L}+k_{\rm R}^{\mu})$.  It is therefore clear that we are super-imposing mass eigenstates with asymmetric left-right momenta and so the manifold in which the coherent states live is in fact \emph{compact}. This is an $S^1$ compactification in a direction specified by the \emph{null} vector $q^{\mu}$. We can read off the radius of compactification directly from $k_{\rm L}-k_{\rm R}$ or equivalently one may compute it by applying the operator, $\oint \big(\dslash z\, \partial X^{\mu}+\dslash \bar{z}\,\bar{\partial} X^{\mu}\big)$, (that measures the total change in $X^{\mu}(z,\bar{z})$ in going once around the string \cite{Polchinski_v1}) to a mass eigenstate and identify the corresponding eigenvalue with $ R^{\mu}w$, with $w$ the winding number. This leads to $w = N-\bar{N}$ and $R^{\mu}=-\frac{\alpha'}{2}\,q^{\mu}$ and therefore: $R^2=0.$ We learn that the underlying  spacetime manifold is compactified in a \emph{light-like} spacetime direction, that is we are considering the DLCQ \cite{PauliBrodsky85} of string theory.  Lightlike compactifications show up in the connection of M(atrix) models to string theories: DLCQ of M-theory has been conjectured \cite{Susskind97} to be equivalent to U(N) super Yang-Mills at \emph{finite} N. (See for example, \cite{PauliBrodsky85,Susskind97,Seiberg97,HellermanPolchinski99} and also \cite{GrignaniOrlandPaniakSemenoff00,Semenoff00, Semenoff02}.) Although lightlike compactifications are in general rather non-trivial \cite{HellermanPolchinski99}, various properties of a vertex operator in a lightlike compactified spacetime can be extracted rather straightforwardly as we show next.

\begin{figure*}
\includegraphics[width=.7\textwidth]{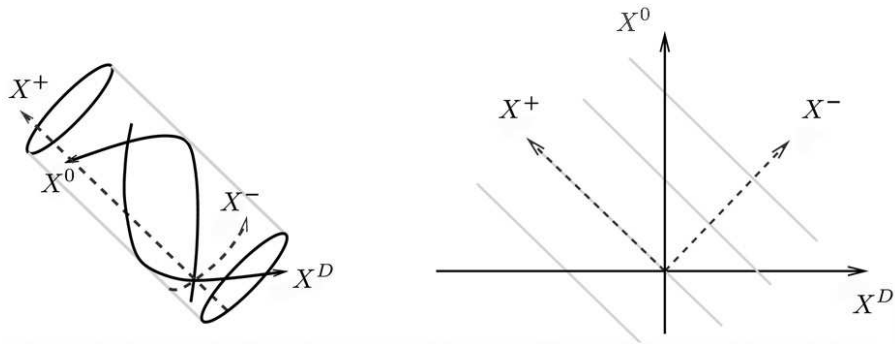}\caption{Lightlike spacetime compactification. The two-dimensional plane $X^0$-$X^D$ is shown. In the figure on the right we are to identify the parallel grey lines such that $X^-\sim X^-+2\pi R^-$. The future lightcone of a given spacetime event is specified by the dashed lines. The aforementioned identification leads to the equivalent $S^1\times \mathbb{R}$ spacetime cylinder on the left. Signals slower than the speed of light and lightlike signals in the negative $X^D$ direction always propagate up the cylinder in the positive $X^+$ direction. Lightlike signals in the positive $X^D$ direction are stuck at $X^+={\rm const}$ hypersurfaces. Causality is not violated (the spacetime is marginally causal).\label{fig:nullcyl}}
\end{figure*}

To become more explicit go to a frame where $q^+=q^i=0$ and $q^-=-\frac{2}{\alpha'}R^-$ which implies the identification (with $X^+$ non-compact),
\begin{equation}
X^-\sim X^-+2\pi R^-.
\end{equation}
This is shown schematically in Fig.~\ref{fig:nullcyl}. 
Let us go to the rest frame (in the lightcone gauge sense) where in addition, $p^i=0$. With this and the above ansatz for $q^{\mu}$ we can solve the constraints $p^2=4/\alpha'$, $p\cdot q=2/\alpha'$ and $q^2=0$ which lead to the following expressions for the \emph{total} momentum of a lightlike compactified mass eigenstate,
\begin{equation}\label{eq:massshell}
k^0 = \frac{1}{\sqrt{2}}\Big(\frac{1}{R^-}+m^2R^-\Big),\quad k^D = \frac{1}{\sqrt{2}}\Big(\frac{1}{R^-}-m^2R^-\Big),
\end{equation}
and $k^i=0$, with $m^2= \frac{2}{\alpha'}(N+\bar{N}-2)$, the mass squared of the particular mass eigenstate in the superposition (\ref{eq:mom op string coh_closed}). That $m^2$ does not depend on $R^-$ naively seems to imply that lightlike compactification does not change the mass spectrum of the uncompactified theory. However, the $L_0-\bar{L}_0$ Virasoro constraint is already satisfied by the above state and so $N$ need \emph{not} equal $\bar{N}$: the Hilbert space, $\mathcal{H}$, contains all the usual states where $N=\bar{N}$ (and hence $w=0$) but also includes additional states for which $N\neq\bar{N}$ (and $w\neq 0$) without breaking conformal invariance. 

The Hilbert space $\mathcal{H}$ admits the orthogonal decomposition, 
$$
\mathcal{H}=\bigoplus_{w\in\mathbb{Z}}\mathcal{H}_w,
$$ 
such that vertices $V_w\in \mathcal{H}_w$ wind around the lightlike direction with winding number $w$ \footnote{The decomposition is orthogonal in the sense that $\langle V_m|V_n\rangle = \delta_{m,n}$.}. Given that winding number is conserved (i.e.~commutes with the worldsheet Hamiltonian, $[L_0+\bar{L}_0-2,\hat{W}]\cdot V_w\cong0$), suggests that we can project out the winding states and thus obtain a  vertex operator, $V_0\in \mathcal{H}_0$, with (as we show below, see p.~\pageref{eq:V_0}) coherent state properties which can be embedded in fully non-compact spacetime \footnote{DS would like to thank Joe Polchinski for suggesting that the projected states should also have coherent state properties.}.

Given that (\ref{eq:massshell}) is not of the standard form, $k=n/R$, for the total momentum in a compact dimension of radius $R$ \cite{Polchinski_v1}, one may wonder whether the corresponding wavefunctions are still single-valued \footnote{The authors would like to thank Diego Chialva for raising this question.} -- single-valuedness of the wavefunction is the reason as to why one enforces $k=n/R$ in the first place. That they are single valued can be seen as follows. Translations along a compact dimension whose direction is specified by the vector $R^{\mu}$ are generated by,
$\exp\big(2\pi iR\cdot \hat{p}\big):X^{\mu}(z,\bar{z})\rightarrow X^{\mu}(z,\bar{z})+2\pi R^{\mu},$ with $\hat{p}^{\mu}$ the total Noether momentum, 
$\hat{p}^{\mu}=\frac{1}{\alpha'}\oint \big(\dslash z\,\partial X^{\mu}(z)-\dslash \bar{z}\,\bar{\partial}X^{\mu}(\bar{z})\big)$. The excitations that appear in $V(\lambda,\bar{\lambda})$ (i.e.~the polynomials of $\partial^{\#}X,\bar{\partial}^{\#}X$) commute with $\hat{p}$ and so single-valuedness of the vertex operator amounts to showing that:
\begin{equation*}
\begin{aligned}
\exp\big(2\pi i R&\cdot \hat{p}\big) \exp\big(ik_{\rm L}\cdot X(z)+ik_{\rm R}\cdot X(\bar{z})\big)\\
= &\exp\big(ik_{\rm L}\cdot X(z)+ik_{\rm R}\cdot X(\bar{z})\big),
\end{aligned}
\end{equation*}
for any mass eigenstate in the superposition. Carrying out the operator products on the left hand side (with the contour integrals encircling $z,\bar{z}$ and $k_{\rm L}=p-Nq$, $k_{\rm R}=p-\bar{N}q$) it follows that the above statement holds true for the individual mass eigenstates with lightlike winding and hence is also true for the closed string coherent states. We conclude that $V(\lambda,\bar{\lambda})$ is indeed single-valued under translations around the compact direction \footnote{This proves that the solution to the single-valuedness requirement that one normally considers, $k=n/R$, must be generalized in lightlike compactified spacetimes.}.

Curiously, lightlike compactification seems to be invisible at the classical level when,
$$
\sum_{n=1}^{\infty}|\lambda_n|^2=\sum_{n=1}^{\infty}|\bar{\lambda}_n|^2,
$$
which is none other than the statement of \textquotedblleft classical level matching\textquotedblright, $\langle N\rangle = \langle\bar{N}\rangle$, because 
$$
\langle N\rangle=\sum_{n=1}^{\infty}|\lambda_n|^2\qquad{\rm  and} \qquad\langle\bar{N}\rangle = \sum_{n=1}^{\infty}|\bar{\lambda}_n|^2
$$ 
are none other that the expectation values of the number operators, 
$
N=\sum_{n>0}\alpha_{-n}\cdot\alpha_n$ 
and 
$
\bar{N}=\sum_{n>0}\tilde{\alpha}_{-n}\cdot\tilde{\alpha}_n.
$ 
Furthermore, classical level matching is required for consistency (see below). One way of seeing that lightlike compactification is invisible at the classical level is by directly computing the expectation value $\big\langle \hat{p}^-_{\rm L}\big\rangle-\big\langle\hat{p}^-_{\rm R}\big\rangle$ (with respect to the state (\ref{eq:DDF_coherent_closed})) and showing that it vanishes, as this would imply that $\langle X^-(z,\bar{z})-x^-\rangle= -i\langle\hat{p}^-_{\rm L}\rangle\ln z-i\langle\hat{p}^-_{\rm R}\rangle\ln \bar{z}+\dots$ is single valued as one traverses a spacelike direction of the worldsheet which is classically only possible if $X^-$ is non-compact, i.e.~if $\langle X^-(z,\bar{z})-x^-\rangle= -i\langle\hat{p}^-\rangle\ln |z|^2+\dots$.

On account of (\ref{eq:<VV> coh closed}), it follows that \footnote{For completeness we note also that $\frac{\alpha'}{2}\big\langle\hat{p}^-_{\rm L}\big\rangle=\left[\big\langle N\big\rangle-1\right]R^-$, $\frac{\alpha'}{2}\big\langle\hat{p}^-_{\rm R}\big\rangle=\big[\big\langle \bar{N}\big\rangle-1\big]R^-$ and $ \big\langle\hat{p}^+_{\rm L}\big\rangle=\big\langle\hat{p}^+_{\rm R}\big\rangle = 1/R^-$ with $\hat{p}^{\mu}=\frac{1}{2}\big(\hat{p}_{\rm L}^{\mu}+\hat{p}_{\rm R}^{\mu}\big)$ and $\hat{p}^{\pm}_{\rm L,R}=\frac{1}{\sqrt{2}}\big(\hat{p}_{\rm L,R}^0\pm \hat{p}_{\rm L,R}^D\big)$.},
\begin{equation}\label{eq:X^-cov}
\begin{aligned}
\big\langle &X^-(z,\bar{z})-x^-\big\rangle\\
&= -i\Big(\big\langle N\big\rangle-1\Big) R^-\ln z-i\Big(\big\langle \bar{N}\big\rangle-1\Big) R^-\ln \bar{z}.
\end{aligned}
\end{equation}
Notice that only zero modes contribute to the position expectation value in the covariant gauge version of the state (\ref{eq:DDF_coherent_closed}) for a reason that was first realized in \cite{Blanco-PilladoIglesiasSiegel07}, and which we expand on in the following paragraph. For the $X^+$ direction we find,
\begin{equation}\label{eq:X^+cov}
\begin{aligned}
\big\langle X^+(z,\bar{z})-x^+\big\rangle= -\frac{\alpha'}{2}\frac{i}{R^-}\ln |z|^2.
\end{aligned}
\end{equation}

Recall that the operator $L_0-\bar{L}_0$ generates spacelike worldsheet translations, 
\begin{equation}\label{eq:[Lo-L0bar]..}
\big[L_0-\bar{L}_0,X^{\mu}(z,\bar{z})\big]=\big(z\partial -\bar{z}\bar{\partial}\big)X^{\mu}(z,\bar{z}),
\end{equation}
and that one of the physical state conditions is that states be invariant under such translations, $$
\exp [-i\epsilon(L_0-\bar{L}_0)]\cdot V\cong V;
$$
infinitesimally, $|\epsilon|\ll1$, we have $(L_0-\bar{L}_0)\cdot V\cong 0$. Computing the expectation value, $\big\langle\big[L_0-\bar{L}_0,X^{\mu}(z,\bar{z})\big]\big\rangle=\big(z\partial -\bar{z}\bar{\partial}\big)\big\langle X^{\mu}(z,\bar{z})\big\rangle$, with respect to a physical state $V$ it then follows that 
$$
\big(z\partial -\bar{z}\bar{\partial}\big)\big\langle X^{\mu}(z,\bar{z})\big\rangle=0
$$
must be satisfied by any such state.  This in turn explains why there are only zero mode contributions in (\ref{eq:X^-cov}) and (\ref{eq:X^+cov}) (non-zero mode contributions would violate this condition), and secondly enforces \emph{classical level matching}, 
\begin{equation}\label{eq:N=Nbar}
\langle N\rangle = \langle \bar{N}\rangle,
\end{equation}
so as to ensure that the operator $\big(z\partial -\bar{z}\bar{\partial}\big)$ annihilates (\ref{eq:X^-cov}). Given that $V(\lambda,\bar{\lambda})$ has an effective mass given by 
$\langle m^2\rangle = \frac{2}{\alpha'}(\langle N\rangle+\langle\bar{N}\rangle-2)$ it follows that the full momenta are given by, $\langle \hat{p}^-\rangle = \frac{1}{2}\langle m^2\rangle R^-$, $\langle \hat{p}^+\rangle = 1/ R^-$, enabling one to write: 
\begin{equation}\label{eq:<Xpm>}
\big\langle X^{\pm}(z,\bar{z})-x^{\pm}\big\rangle=-i\frac{\alpha'}{2}\langle p^{\pm}\rangle\ln|z|^2.
\end{equation}
This implies that indeed as claimed above lightlike compactification seems to be invisible at the classical level. However, this result is not unique to lightlike compactifications. In particular, notice that the reasoning following (\ref{eq:[Lo-L0bar]..}) also applies in the case of spacelike compactifications, $x^i\sim x^i+2\pi R$. In this latter case, in particular, one finds the consistency requirement: 
$$
\langle p_L^i\rangle=  \langle p_R^i\rangle.
$$ 
Curiously, this seems to imply that toroidal compactification in general is invisible in such expectation values. 

That only zero modes contribute to the expectation values (\ref{eq:X^-cov}) and (\ref{eq:X^+cov}) of course does not mean that the coherent state (\ref{eq:DDF_coherent_closed}) does not have a classical interpretation, but rather implies that the condition for classicality, $\big\langle X^{\mu}(z,\bar{z})\big\rangle=X^{\mu}_{\rm cl}(z,\bar{z})$, with $\partial\bar{\partial} X^{\mu}_{\rm cl}(z,\bar{z})=0$ is not compatible with the symmetries of closed string theory when the gauge choice (covariant gauge in this example) does not fix the invariance under spacelike worldsheet translations \cite{Blanco-PilladoIglesiasSiegel07}. Note that \emph{any} covariant vertex operator must satisfy $\big(z\partial -\bar{z}\bar{\partial}\big)\big\langle X^{\mu}(z,\bar{z})\big\rangle=0$, whether or not it has a classical interpretation. To get round this, one may fix the invariance of the state under such translations (as done in \cite{Blanco-PilladoIglesiasSiegel07}) but this is somewhat messy and not practical for general states. Alternatively, one may pick a gauge that explicitly breaks the invariance under such translations from the outset, e.g.~static gauge. To see this notice that \footnote{The following was suggested by Ashoke Sen and DS would like to thank him for extensive very helpful discussions of these issues.}  in static gauge, e.g.~$X^0=\alpha'p^0\tau$, $X^D=R\sigma$ and $X^D\sim X^D+2\pi R$, from the outset where it is manifest that spacelike worldsheet translation invariance, $\sigma\rightarrow \sigma+s$, is broken by the gauge choice. Here $\langle X^i(\sigma,\tau)\rangle=X_{\rm cl}^i(\sigma,\tau)$ can be satisfied non-trivially because in static gauge states of the form $e^{\lambda_n^i \alpha^i_{-n}}e^{\bar{\lambda}_n^i\tilde{\alpha}^i_{-n}}|0,0;p^i,p^D_{\rm L},p_{\rm R}^D\rangle$ are physical without requiring the existence of a lightlike compactification. Unfortunately, it is not known how to quantize the string in static gauge unless (starting from the Nambu-Goto action) one restricts to \emph{small} fluctuations transverse to $X^0,X^D$ with $R$ large, in which case the leading term in the action becomes quadratic in the fields $X^i$ and the path integral can be carried out perturbatively in $1/R$. We would like to discuss the construction of quantum states which correspond to arbitrary classical solutions (e.g.~solutions with cusps where the above expansion would presumably not suffice) and so this is not the approach we shall take here. A better solution is possibly to instead replace the definition of classicality, $\big\langle X^{\mu}(z,\bar{z})\big\rangle=X^{\mu}_{\rm cl}(z,\bar{z})$, with \footnote{This was suggested by Joe Polchinski and the authors are very grateful to him for this suggestion.},
\begin{equation}\label{eq:class_defn_}
\big\langle \!:\!X^{\mu}(\sigma',\tau)X^{\nu}(\sigma,\tau)\!:\!\big\rangle =\int_0^{2\pi} \!\!\!\dslash s\,X^{\mu}_{\rm cl}(\sigma'-s,\tau)X^{\nu}_{\rm cl}(\sigma-s,\tau),
\end{equation}
modulo zero mode contributions (recall that $z=e^{-i(\sigma+i\tau)}$, $\bar{z}=e^{i(\sigma-i\tau)}$). Rather than fixing the invariance under $\sigma$-translations on the quantum side (as done in \cite{Blanco-PilladoIglesiasSiegel07}) we average over $\sigma$-translations on the classical side. 

The definition for classicality (\ref{eq:class_defn_}) is appropriate for states in any gauge (e.g.~covariant or lightcone gauge) that does not fix the invariance under spacelike worldsheet translations and we will be making use of it when we present the construction of coherent states in non-compact spacetimes. For the states (\ref{eq:DDF_coherent_closed}) however there is yet another solution which is even simpler -- the solution is to go to lightcone gauge, because in lightcone gauge the presence of lightlike compactification breaks the invariance under such translations thus making the classical-quantum map, $\big\langle X^{\mu}(z,\bar{z})\big\rangle=X^{\mu}_{\rm cl}(z,\bar{z})$, possible.\label{classicality discussion}

Before we elaborate on the lightcone gauge construction, we would like to point out that one should be careful in drawing conclusions from statements of the form (\ref{eq:<Xpm>}) when the expectation value is evaluated in covariant gauge. One can argue that it is not permissible to compute the expectation value of (\ref{eq:[Lo-L0bar]..}) given that $X^{\mu}(z,\bar{z})$ is not a well defined conformal operator \cite{Polchinski_v1}. In lightcone gauge there is no such subtlety because the constraints associated to quantum conformal symmetry are satisfied automatically by the gauge choice.

Above we mentioned that lightlike compactification breaks the invariance under worldsheet spacelike translations. To understand why this is the case recall that \cite{GSW1} in lightcone gauge the constraints $(\partial X)^2=(\bar{\partial}X)^2=0$ reduce to the operator equations 
$\al_0^- = \sqrt{\frac{2}{\alpha'}}\frac{1}{p^+}\left( {L}_0^\perp - 1 \right)$, and 
$\tilde\al_0^- = \sqrt{\frac{2}{\alpha'}}\frac{1}{p^+}\left( \bar{L}_0^\perp - 1 \right)$,
with $L_0^{\perp}$, $\bar{L}_{0}^{\perp}$ the transverse Virasoro generators \footnote{Recall that the transverse Virasoro generators read, ${L}_0^\perp = \frac{\alpha'}{4}\hat{\bf p}^2_{\rm L} +  N^{\perp}$,  $\bar{L}_0^\perp =\frac{\alpha'}{4} \hat{\bf p}^2_{\rm R}+  \bar{N}^{\perp}$, and $N^{\perp}=\sum_{n > 0} {\alpha_{-n}^i\alpha_n^i}$, $\bar{N}^{\perp}=\sum_{n > 0} {\tilde{\alpha}_{-n}^i\tilde{\alpha}_n^i}$.}. Therefore, level matching in lightcone gauge corresponds to the statement, 
\begin{equation}\label{eq:Vira lcg}
(\alpha_0^--\tilde{\alpha}_0^-)|V\rangle_{\rm lc}=\sqrt{\frac{2}{\alpha'}}\frac{1}{p^+}(L_0^{\perp}-\bar{L}_0^{\perp})|V\rangle_{\rm lc},
\end{equation}
from which it follows that states compactified in a lightlike spacetime direction, for which $\alpha^-_0\neq \tilde{\alpha}_0^-$ (recall that $\alpha_0^-$ and $\tilde{\alpha}_0^-$ are the left- and right-moving momentum operators, $\sqrt{\frac{\alpha'}{2}}p^-_{\rm L}$ and $\sqrt{\frac{\alpha'}{2}}p_{\rm R}^-$ repsectively), are not invariant under spacelike worldsheet shifts, $(L_0^{\perp}-\bar{L}_0^{\perp})|V\rangle_{\rm lc}\neq 0$. Therefore, the above argument which led to $\big(z\partial -\bar{z}\bar{\partial}\big)\big\langle X^i(z,\bar{z})\big\rangle=0$ does not apply in lightlike compactified spacetimes, $X^-\sim X^-+2\pi R^-$, thus implying that the classical-quantum map, $\big\langle X^{\mu}(z,\bar{z})\big\rangle=X^{\mu}_{\rm cl}(z,\bar{z})$, \emph{may} be realized. We show next that indeed the lightcone gauge realization of the coherent states (\ref{eq:DDF_coherent_closed}) can be mapped in this way to arbitrary general classical solutions.

According to the discussions in Sec.~\ref{DDF} the lightcone gauge version, $|V(\lambda,\bar{\lambda})\rangle_{\rm lc}$, of the vertex (\ref{eq:DDF_coherent_closed}) is obtained by the mapping, $A_{-n}^i\rightarrow \alpha^i_{-n}$ and $g_c\,e^{ip\cdot X(z,\bar{z})}\rightarrow |0,0;p^+,p^i\rangle$, so that
\begin{equation}\label{eq:DDF_coherent_closed_lcg}
\begin{aligned}
\big|V(\lambda,\bar{\lambda})\big\rangle_{\rm lc}&=\frac{1}{\sqrt{2p^+\mathcal{V}_{d-1}}}\,C_{\lambda\bar{\lambda}}\exp\Big(\sum_{n=1}^{\infty}\frac{1}{n}\lambda_n\cdot \alpha_{-n}\Big)\\
&\times\exp\Big(\sum_{m=1}^{\infty}\frac{1}{m}\bar{\lambda}_m\cdot \tilde{\alpha}_{-m}\Big)\,|0,0;p^+,p^i\rangle.
\end{aligned}
\end{equation}
This is similar to the open string case (\ref{eq:lc_coherent open}); it is an eigenstate of the annihilation operators, $\alpha_{n>0}^i,\tilde{\alpha}_{n>0}^i$, with eigenvalues $\lambda^i_n, \bar{\lambda}_n^i$, and of the momenta $\hat{p}^+,\hat{p}^i$ with eigenvalues $p^+,p^i$, respectively. The vacuum is normalized as in (\ref{eq:tach and vac norm lcc}),
$$ 
\langle 0,0;p'|0,0;p\rangle = 2p^+(2\pi)\delta({p^+}'-p^+)(2\pi)^{d-2}\delta^{d-2}({\bf p}'-{\bf p}).
$$
The position expectation value in the transverse directions is therefore given by, 
$$
\big\langle X^i(z,\bar{z})-\hat{x}^i\big\rangle_{\rm lc}=\big(X^i(z,\bar{z})-x^i\big)_{\rm cl},
$$ 
 with
\begin{equation}\label{eq:<X>_coherent_lc closed}
\begin{aligned}
\big(&X^i(z,\bar{z})-x^i\big)_{\rm cl}\\
&=-i\frac{\alpha'}{2}p^i\ln |z|^2+i\sqrt{\frac{\alpha'}{2}}\sum_{n\neq0}\frac{1}{n}\,\big(\lambda_n^i\,z^{-n}+\bar{\lambda}_n^i\,\bar{z}^{-n}\big).
\end{aligned}
\end{equation}
Furthermore, from the operator equations, $\alpha_n^-=\frac{1}{p^+}\big(L_n^{\perp}-\delta_{n,0}\big)$, we learn that in the longitudinal directions \footnote{Recall that $L_n^{\perp}=\frac{1}{2}\sum_{r\in \mathbb{Z}}:\alpha_{n-r}^i\alpha_{r}^i:$.},
$$
\big\langle X^-(z,\bar{z})-\hat{x}^-\big\rangle_{\rm lc}=\big(X^-(z,\bar{z})-x^-\big)_{\rm cl},
$$ 
with
\begin{widetext}
\begin{equation}\label{eq:<X^->_coherent_lc closed}
\begin{aligned}
\big(&X^-(z,\bar{z})-x^-\big)_{\rm cl}=\\
& -i\frac{1}{p^+}\Big(\frac{\alpha'}{4}{\bf p}^2+\big\langle N\big\rangle-1\Big)\ln z-i\frac{1}{p^+}\Big(\frac{\alpha'}{4}{\bf p}^2+ \big\langle\bar{N}\big\rangle-1\Big)\ln \bar{z}+i\sum_{n\neq0}\frac{1}{n}\sum_{r\in \mathbb{Z}}\frac{1}{2p^+}\Big(\lambda_{n-r}\cdot\lambda_{r}z^{-n}+\bar{\lambda}_{n-r}\cdot\bar{\lambda}_{r}\bar{z}^{-n}\Big),\\
\end{aligned}
\end{equation}
\end{widetext}
with the definitions $$\lambda_0^i\equiv \sqrt{\frac{\alpha'}{2}}p^i, \qquad \bar{\lambda}_0^i\equiv \sqrt{\frac{\alpha'}{2}}p^i,$$ and $p^ip^i={\bf p}^2$, and as discussed above we are to enforce classical level matching, $\langle N\rangle=\langle \bar{N}\rangle$. For completeness we note also that (in lightcone gauge),
$$
X^+(z,\bar{z})=-i\frac{\alpha'}{2}p^+\ln|z|^2.
$$ 
Notice that in the rest frame, $p^i=0$, the zero mode contribution in (\ref{eq:<X^->_coherent_lc closed}) is identical to that found in the covariant gauge (\ref{eq:X^-cov}) when $p^+=1/R^-$. The quantities (\ref{eq:<X>_coherent_lc closed}) and (\ref{eq:<X^->_coherent_lc closed}) are none other than the general solutions to the equations of motion, $\partial\bar{\partial}X^{\mu}=0$, in lightcone gauge \cite{Zwiebach04}. We therefore conclude that indeed the classical-quantum map, $\langle X^{\mu}(z,\bar{z})\rangle_{\rm lc} = X_{\rm cl}^{\mu}(z,\bar{z})$, can be realized in a spacetime with lightlike compactification when this map is carried out in lightcone gauge. This is in accordance with the above considerations. Note that this is specific to \emph{lightlike}-compactified spacetimes and does not apply in spacelike compactifications, because this conclusion relied on the left-hand-side of (\ref{eq:Vira lcg}) being non-vanishing. 

Finally, before we construct closed string coherent states in fully non-compact spacetime let us show that the angular momentum of the covariant gauge, lightcone gauge and classical descriptions are all identical, as we did in the open string case (\ref{eq:Jcorr}) above. For the closed string,\begin{equation}\label{eq:Jmunu_closed}
\begin{aligned}
J^{\mu\nu}& =\frac{2}{\alpha'}\Big(\oint \dslash z X^{[\mu}\partial X^{\nu]}-\oint \dslash \bar{z} X^{[\mu}\bar{\partial} X^{\nu]}\Big),\\
&= L^{\mu\nu}+S^{\mu\nu},
\end{aligned}
\end{equation}
with the zero mode contribution denoted by $L_{\mu\nu}$ (given in a footnote on p.~\pageref{L_munu}) and $S^{\mu\nu}=S^{\mu\nu}(\alpha)+S^{\mu\nu}(\tilde{\alpha})$ with $S^{\mu\nu}(\alpha)=-i\sum_{\ell=1}^{\infty}\big(\alpha_{-\ell}^{\mu}\alpha^{\nu}_{\ell}-\alpha^{\nu}_{-\ell}\alpha_{\ell}^{\mu}\big)$ and a similar expression for the antiholomorphic sector, $S^{\mu\nu}(\tilde{\alpha})$. We shall concentrate on the non-zero mode part: $S^{\mu\nu}$. The derivation is almost identical to the open string case and so we do not repeat it here, the only difference being that the open string normalization of the momentum is half that of the closed string: $\frac{1}{2}p_c=p_o$ (although we don't bother to keep the subscripts when the context is clear). We find that for the transverse directions,
\begin{equation}\label{eq:Jcorr_closed}
\begin{aligned}
&\langle S^{ij}\rangle_{\rm cov}=\langle S^{ij}\rangle_{\rm lc}=S_{\rm cl}^{ij}=\sum_{n>0}\frac{2}{n}{\rm Im}\big(\lambda_n^{*i}\lambda_n^j+\bar{\lambda}_n^{*i}\bar{\lambda}_n^j\big),
\end{aligned}
\end{equation}
and for the longitudinal components,
\begin{equation}\label{eq:Jcorr_closed_long}
\begin{aligned}
&\langle S^{-i}\rangle_{\rm cov}=\langle S^{-i}\rangle_{\rm lc}=S_{\rm cl}^{-i}=\\
&\sqrt{\frac{2}{\alpha'}}\sum_{m>0}\sum_{\ell\in\mathbb{Z}}\frac{1}{mp^+}{\rm Im}\big(\lambda_{m-\ell}^*\cdot \lambda_{\ell}^{*}\,\lambda_m^i+\bar{\lambda}_{m-\ell}^*\cdot\bar{\lambda}_{\ell}^{*}\,\bar{\lambda}_m^i\big),
\end{aligned}
\end{equation}
with in addition all components involving the $+$ direction equal to zero.
This correspondence provides further evidence for the conjecture that the covariant gauge vertex operator (\ref{eq:DDF_coherent_closed}) and the lightcone gauge state (\ref{eq:DDF_coherent_closed_lcg}) describe the same physics (share identical correlation functions) and are different manifestations of the same state which classically have a lightcone gauge description given by (\ref{eq:<X>_coherent_lc closed}) and (\ref{eq:<X^->_coherent_lc closed}).

Before delving into the coherent state construction in non-compact spacetimes it is worth noting that the requirement of a lightlike compactified background in the naive construction of the current section is the cost of working in a standard gauge, namely lightcone or covariant gauge where all the string technology for amplitude computations is well developed. It is also possible to construct closed string coherent states in a modified lightcone gauge \cite{KanitscheiderSkenderisTaylor07}, where the requirement of a lightlike compactified background, $X^-\sim X^-+2\pi R^-$, gets replaced by the requirement of a spacelike compactified background, $X^D\sim X^D+2\pi R^D$. Here, instead of making the lightcone gauge identification $X^+(z,\bar{z})=-i\frac{\alpha'}{2}\,p^+\ln|z|^2$, one chooses $X^+(z,\bar{z})=-i\frac{\alpha'}{2}\,p^+_{\rm L}\ln z-i\frac{\alpha'}{2}\,p^+_{\rm R}\ln\bar{z}$, which in turn solves the constraints in a manner similar to the lightcone gauge case. Here however, with the additional freedom of choosing $p^+_{\rm L}$ and $p_{\rm R}^+$ independently it becomes possible to rotate the spacetime coordinate system in such a way that the resulting coherent states propagate in a spacelike rather than a lightlike compactified spacetime \footnote{The authors would like to thank Kostas Skenderis and Marika Taylor for bringing \cite{KanitscheiderSkenderisTaylor07} to their attention.}.

\subsection{Coherent States in Minkowski Space}\label{CSNCB}
We next construct coherent states in fully non-compact spacetimes. We showed above that the coherent state (\ref{eq:DDF_coherent_closed}),
\begin{equation}\label{eq:DDF_coherent_closed*}
\begin{aligned}
V(\lambda,\bar{\lambda},p)&=\frac{g_c}{\sqrt{2p^+\mathcal{V}_{d-1}}}\,C_{\lambda\bar{\lambda}}\exp\Big(\sum_{n=1}^{\infty}\frac{1}{n}\lambda_n\cdot A_{-n}\Big)\\
&\times\exp\Big(\sum_{m=1}^{\infty}\frac{1}{m}\bar{\lambda}_m\cdot \bar{A}_{-m}\Big)\,e^{ip\cdot X(z,\bar{z})},
\end{aligned}
\end{equation}
satisfies all the coherent state defining properties but only when the underlying spacetime manifold is compactified in a lightlike direction of spacetime. Below (\ref{eq:massshell}) we concluded that in addition to the usual states in the underlying Hilbert space which satisfy $N=\bar{N}$, there were additional states for which $N\neq \bar{N}$ and these correspond to states with lightlike winding. This suggests that starting from (\ref{eq:DDF_coherent_closed*}) we may truncate the underlying Hilbert space and project out all states with $N\neq \bar{N}$. The resulting states will be manifestly level-matched and will propagate consistently in fully non-compact (but also compact) spacetimes.

To project out all states with $N\neq \bar{N}$, thus leaving only $N=\bar{N}$ states in the underlying spectrum, we define a projection operator,
\begin{equation}\label{eq:Gw}
G_w= \int_0^{2\pi}\!\!\!\dslash s\,e^{ is(\hat{W}-w)},
\quad \hat{W}\equiv\frac{\alpha'}{2}(\hat{p}^+_{\rm L}\hat{p}_{\rm L}^--\hat{p}^+_{\rm R}\hat{p}_{\rm R}^-),
\end{equation}
with $\hat{p}_{\rm L}^{\mu}=\frac{2}{\alpha'}\oint \dslash z\partial X^{\mu}$, $\hat{p}_{\rm R}^{\mu}=-\frac{2}{\alpha'}\oint \dslash \bar{z}\bar{\partial} X^{\mu}$, and $\hat{W}$ the lightlike winding number operator. The Virasoro constraints associated to level matching read, 
\begin{equation}
\begin{aligned}
L_0-&\bar{L}_0=\Big(\frac{\alpha'}{4}\hat{p}_{\rm L}^2+N\Big)-\Big(\frac{\alpha'}{4}\hat{p}_{\rm R}^2+\bar{N}\Big)\\
&=-\frac{\alpha'}{2}(\hat{p}^+_{\rm L}\hat{p}_{\rm L}^--\hat{p}^+_{\rm R}\hat{p}_{\rm R}^-)+\frac{\alpha'}{4}\big(\hat{\bf p}_{\rm L}^2-\hat{\bf p}_{\rm R}^2\big)+N-\bar{N}\\
&=0,
\end{aligned}
\end{equation}
 from which the origin of the projector, $G_w$, becomes clear: when $G_w$ is applied to arbitrary vertices it projects out all states in the underlying Hilbert space except for those with lightlike winding number $w$. In the case of interest when there are no transverse compact directions, ${\bf p}^2_{\rm L}={\bf p}^2_{\rm R}={\bf p}^2$, we can equivalently write the covariant expression 
 $$
 \hat{W}=-\alpha'p\cdot \hat{w},
$$ 
where $p^{\mu}=\frac{1}{2}(p_{\rm L}^{\mu}+p_{\rm R}^{\mu})$ is the momentum of the vacuum, $p^2=4/\alpha'$, $p\cdot q=2/\alpha'$, and $\hat{w}^{\mu}=\frac{1}{2}\big(\hat{p}_{\rm L}^{\mu}-\hat{p}_{\rm R}^{\mu}\big)$ is the winding vector (see Appendix \ref{C}). Notice for example that for some generic vertex operator,
\begin{equation}
\begin{aligned}
&\hat{W}\cdot P(\partial^{\#}X,\bar{\partial}^{\#}X)e^{ i(p-Nq)\cdot X(z)}e^{i(p-\bar{N}q)\cdot X(\bar{z})}\\
&=(N-\bar{N})P(\partial^{\#}X,\bar{\partial}^{\#}X)e^{i(p-Nq)\cdot X(z)}e^{i(p-\bar{N}q)\cdot X(\bar{z})},
\end{aligned}
\end{equation}
with $P(\partial^{\#}X,\bar{\partial}^{\#}X)$ the oscillator contribution that commutes with $\hat{W}$. Then, covariant vertex operators without lightlike winding will be given by, 
\begin{equation}\label{eq:V_0}
V_0(\lambda,\bar{\lambda})\cong G_0\cdot V(\lambda,\bar{\lambda}),
\end{equation}
the dot denoting operator product contractions. Taking $V(\lambda,\bar{\lambda})$ to be the coherent state (\ref{eq:DDF_coherent_closed*}) we are to commute $G_0$ through the DDF operators, the relevant term giving $e^{is\hat{W}}e^{\sum_{n=1}^{\infty}\frac{1}{n}\lambda_n\cdot A_{-n}}=e^{\sum_{n=1}^{\infty}\frac{1}{n}e^{ins}\lambda_n\cdot A_{-n}}e^{is\hat{W}}$ with a similar relation for the anti-holomorpic sector, with $e^{-ins}$ replacing $e^{ins}$. This follows from the Baker-Campbell-Hausdorff formula, the commutators $\big[\hat{W},A_{-n}^i\big]=nA^i_{-n}$, $\big[\hat{W},\bar{A}_{-n}^i\big]=-n\bar{A}^i_{-n}$ , and the elementary Schur polynomial representation (\ref{eq:S_m(a)a}) with $a_s=\frac{1}{s!}\sum_{n=1}^{\infty}(ins)^{s}\frac{1}{n}\lambda_n\cdot A_{-n}$. The resulting vertex operators are then the candidate quantum states to represent arbitrary classical loops in non-compact Minkowski spacetime:
\begin{widetext}
\begin{equation}\label{V_0A}
\begin{aligned}
V_0(\lambda,&\bar{\lambda};p)=\frac{g_c}{\sqrt{2p^+\mathcal{V}_{d-1}}}\,\mathcal{C}_{\lambda\bar{\lambda}}\int_0^{2\pi}\!\!\!\dslash s\exp\Big\{\sum_{n=1}^{\infty}\frac{1}{n}e^{ins}\lambda_n\cdot A_{-n}\Big\}\exp\Big\{\sum_{m=1}^{\infty}\frac{1}{m}e^{-ims}\bar{\lambda}_m\cdot \bar{A}_{-m}\Big\}\,e^{ip\cdot X(z,\bar{z})},
\end{aligned}
\end{equation}
\end{widetext}
with:
\begin{equation}\label{eq:C_llbar coh norm}
\mathcal{C}_{\lambda\bar{\lambda}}=\Big[\int_{0}^{2\pi}\!\! \dslash s\exp\Big(\sum_{n=1}^{\infty}\frac{1}{n}|\lambda_{n}|^2e^{ins}+\frac{1}{n}|\bar{\lambda}_{n}|^2e^{-ins}\Big)\Big]^{-1/2}
\end{equation} 
a normalization constant. 
The normalization as usual fixed by the `one string in volume $\mathcal{V}_{d-1}$' condition, which leads to a unitary $S$-matrix. As discussed in Sec.~\ref{sec:NSW}, this is equivalent to fixing the most singular term in the operator product expansion as in (\ref{eq:norm V coh}), which in our conventions, as discussed there, is equivalent to requiring that the state have unit norm,
$$
\langle V_0(\lambda,\bar{\lambda};p)|V_0(\lambda,\bar{\lambda};p)\rangle = 1,\qquad |0,0;p\rangle \cong g_c\,e^{ip\cdot X}.
$$
Note that the out state $V_0(\lambda,\bar{\lambda})^{\dagger}$ is given by $V_0(\lambda,\bar{\lambda})$ with $\{\lambda^*_n\}$, $A_n$ and $-p$ replacing $\{\lambda_n\}$, $A_{-n}$ and $p$ respectively (corresponding to Hermitian conjugation in Minkowski signature worldsheet), and similarly for the anti-holomorphic sector.

We first check that (\ref{V_0A}) satisfies the defining properties (a-c) of a string coherent state as laid out in the beginning of this section. The properties (a,c) are trivially satisfied because the state is still specified by a set of continuous labels and the projection operator (\ref{eq:Gw}) does not alter the states in the underlying Hilbert space, $\mathcal{H}$. The Hilbert space is instead truncated \footnote{The picture we have in mind here is, $\mathcal{H}=\bigoplus_{w\in\mathbb{Z}} \mathcal{H}_w$, with $G_w$ such that, $G_w:\mathcal{H}\rightarrow \mathcal{H}_w$, and $G_w:\mathcal{H}_w\rightarrow \mathcal{H}_w$.} and so, given that any linear combination of physical states is also a physical state, the vertex (\ref{V_0A}) must be physical. To check that (b) is satisfied, i.e.~that a completeness relation exists for the projected states, we start from the completeness relation associated to the unprojected states \footnote{We occasionally write $V_{\lambda\bar{\lambda}}(p)$, $V(\lambda,\bar{\lambda})$, $V(\lambda,\bar{\lambda};p)$, or even $V(z,\bar{z})$ (with $z,\bar{z}$ the worldsheet location where the vertex is inserted) to denote the same object $V(\lambda,\bar{\lambda},p)$.}, the existence of which was established on p.~\pageref{eq:completeness}, 
\begin{equation*}\label{eq:completeness closed}
\begin{aligned}
\mathds{1} &=\mathcal{V}_{d-1}\int_0^{\infty} \frac{dp^+}{2\pi}\int_{\mathbb{R}^{24}}\frac{d^{24}{\bf p}}{(2\pi)^{24}}\\
&\times \int \bigg(\prod_{n,A}\frac{d^2\lambda_n^A}{2\pi n}\bigg)\bigg(\prod_{n,A}\frac{d^2\bar{\lambda}_n^A}{2\pi n}\bigg) \big|V(\lambda,\bar{\lambda})\big\rangle \big\langle V(\lambda,\bar{\lambda})\big|.
\end{aligned}
\end{equation*}
Apply a projection operator, $G_w$, on either side of this expression to find that:
\begin{equation}\label{eq:completeness w}
\mathds{1}_w=\int d\mu(p)d\lambda d\bar{\lambda} \big|V_w(\lambda,\bar{\lambda})\big\rangle\big\langle V_w(\lambda,\bar{\lambda})\big|,
\end{equation}
where we write $d\mu(p)=\mathcal{V}_{d-1}\frac{dp^+}{2\pi}\frac{d^{24}{\bf p}}{(2\pi)^{24}}$, $d\lambda =\prod_{n,i}\frac{d^2\lambda_n^i}{2\pi n}$, $d^2\lambda_n^i=id\lambda_n^i\wedge d\lambda_n^{*i}$, and similarly for the anti-holomorphic sector, $d\bar{\lambda} $, with $\bar{\lambda}_n^i$ replacing $\lambda_n^i$. We have defined, $G_w\equiv\mathds{1}_w$, as $G_w$ is none other than the unit operator, $\mathds{1}_w$, with respect to the truncated Hilbert space, $\mathcal{H}_w$, which consists of all states with lightlike winding number $w$.  To show this note that $|V_w\rangle = G_w|V\rangle$ and $G_w^2=G_w$ (recall that $G_w$ is Hermitian). From the latter two expressions it follows that 
$$
G_w|V_w\rangle=|V_w\rangle,
$$
and so indeed $G_w=\mathds{1}_w$.  Thus, there exists a completeness relation for the projected states also, as required from the definition of a coherent state.

Note that if we sum over $w$ in (\ref{eq:completeness w}) we learn that,  
$$
\mathds{1}=\int d\mu(p)d\lambda d\bar{\lambda} \sum_{w=-\infty}^{\infty}\big|V_w(\lambda,\bar{\lambda})\big\rangle\big\langle V_w(\lambda,\bar{\lambda})\big|,
$$ 
with $\mathds{1}$ the unit operator with respect to the larger Hilbert space $\mathcal{H}$, and this serves as a consistency check.  

The Hilbert space of interest here is $\mathcal{H}_0$ which is the \emph{coherent state Hilbert space} associated to non-compact spacetimes.  From the above considerations we conclude that (\ref{eq:completeness w}) is indeed a resolution of unity with respect to $\mathcal{H}_w$, and have thus shown that the string coherent state defining properties are satisfied by the states $V_w(\lambda,\bar{\lambda};p)$. Next notice that because winding number is conserved, $\big[\hat{H},\hat{W}\big]\cdot V_w(\lambda,\bar{\lambda};p)\cong 0$, with $\hat{H}=L_0+\bar{L}_0-2$ the worldsheet Hamiltonian, the Hilbert space decomposition, $\mathcal{H}=\bigoplus_{w\in\mathbb{Z}} \mathcal{H}_w$, is indeed orthogonal; when all quantum numbers other than winding number are equal, $\langle V_m|V_n\rangle = \delta_{m,n}$ for vertices, $V_m\in \mathcal{H}_m$. We conclude that vertex operators, 
$$
V_0(\lambda,\bar{\lambda};p)\in\mathcal{H}_0,
$$
can propagate in fully non-compact spacetimes, and have shown in particular that the vertex operator (\ref{V_0A}) is a closed string coherent state that can be consistently embedded in non-compact flat Minkowski spacetime.

In a scattering amplitude that involves say $n$ coherent states $V_0$ and any number of non-coherent states, one can drop the $G_0$'s in $n-1$ of these vertices. To see this let us look at an example, say the elastic massive string forward scattering amplitude from an arbitrary closed string coherent state, $V_0$,
\begin{equation}\label{eq:one s-int}
\begin{aligned}
\big\langle V_0^{\dagger}U^{\dagger}UV_0\big\rangle&=\big\langle (G_0V)^{\dagger}U^{\dagger}U(G_0 V)\big\rangle\\
&=\big\langle V^{\dagger}U^{\dagger}U G_0^2V\big\rangle\\
&=\big\langle V^{\dagger}U^{\dagger}U V_0\big\rangle,
\end{aligned}
\end{equation}
with, $U=P(\partial^{\#}X,\bar{\partial}^{\#}X)e^{ik\cdot X(z,\bar{z})}$, a vertex operator without lightlike winding, and we have used the fact that $G_0$ is Hermitian, commutes with $U$ and squares to itself.

The inner product associated to the projected states can be derived from the properties, 
$$
A_n^i \cdot V_0\cong\lambda_n^iV_n,\qquad\bar{A}_n^i\cdot V_0\cong\bar{\lambda}_n^iV_n,
$$ 
valid for $n>0$, and, 
$$
\langle V_n^{\dagger}V_m\rangle = \delta_{n,m},
$$
which follow from the DDF operator commutation relations. From these it follows that the constructed coherent states are as usual over-complete,
\begin{equation}
\begin{aligned}
\big\langle &V_0(\lambda,\bar{\lambda};p')|V_0(\xi,\bar{\xi};p)\big\rangle =\delta_{p',p}\mathcal{C}_{\lambda\bar{\lambda}}\mathcal{C}_{\xi,\bar{\xi}} \\
&\times\int_0^{2\pi}\dslash s\exp\Big(\sum_{n>0}\frac{1}{n}\lambda^*_{n}\cdot\xi_{n}\,e^{ins}+\frac{1}{n}\bar{\lambda}^*_{n}\cdot\bar{\xi}_{n}\,e^{-ins}\Big),
\end{aligned}
\end{equation}
and this reduces to unity when $(\lambda,\bar{\lambda})=(\xi,\bar{\xi})$. We have again made use of the fact that $G_0^2=G_0$. Note that $\delta_{p',p}$ is a \emph{Kronecker} delta which reduces to unity when ${p^+}'=p^+$ and ${\bf p}'={\bf p}$, with $p$ and $p'$ the momenta of the vacua associated to the in and out states, as above.

\subsubsection{Functional Representation}
 The normal ordered version of $V_0(\lambda,\bar{\lambda})$ analogous to (\ref{eq:DDF_coherent_cov_closed_transverse}) can be derived from (\ref{eq:DDF_coherent_cov_closed_transverse}) by computing the operator product, $V_0(\lambda,\bar{\lambda})\cong G_0\cdot V(\lambda,\bar{\lambda})$. In the particular case that $\lambda_{n>0}\cdot \lambda_{m>0}=0$, one finds an expression identical to (\ref{V_0A}) with $H_n^i(z)e^{-inq\cdot X(z)}$, $\bar{H}_n^i(\bar{z})e^{-inq\cdot X(\bar{z})}$ replacing $A_{-n}^i$, $\bar{A}_{-n}^i$ respectively, with an overall integral over $s$,
 \begin{widetext}
\begin{equation}\label{V_0P}
\begin{aligned}
V_0&(\lambda,\bar{\lambda})=\frac{g_c}{\sqrt{2p^+\mathcal{V}_{d-1}}}\,\mathcal{C}_{\lambda\bar{\lambda}}\int_0^{2\pi}\!\!\!\dslash s\exp\Big(\sum_{n=1}^{\infty}\frac{1}{n}e^{ins}\lambda_n\cdot H_n\,e^{-inq\cdot X(z)}\Big)
\exp\Big(\sum_{m=1}^{\infty}\frac{1}{m}e^{-ims}\bar{\lambda}_m\cdot \bar{H}_m\,e^{-imq\cdot X(\bar{z})}\Big)\,e^{ip\cdot X(z,\bar{z})}.
\end{aligned}
\end{equation}
\end{widetext}
This follows from the general result (\ref{eq:xiA^g e^ipX1}) and (\ref{V_0A}), and the polynomials $H_n(z)$ have been defined in (\ref{eq:H_n}), see also (\ref{eq:Sm(nqz)}), (\ref{eq:S_m(a)mt}) and (\ref{eq:PPbardfn}). Notice that this is still an eigenstate of $\hat{p}^+, \hat{p}^i$ if we make the choice $q^+,q^i=0$ and $q^-$ non-zero, as was the unprojected state $V(\lambda,\bar{\lambda})$. Recall also that in the rest frame in addition to taking $p^i=0$ we are to take $H_n(z)\rightarrow P_n(z)$ as discussed in Sec.~\ref{DDF}. When the polarization tensors are arbitrary, subject only to the constraints $\lambda_n\cdot q=0$ (for all $n\in \mathbb{Z}$), we have instead:
\begin{equation}\label{eq:V_0 closed full functional}
V_0(\lambda,\bar{\lambda})=\frac{g_c}{\sqrt{2p^+\mathcal{V}_{d-1}}}\,\mathcal{C}_{\lambda\bar{\lambda}}\int_0^{2\pi} \!\!\!\dslash s\,U_0(\lambda)\bar{U}_0(\bar{\lambda}) e^{ip\cdot X(z,\bar{z})},
\end{equation} 
with $U_0(\lambda)$ given by:
\begin{widetext}
$$
U_0(\lambda)=\sum_{g=0}^{\infty}
\sum_{a=0}^{\lfloor g/2\rfloor}\frac{1}{a!(g-2a)!}\Big(\sum_{n,m>0}\frac{e^{i(n+m)s}}{2nm}\lambda_n\cdot \lambda_m\,\mathbb{S}_{n,m}\,e^{-i(n+m)q\cdot X(z)}\Big)^a\Big(\sum_{n>0}\frac{e^{ins}}{n}\lambda_n\cdot H_n\,e^{-inq\cdot X(z)}\Big)^{g-2a},
$$
\end{widetext}
and $\bar{U}_0(\bar{\lambda})$ given by a similar expression with $\bar{\lambda}_m^i$, $\bar{z}$, $\bar{\mathbb{S}}_{m,m}(\bar{z})$ and $\bar{H}_m^i(\bar{z})$ replacing the corresponding holomorphic quantities, and $e^{-iNs}$ replacing $e^{iNs}$ for any integers $N$. The explicit form for $U_0(\lambda)$ has been derived from the general result (\ref{eq:xiA^g e^ipX1}) and (\ref{V_0A}).

It is possibly useful at this point to give an example. The simplest coherent state vertex operator is when only $\lambda^i\equiv\lambda_1^i$ is non-vanishing and $\lambda\cdot \lambda=\bar{\lambda}\cdot \bar{\lambda}=\lambda_{n\neq\pm1}^i=\bar{\lambda}_{n\neq\pm1}^i=0$. From (\ref{V_0P}) and find that,
\begin{equation}\label{eq:cohvert*}
\begin{aligned}
V_0(&z,\bar{z}) = \frac{g_c}{\sqrt{2p^+\mathcal{V}_{d-1}}}\,\mathcal{C}_{\lambda\bar{\lambda}}\\
&\times\int_0^{2\pi}\!\!\dslash s\,\exp\Big(ie^{is}\zeta\cdot \partial X\,e^{-iq\cdot X(z)}\Big)\\
&\times\exp\Big(ie^{-is}\bar{\zeta}\cdot \bar{\partial} X\,e^{-iq\cdot X(\bar{z})}\Big)\,e^{ip\cdot X(z,\bar{z})},
\end{aligned}
\end{equation}
with 
$$
\zeta_{\mu} \equiv \lambda^i(\delta^i_{\mu}-\tfrac{\alpha'}{2}p^iq_{\mu}),\qquad\bar{\zeta}^{\mu} \equiv \bar{\lambda}^i(\delta^i_{\mu}-\tfrac{\alpha'}{2}p^iq_{\mu}),
$$
and 
$$
|\zeta|^2=|\lambda|^2,\qquad |\bar{\zeta}|^2=|\bar{\lambda}|^2.
$$
It is manifest that the $s$-integral serves to set the total number of holomorphic and anti-holomorphic worldsheet derivatives to be equal in every term of the series expansions of the exponentials.

\subsubsection{Closed String Coherent State Properties}
We next derive various properties of the projected coherent states. Proceeding in a similar manner to the open string case, we map to the lightcone gauge states corresponding to (\ref{V_0A}) or equivalently (\ref{eq:V_0 closed full functional}), which are given by,
\begin{equation}\label{eq:DDF_coherent_closed_lcg_proj}
\begin{aligned}
\big|V_0&(\lambda,\bar{\lambda})\big\rangle_{\rm lc}=\frac{1}{\sqrt{2p^+\mathcal{V}_{d-1}}}\,\mathcal{C}_{\lambda\bar{\lambda}}\\
&\times\int_0^{2\pi}\!\!\!\dslash s\exp\Big(\sum_{n=1}^{\infty}\frac{1}{n}e^{ins}\lambda_n\cdot \alpha_{-n}\Big)\\
&\times\exp\Big(\sum_{m=1}^{\infty}\frac{1}{m}e^{-ims}\bar{\lambda}_m\cdot \tilde{\alpha}_{-m}\Big)\,|0,0;p^+,p^i\rangle.
\end{aligned}
\end{equation}
Let us consider the lightcone gauge classical solutions, $X_{\rm cl}^{\mu}(z,\bar{z})$, corresponding to this state. Having projected out the lightlike winding states, worldsheet translation invariance is restored (in both lightcone and covariant gauges) and according to the discussion on p.~\pageref{classicality discussion} the condition for classicality $\langle X^{\mu}(z,\bar{z})\rangle=X_{\rm cl}^{\mu}(z,\bar{z})$ is replaced by (\ref{eq:class_defn_}), rewritten here for convenience in the $(z,\bar{z})=(e^{-i(\sigma+i\tau)},e^{i(\sigma-i\tau)})$ coordinate system with the zero mode contributions explicitly subtracted,
\begin{equation}\label{eq:class_defn_2}
\begin{aligned}
&\big\langle \!:\!\big[X^{\mu}(z',\bar{z}')-\hat{x}^{\mu}\big]\big[X^{\nu}(z,\bar{z})-\hat{x}^{\nu}\big]\!:\!\big\rangle =\\
&\int_0^{2\pi} \!\!\!\dslash s\,\big[X^{\mu}(z'e^{is},\bar{z}'e^{-is})-x^{\mu}\big]_{\rm cl}\big[X^{\nu}(ze^{is},\bar{z}e^{-is})-x^{\nu}\big]_{\rm cl}.
\end{aligned}
\end{equation}
Given that we know the classical solution, i.e.~the right-hand-side of (\ref{eq:class_defn_2}), in lightcone gauge, see (\ref{eq:<X>_coherent_lc closed}) and (\ref{eq:<X^->_coherent_lc closed}), we establish (\ref{eq:class_defn_2}) for the projected states in lightcone gauge. For the transverse directions, $i,j$, to evaluate the left hand side of (\ref{eq:class_defn_2}) in the state (\ref{eq:DDF_coherent_closed_lcg_proj}), we make use of the closed string mode expansion,
\begin{equation}\label{eq:X^i-x^i}
\begin{aligned}
X^i(z,\bar{z})&-\hat{x}^i=-i\frac{\alpha'}{2}\hat{p}^i\ln |z|^2\\
&+i\sqrt{\frac{\alpha'}{2}}\sum_{n\neq0}\frac{1}{n}\,\big(\alpha_n^i\,z^{-n}+\tilde{\alpha}_n^i\,\bar{z}^{-n}\big),
\end{aligned}
\end{equation}
and the fact that:
\begin{equation}\label{eq:a_n|V>=lambda_n|V-n>}
\begin{aligned}
&\alpha_{n>0}^i |V_w\rangle_{\rm lc}=\lambda_n^i|V_{w-n}\rangle_{\rm lc},\\ 
&\tilde{\alpha}_{n>0}^i |V_w\rangle_{\rm lc}=\bar{\lambda}_n^i|V_{w+n}\rangle_{\rm lc},
\end{aligned}
\end{equation}
and $\langle V_n|V_m\rangle_{\rm lc} = \delta_{n,m}$, which follow from the oscillator commutation relations, $[\alpha_n^i,\alpha_m^j]=n\delta_{n+m,0}\delta^{ij}$, $[\tilde{\alpha}_n^i,\tilde{\alpha}_m^j]=n\delta_{n+m,0}\delta^{ij}$ and $G^{\dagger}_w=G_w$, $G^{\dagger}_wG_m=\delta_{w,m}G_m$. Furthermore, we have $\langle V_w|\alpha_{-m}^i=\lambda^{*i}_m\langle V_{w- m}|$ and $\hat{p}^i|V_0\rangle=p^i|V_0\rangle$. From these expressions we learn that,
\begin{equation}\label{eq:<XX>cov}
\begin{aligned}
\big\langle \!:\!&\big[X^{i}(z',\bar{z}')-x^i\big]\big[X^{j}(z,\bar{z})-x^j\big]\!:\!\big\rangle=\\
&-\Big(\frac{\alpha'}{2}\Big)^2p^ip^j\ln|z'|^2\ln|z|^2 +\frac{\alpha'}{2}\sum_{n\neq0}\frac{1}{n^2}\Big[\lambda_n^i\lambda^{*j}_n\Big(\frac{z}{z'}\Big)^n\\
&+\bar{\lambda}_n^i\bar{\lambda}^{*j}_n\Big(\frac{\bar{z}}{\bar{z}'}\Big)^n-\lambda_n^i\bar{\lambda}^{j}_n\Big(\frac{1}{z'\bar{z}}\Big)^n-\bar{\lambda}_n^i\lambda^j_n\Big(\frac{1}{\bar{z}'z}\Big)^n\Big]
\end{aligned}
\end{equation}
It is trivial to show that this expression is identical to the right-hand side of (\ref{eq:class_defn_2}) when,
\begin{equation}\label{eq:X_cl^i}
\begin{aligned}
\big(X^i(z,\bar{z})&-x^i\big)_{\rm cl}=-i\frac{\alpha'}{2}p^i\ln |z|^2\\
&+i\sqrt{\frac{\alpha'}{2}}\sum_{n\neq0}\frac{1}{n}\,\big(\lambda_n^i\,z^{-n}+\bar{\lambda}_n^i\,\bar{z}^{-n}\big),
\end{aligned}
\end{equation}
thus proving that the definition of classicality (\ref{eq:class_defn_}) is satisfied by the projected coherent states, at least for the transverse directions. For the longitudinal directions, to evaluate the left-hand side of (\ref{eq:class_defn_2}) in the state (\ref{eq:DDF_coherent_closed_lcg_proj}), we make use of mode expansions,
\begin{equation}
\begin{aligned}
&X^-(z,\bar{z})-\hat{x}^- = -i\frac{\alpha'}{2}\hat{p}^-\ln|z|^2 \\
&\hspace{1cm}+ i\sqrt{\frac{\alpha'}{2}}\sum_{n\neq 0}\frac{1}{n}\big(\alpha^-_nz^{-n}+\tilde{\alpha}^-_n\bar{z}^{-n}\big)\\
&X^{+}(z,\bar{z})=-i\frac{\alpha'}{2}\hat{p}^+\ln |z|^2
\end{aligned}
\end{equation}
We find that,
\begin{equation}\label{eq:<X^-X^i>cov}
\begin{aligned}
\big\langle \!:\!\big[&X^{-}(z',\bar{z}')-x^-\big]\big[X^{j}(z,\bar{z})-x^j\big]\!:\!\big\rangle\\
& =-\Big(\frac{\alpha'}{2}\Big)^2\langle \hat{p}^-\rangle p^j\ln|z'|^2\ln|z|^2\\
&+\frac{\alpha'}{2}\sum_{n\neq0}\frac{1}{n^2}\Big[\lambda_n^-\lambda^{*j}_n\Big(\frac{z}{z'}\Big)^n+\bar{\lambda}_n^-\bar{\lambda}^{*j}_n\Big(\frac{\bar{z}}{\bar{z}'}\Big)^n\\
&-\lambda_n^-\bar{\lambda}^{j}_n\Big(\frac{1}{z'\bar{z}}\Big)^n-\bar{\lambda}_n^-\lambda^j_n\Big(\frac{1}{\bar{z}'z}\Big)^n\Big],
\end{aligned}
\end{equation}
where we have found it convenient to write,
$$
\lambda_n^- = \frac{1}{\sqrt{2\alpha'}}\sum_{r\in \mathbb{Z}}\frac{1}{p^+}\lambda_{n-r}\cdot\lambda_{r},
$$
with a similar expression for $\bar{\lambda}_n^-$ with $\bar{\lambda}_n$ replacing $\lambda_n$. 
This is computed using the fact that the $\alpha^-_n$ are determined entirely in terms of the $\alpha_n^i$, according to (for $n\neq 0$), 
$$
\alpha_n^-=\frac{1}{\sqrt{2\alpha'}}\frac{1}{p^+}\sum_{r\in \mathbb{Z}}:\alpha_{n-r}^i\alpha_{r}^i:,
$$
and similarly for $\tilde{\alpha}_n^-$ with $\alpha_n$ replacing $\tilde{\alpha}_n$, which follows from the relations (\ref{eq:a_n|V>=lambda_n|V-n>}), and from the commutation relations $[L_n^{\perp},\alpha_m^i]=-n\alpha^i_{n+m}$ and $[\bar{L}_n^{\perp},\tilde{\alpha}_m^i]=-n\tilde{\alpha}^i_{n+m}$ with $L_n^{\perp}=\sqrt{\frac{\alpha'}{2}}p^+\alpha^-_n$ and $\bar{L}_n^{\perp}=\sqrt{\frac{\alpha'}{2}}p^+\tilde{\alpha}^-_n$ (for $n\neq0$). The $n=0$ term yields the lightcone gauge Hamiltonian, $\hat{p}^-=\frac{1}{\sqrt{2\alpha'}}\big(\al_0^- +\tilde{\alpha}_0^-\big)$, with $\al_0^- = \sqrt{\frac{2}{\alpha'}}\frac{1}{p^+}\left( {L}_0^\perp - 1 \right)$, and 
$\tilde\al_0^- = \sqrt{\frac{2}{\alpha'}}\frac{1}{p^+}\left( \bar{L}_0^\perp - 1 \right)$, or:
$$
\hat{p}^- = \frac{1}{\alpha'p^+}\big(L_0^{\perp}+\bar{L}_0^{\perp}-2\big).
$$
The expectation value of the lightcone gauge Hamiltonian is in turn given by,
$$
\langle \hat{p}^-\rangle = \frac{1}{\alpha'p^+}\Big(\frac{\alpha'}{2}{\bf p}^2+\sum_{n>0}|\lambda_n|^2+\sum_{n>0}|\bar{\lambda}_n|^2-2\Big),
$$
exactly as for the DLCQ coherent states, and there is again thus an effective level number for the left- and right-movers $\langle N\rangle\sum_{n>0}|\lambda_n|^2$ and $\langle\bar{N}\rangle=\sum_{n>0}|\bar{\lambda}_n|^2$ respectively. For the right-hand-side of (\ref{eq:class_defn_2}), the computation is the same as for the transverse directions, given that the integrals do not see the polarization dependence, and so the result is as in (\ref{eq:<XX>cov}) but with $\lambda_n^-$ replacing $\lambda^i_n$ in accordance with the above result.

Similarly, for the $X^-X^-$ directions, the result is:
\begin{equation}\label{eq:<X^-X^->lc}
\begin{aligned}
\big\langle \!:\!\big[X^{-}&(z',\bar{z}')-x^-\big]\big[X^{-}(z,\bar{z})-x^-\big]\!:\!\big\rangle\\
&=-\Big(\frac{\alpha'}{2}\Big)^2\langle :\!(\hat{p}^-)^2\!\!:\rangle \ln|z'|^2\ln|z|^2\\
& +\frac{\alpha'}{2}\sum_{n\neq0}\frac{1}{n^2}\Big[\lambda_n^-\lambda^{*-}_n\Big(\frac{z}{z'}\Big)^n+\bar{\lambda}_n^-\bar{\lambda}^{*-}_n\Big(\frac{\bar{z}}{\bar{z}'}\Big)^n\\
&-\lambda_n^-\bar{\lambda}^{-}_n\Big(\frac{1}{z'\bar{z}}\Big)^n-\bar{\lambda}_n^-\lambda^-_n\Big(\frac{1}{\bar{z}'z}\Big)^n\Big],
\end{aligned}
\end{equation}
whereas for the $X^-X^+$ and $X^iX^+$ directions only the zero modes contribute, because $\langle X^{\mu}-x^{\mu}\rangle=-i\frac{\alpha'}{2}\langle \hat{p}^{\mu}\rangle\ln|z|^2$ (with $\mu=\{\pm,i\}$),
\begin{equation}\label{eq:<X^-X^+>lc}
\begin{aligned}
\big\langle \!:\!\big[X^{\mu}(z',\bar{z}')&-x^{\mu}\big]\big[X^{+}(z,\bar{z})\big]\!:\!\big\rangle\\
&=-\Big(\frac{\alpha'}{2}\Big)^2\langle \hat{p}^{\mu}\rangle p^+\ln|z'|^2\ln|z|^2.
\end{aligned}
\end{equation}
We have thus proven that (\ref{eq:class_defn_2}) is indeed satisfied for the lightcone gauge coherent states (\ref{eq:DDF_coherent_closed_lcg_proj}), in all spacetime directions.

Furthermore, from (\ref{eq:<XX>cov}) it follows that the \emph{rms} transverse distance from the center of mass to an arbitrary point on the string, $r=\sqrt{\langle ({\bf X}(z,\bar{z})-{\bf x})^2\rangle}$, in the rest frame, ${\bf p}=0$, is given by,
\begin{equation}\label{eq:rms r}
r^2 = \frac{\alpha'}{2}\sum_{n>0}\frac{1}{n^2}\Big(|\lambda_n|^2+|\bar{\lambda}_n|^2-2{\rm Re}\big(\lambda_n\cdot \bar{\lambda}_ne^{-2in\tau_{\rm M}}\big)\Big),
\end{equation}
where we have Wick rotated back to a Minkowski signature worldsheet, $\tau=i\tau_{\rm M}$. The vertex operator (\ref{eq:DDF_coherent_closed_lcg_proj}) and by extension (\ref{V_0P}) clearly represents a macroscopic string when $\lambda_n$ and $\bar{\lambda}_n$ satisfy,
$$\sum_{n>0}\frac{1}{n^2}\Big(|\lambda_n|^2+|\bar{\lambda}_n|^2-2{\rm Re}\big(\lambda_n\cdot \bar{\lambda}_ne^{-2in\tau_{\rm M}}\big)\Big)\gg1.$$
Recall that we are to enforce $\sum_{n>0}|\lambda_n|^2<\infty$ and similarly for the antiholomorphic sector in order to ensure that the coherent state vertex operators are well behaved \cite{Calucci89}.

Let us compare the result (\ref{eq:rms r}) for the size of a string with the naive estimate for the length or size of a string, $\ell\sim \sqrt{\alpha'\langle N\rangle}$, which follows from, $m_{\rm eff}^2\sim 4\langle N\rangle/\alpha'$ and $m_{\rm eff}\sim \mu \ell$ (with $m_{\rm eff} = \langle m\rangle$, $\mu=1/(2\pi\alpha')$ the string tension and $\ell$ its length). Recall that $\langle N\rangle = \sum_{n>0}|\lambda_n|^2$, and therefore, $$\frac{r^2}{\alpha'\langle N\rangle}\sim\frac{\sum_{n>0}\frac{1}{n^2}|\lambda_n|^2}{\sum_{n>0}|\lambda_n|^2}\leq 1.$$ For an arbitrarily excited cosmic string where arbitrarily large harmonics, $n$, contribute to $\langle N\rangle$,
$$\ell \ll \sqrt{\alpha'\langle N\rangle},$$ and so the naive estimate $\ell\sim \sqrt{\alpha'\langle N\rangle}$ breaks down when the contribution of high harmonics is significant. This is of course to be expected, because the presence of high harmonics implies also that greater amounts of energy are concentrated in a smaller region of space.

We next show that the non-zero mode components of the angular momentum, $S^{ij}$, and $S^{i-}$ associated to the covariant gauge coherent vertex operator (\ref{V_0A}), that associated to the corresponding lightcone gauge state (\ref{eq:DDF_coherent_closed_lcg_proj}) and that of the classical solutions (\ref{eq:X_cl^i}) are all equal to the expressions found for lightlike compactified states (\ref{eq:Jcorr_closed}) and (\ref{eq:Jcorr_closed_long}), re-written here for convenience: for the transverse directions,
\begin{equation}\label{eq:Jcorr_closed2}
\langle S^{ij}\rangle_{\rm cov}=\langle S^{ij}\rangle_{\rm lc}=\sum_{n>0}\frac{2}{n}{\rm Im}\big(\lambda_n^{*i}\lambda_n^j+\bar{\lambda}_n^{*i}\bar{\lambda}_n^j\big)=S_{\rm cl}^{ij},
\end{equation}
and for the longitudinal components,
\begin{equation}\label{eq:Jcorr_closed_long2}
\begin{aligned}
\langle &S^{-i}\rangle_{\rm cov}=\langle S^{-i}\rangle_{\rm lc}=S_{\rm cl}^{-i}\\
&=\sqrt{\frac{2}{\alpha'}}\sum_{m>0}\sum_{\ell\in\mathbb{Z}}\frac{1}{np^+}{\rm Im}\big(\lambda_{m-\ell}^*\cdot \lambda_{\ell}^{*}\,\lambda_m^i+\bar{\lambda}_{m-\ell}^*\cdot\bar{\lambda}_{\ell}^{*}\,\bar{\lambda}_m^i\big),
\end{aligned}
\end{equation}
with in addition all components involving the $+$ direction equal to zero. The derivation of these expressions is almost identical to that described in the open string coherent state section. The three modifications that are worth mentioning are: (a) the covariant and lightcone gauge projected vertex operators are not eigenstates of the annihilation operators, there being instead the relations (\ref{eq:a_n|V>=lambda_n|V-n>}) for the lightcone gauge and,
\begin{equation}\label{eq:aiVs}
\alpha_{m>0}^i\cdot V_0(\lambda)\cong\sum_{n=1}^{\infty}\frac{m}{n}\lambda_n^iB^{-n}_{m}\cdot V_{-m}(\lambda).
\end{equation}
for the covariant gauge; (b) there is a single $s$-integral due to the property mentioned with an example in (\ref{eq:one s-int}) and so we do not need the relation analogous to (\ref{eq:aiVs}) for the longitudinal direction; and (c) there exist the orthogonality relations, $\langle V_n^{\dagger}V_m\rangle_{\rm cov} = \delta_{n,m}$, $\langle V_n|V_m\rangle_{\rm lc}=\delta_{n,m}$ in covariant and lightcone gauge respectively.

\section{Consistency Check}
In this section we would like to check that the coherent state vertex operators (\ref{V_0P}) have the correct singularity structure (i.e.~that required by conformal invariance) when two vertices approach on the worldsheet, namely:
\begin{equation}\label{eq:<<VV>> norm}
\begin{aligned}
\lim_{z\rightarrow w}\llangle V&^{\dagger}(z,\bar{z})V(w,\bar{w})\rrangle \sim\\
&\sim i(2\pi)^d\delta^d(0)\,\Big(\frac{g_c^2}{2p^+\mathcal{V}_{d-1}}\Big)\,\frac{1}{|z-w|^4},
\end{aligned}
\end{equation}
with the expectation value defined in terms of a path integral over embeddings,
\begin{equation}\label{eq: <<VV>>}
\begin{aligned}
\llangle V^{\dagger}(z,&\bar{z})V(w,\bar{w})\rrangle\equiv\left(\frac{4\pi^2\alpha^{\prime}}{\int_{\Sigma}d^2z\sqrt{g}}\,{\rm det}^{\prime}\Delta_{(0)}\right)^{d/2}\\
&\times\int_{\mathcal{E}}\mathcal{D}Xe^{-S[X]}\,\,V^{\dagger}(z,\bar{z})V(w,\bar{w}),
\end{aligned}
\end{equation}
with $S[X]=\frac{1}{2\pi\alpha'}\int d^2z\partial_zX\cdot \partial_{\bar{z}}X$ the usual Polyakov action,
the worldsheet Laplacian, $\Delta_{(0)}=-2g^{z\bar{z}}\partial_z\partial_{\bar{z}}$, and the measure defined according to: $\int\mathcal{D}Xe^{-\|X\|^2/4\pi\alpha^{\prime}}\equiv 1$ with $
\|\delta X\|^2= \int_{\Sigma}d^2z\sqrt{g}\delta X\cdot \delta X$. This will in turn confirm that  the normalization of the functional representation of the coherent state (\ref{V_0A}), namely (\ref{V_0P}), is consistent with that obtained by operator methods, namely (\ref{eq:C_llbar coh norm}). Note furthermore, that the statement (\ref{eq:<<VV>> norm}) is equivalent to the CFT statement (\ref{eq:VVope normalization}) that was derived by the `one string in volume $\mathcal{V}_{d-1}$' requirement that leads to correctly normalized $S$-matrix elements. To simplify the computation we will consider the case when there is only a single harmonic present (in the corresponding lightcone gauge state) and work in the rest frame, where the vertex operator interest is of the form,
\begin{equation}\label{eq:(n,m) coherent vertex}
\begin{aligned}
V(z,&\bar{z})= C\int_0^{2\pi} \dslash s \exp\Big(\frac{1}{n}\,e^{ins}\lambda_n\cdot P_n\,e^{-inq\cdot X(z)}\Big)\\
&\times\exp\Big(\frac{1}{m}\,e^{-ims}\bar{\lambda}_m\cdot \bar{P}_m\,e^{-imq\cdot X(\bar{z})}\Big)\,e^{ip\cdot X(z,\bar{z})}.
\end{aligned}
\end{equation}
We are to choose the normalization constant $C$ such that (\ref{eq:<<VV>> norm}) is satisfied. From the definitions of the dimensionless quantities $P_n(z)$ and $\bar{P}_m(\bar{z})$, defined (we use units where $\alpha'=2$ in this section) in (\ref{eq:PPbardfn}), it follows that we can equivalently write,
\begin{equation}\label{eq:V in terms of J}
\begin{aligned}
V(z,&\bar{z}) = C\int_0^{2\pi} \dslash s \sum_{a,b=0}^{\infty}\frac{e^{i(na-mb)s}}{a!b!}\\
&\times \prod_{i=1}^a\big(\lambda_n\cdot D^{(n)}_iX\big)\prod_{j=1}^b\big(\bar{\lambda}_m\cdot \bar{D}^{(m)}_jX\big)\\
&\times e^{i\int d^2z'J(z',\bar{z}')\cdot X(z',\bar{z}')},
\end{aligned}
\end{equation}
provided we define the operators:
\begin{subequations}\label{eq:D^n Dbar^m}
\begin{align}
&D^{(n)}_j \equiv \frac{1}{n}\oint \frac{dw_j}{2\pi iw_j}\sum_{\ell=1}^n w_j^{-(n-\ell)}\,\frac{i}{(\ell-1)!}\,\partial_z^{\ell},\\
&\bar{D}^{(m)}_j \equiv -\frac{1}{m}\oint \frac{d\bar{w}_j}{2\pi i\bar{w
}_j}\sum_{\ell=1}^m \bar{w}_j^{-(m-\ell)}\,\frac{i}{(\ell-1)!}\,\partial_{\bar{z}}^{\ell},
\end{align}
\end{subequations}
and, taking into account the constraint enforced by the $s$-integral, $na=mb$,
\begin{equation}\label{eq:J}
\begin{aligned}
&J^{\mu}(z',\bar{z}') \equiv  \delta^2(z-z')\Big(p^{\mu}-naq^{\mu}\\
&-naq^{\mu}\sum_{s=1}^{\infty}\frac{\sum_{j=1}^aw_j^s}{s!}\,\partial_z^s-mbq^{\mu}\sum_{s=1}^{\infty}\frac{\sum_{j=1}^b\bar{w}_j^s}{s!}\,\partial_z^s\Big).
\end{aligned}
\end{equation}
Using the generating function, 
$
\llangle e^{i\int d^2z J(z,\bar{z})\cdot X(z,\bar{z})}\rrangle=i(2\pi)^d\delta^d(J_0)e^{-\frac{1}{2}\int d^2z\int d^2z' J(z,\bar{z})\cdot J(z',\bar{z}')G'(z,z')},
$
with $G(z,w)$ the closed string propagator, it follows that the path integral over embeddings reads:
\begin{widetext}
\begin{equation}\label{eq: <<VV>>*}
\begin{aligned}
\llangle V^{\dagger}(z,\bar{z})&V(w,\bar{w})\rrangle= |C|^2\int_0^{2\pi} \dslash s \sum_{a,b=0}^{\infty}\frac{e^{i(na-mb)s}}{(a!b!)^2}\\
&\times \sum_{\pi\in S_{2a+2b}/\!\sim}i(2\pi)^d\delta^d(0)\prod_{q=1}^{2a+2b}\mathcal{D}_{\pi(2q-1)}\cdot \mathcal{D}_{\pi(2q)}G(z_{\pi(2q-1)},z_{\pi(2q)}) e^{(2-na-na')G(z,z')},
\end{aligned}
\end{equation}
\end{widetext}
where we have taken into account the fact that in the rest frame $\bar{\lambda}_m\cdot p=\lambda_n\cdot p=0$, and of course $\bar{\lambda}_m\cdot q=\lambda_n\cdot q=0$ holds in all frames.  $S_{\mathcal{I}}$ is the \emph{symmetric group} of degree $\mathcal{I}$ \cite{Hamermesh}, the group of all permutations of $\mathcal{I}=2a+2b$ elements, and the equivalence relation $\sim$ is such that $\pi_i\sim\pi_j$ with $\pi_{i},\pi_j\in S_{\mathcal{I}}$ when they define the same element in (\ref{eq: <<VV>>*}). We have also defined:
\begin{equation}
\begin{aligned}
\mathcal{D}_j \equiv \Big\{&\underbrace{\mathcal{P}_n^{\dagger}(z),\dots,\mathcal{P}_n^{\dagger}(z)}_{a},\underbrace{\bar{\mathcal{P}}_m^{\dagger}(\bar{z}),\dots,\bar{\mathcal{P}}_m^{\dagger}(\bar{z})}_{b},\\
&\underbrace{\mathcal{P}_n(w),\dots,\mathcal{P}_n(w)}_{a},\underbrace{\bar{\mathcal{P}}_m(\bar{w}),\dots,\bar{\mathcal{P}}_m(\bar{w})}_{b}\Big\}.
\end{aligned}
\end{equation}
with the operators $\mathcal{P}_n(w)$ and $\bar{\mathcal{P}}_n(\bar{w})$, given by:
\begin{equation}
\begin{aligned}
&\mathcal{P}_n(w)\equiv \lambda_n\frac{1}{n}\sum_{\ell=1}^n\mathcal{S}_{n-\ell}(n,w)\,\frac{i}{(\ell-1)!}\,\partial_w^{\ell}\\
&\bar{\mathcal{P}}_m(\bar{w})\equiv \bar{\lambda}_m\frac{1}{m}\sum_{\ell=1}^m\bar{\mathcal{S}}_{m-\ell}(m,\bar{w})\,\frac{i}{(\ell-1)!}\,\partial_{\bar{w}}^{\ell},
\end{aligned}
\end{equation}
with similar expressions for $\mathcal{P}_n^{\dagger}(z)$ (and $\bar{\mathcal{P}}_m^{\dagger}(\bar{z})$) with $-\lambda_n^*$ (and $-\bar{\lambda}_n^*$) replacing $\lambda_n$ (and $\bar{\lambda}_m$) in $\mathcal{P}_n(z)$ ($\bar{\mathcal{P}}_m(\bar{z})$).
The modified elementary Schur polynomials are defined in (\ref{eq:S_n(n,z) and bar{S}_m(m,zbar)}). Carrying out the sum over permutations, and using the fact that the modified elementary Schur polynomials (\ref{eq:S_n(n,z) and bar{S}_m(m,zbar)}) in the appropriate limit are of the form,
\begin{equation}\label{eq:S_n(n,z),bar{S}_m(m,zbar) z->w limits}
\begin{aligned}
&\mathcal{S}_{n-\ell}(n,z)\big|_{z\rightarrow w}\simeq \frac{n!}{(n-\ell)!\ell!}\,(z-w)^{-(n-\ell)},\\
& \mathcal{S}_{n-\ell}(n,w)=(-)^{n+\ell}\mathcal{S}_{n-\ell}(n,z),
\end{aligned}
\end{equation}
with a similar result for the antiholomorphic quantities, with $\bar{z}$, $\bar{w}$ replacing $z$, $w$, one can show that 
the most singular term of the two-point function (\ref{eq: <<VV>>*}) in the limit $z\rightarrow w$ is:
\begin{equation}\label{eq: <<VV>>***}
\begin{aligned}
\llangle &V^{\dagger}(z,\bar{z})V(w,\bar{w})\rrangle\big|_{z\rightarrow w}\simeq i(2\pi)^d\delta^d(0) |C|^2\\
&\times\sum_{a,b=0}^{\infty}\frac{\delta_{na,mb}}{a!b!}\bigg(\frac{|\lambda_n|^2}{n}\bigg)^{a}\bigg(\frac{|\bar{\lambda}_m|^2}{m}\bigg)^{b}\,
|z-w|^{-4}.
\end{aligned}
\end{equation}
Recall that the multi-loop scalar propagator is of the form,
\begin{equation}\label{eq: Green's Function+regular}
G(z,w) = -\ln \left|E(z,w)\right|^2 + 2\pi\, {\rm Im}\int\limits_{z}^{w}\omega_I\left({\rm Im}\Omega\right)_{IJ}^{-1}{\rm Im}\int\limits_{z}^{w}\omega_J,
\end{equation}
where the prime form, $E(z,w)$, has the unique property that for any two points on a genus $h$ Riemann surface, $\lim_{z\rightarrow w}E(z,w)\simeq z-w+\dots$; the zero mode contribution is non-singular in this limit. Furthermore, we have made use of the identity:
$$
\sum_{\ell,r=1}^{n} (-)^{\ell+r}\frac{n!}{\ell!(n-\ell)!}\frac{(n-1)!}{r!(n-r)!}\frac{(\ell+r-1)!}{(\ell-1)!(r-1)!}=1.
$$

From (\ref{eq: <<VV>>***}) we see that if we choose the vertex operator normalization,
\begin{equation}\label{eq:Cnormalization}
C = g_c \bigg[\int_0^{2\pi}\!\! \dslash s\,\exp\Big(\frac{e^{ins}}{n}\,|\lambda_n|^2+\frac{e^{-ims}}{m}\,|\bar{\lambda}_m|^2\Big)\bigg]^{-1/2},
\end{equation}
the fundamental requirement (\ref{eq:<<VV>> norm}) is satisfied. This is precisely the vertex operator normalization expected from the operator formalism (\ref{eq:C_llbar coh norm}) when we identify $g_c\,e^{ip\cdot X(z,\bar{z})}\simeq |0,0;p\rangle$, when the relativistic normalization $\langle 0,0;p'|0,0;p\rangle = 2p^+(2\pi)\delta(p^{'+}-p^+)(2\pi)^{d-2}({\bf p}'-{\bf p})$ is used, and the vertex operator is written in terms of DDF operators, as discussed above. With these conventions, the corresponding state has unit norm, $\langle V|V\rangle=1$.

Note that we are required to interpret the quantity, 
$$
\lim_{z\rightarrow w}\, 2\pi\, {\rm Im}\int\limits_{z}^{w}\omega_I\left({\rm Im}\Omega\right)_{IJ}^{-1}{\rm Im}\int\limits_{z}^{w}\omega_J,
$$
in order to reach the conclusion (\ref{eq: <<VV>>***}). This limit is not single-valued and it is to be understood that we choose to take the limit such that it vanishes (which is trivially true at one-loop, but not in general true at higher genus, and the corresponding result in that case is path dependent).

\section{Discussion}

We have presented a construction of a complete set of mass eigenstate covariant normal ordered vertex operators and a complete set of (open and closed string) covariant coherent state vertex operators with all constraints solved completely. The construction became possible by making use of DDF operators which enable one to translate between lightcone gauge states and covariant vertex operators. The coherent state vertex operators are potentially macroscopic and are in one-to-one correspondence with a classically evolving string -- this suggests that they be identified with \emph{fundamental cosmic strings}.

In the next few paragraphs we briefly discuss and elaborate on the underlying structure that has been uncovered. We start with a discussion of the general covariant mass eigenstate vertex operators, and this is followed by a discussion of the more elaborate coherent state vertex operators.

\subsection{Mass eigenstate vertex operators}

One of the key features we have uncovered is that elementary Schur polynomials (equivalently complete Bell polynomials), $S_m(nq;z)$, and the related polynomials, $H_n^i(z)$ and $\mathbb{S}_{m,n}(z)$, all of which are defined in Appendix \ref{C}, play a fundamental role in the construction: arbitrary flat space vertex operators can be represented in terms of elementary Schur polynomials as we have shown explicitly in (\ref{eq:Vgeneral}) and  (\ref{eq:xiA^g e^ipX1}). The traceless subset of these is given by the vertex operators (\ref{eq: many A_-N DDF}). These polynomials have useful integral representations which facilitate path integral computations.

Building on the observations of D'Hoker and Giddings \cite{D'HokerGiddings87}, the use of DDF operators has enabled us to present an explicit one-to-one map between the lightcone gauge states and covariant normal ordered vertex operators. In the case of traceless polarization tensors there is a simple prescription: to construct the normal ordered covariant vertex operator corresponding to a given lightcone gauge state we make the replacements (\ref{eq: alpha map}),
$$
\begin{matrix} 
      \phantom{\Big(}\alpha_{-n}^i \\
      \phantom{\Big(}\tilde{\alpha}_{-\bar{n}}^i  \\
      \phantom{\Big(}|0,0;p^+,p^i\rangle  \\
\end{matrix}
\hspace{0.7cm}
\begin{matrix} 
      \phantom{\Big(}\rightarrow \\
      \phantom{\Big(}\rightarrow  \\
      \phantom{\Big(}\rightarrow  \\
\end{matrix} 
\hspace{0.7cm}
\begin{matrix} 
      \phantom{\Big(}H_n^i(z) \\
      \phantom{\Big(}\bar{H}_{\bar{n}}^i(\bar{z})  \\
      \phantom{\Big(}g_c\,e^{i(p-Nq)^{\mu} X_{\mu}(z,\bar{z})}  \\
\end{matrix} 
$$
In the general case (when the polarization tensors are not traceless), the corresponding map has been identified in Sec.~\ref{arb_vert}. 
The spacetime vectors, $p^{\mu},q^{\mu}$, are defined for the closed string in (\ref{eq:pq cond_closed}) and for the open string in (\ref{eq:pq cond_open}), $q^{\mu}$ is transverse to all oscillator indices and the overall normalization and polarization tensors are then the same on both sides of the correspondence. States on both sides of this map have identical masses, angular momenta and we conjecture that they also share identical interactions. It would be useful to check this conjecture, possibly by performing amplitude computations on both sides of the correspondence and checking that there is agreement. 

Due to the explicit presence of transverse indices on the resulting covariant vertex operators, one may wonder whether these are truly covariant (in the spacetime sense). The answer is that they are covariant but not manifestly so. This is made clear by the two examples (\ref{eq:cov_graviton}) and (\ref{eq:ManiCovVert}) (the first of which has already been given in \cite{D'HokerGiddings87}), which have been re-written in such a way that the resulting polarization tensors and momenta can have all spacetime components non-vanishing, not just the transverse ones. These vertices can be inserted into covariant path integrals \cite{Polchinski_v1,DHokerPhong} and one need not make the covariance manifest in order to do so.

\subsection{Coherent state vertex operators}

The DDF construction has also enabled us to construct a complete set of closed and open string coherent state covariant vertex operators, i.e.~states characterized by \emph{continuous} labels (namely the polarization tensors $\lambda_n^i$, $\bar{\lambda}_n^i$), which transform correctly under all symmetries of bosonic string theory \footnote{We have thus overcome the problems in the covariant coherent state construction encountered by Calucci \cite{Calucci87}, see also \cite{LarsenSanchez00,Blanco-PilladoIglesiasSiegel07} among others.}. The precise definition of a coherent state vertex operator, that we suggest is appropriate in the context of superstring theory, can be found in the opening lines of Sec.~\ref{OS} and Sec.~\ref{CS}, for the open and closed string respectively \footnote{Note that the naive definition that coherent states should be eigenstates of the annihilation operators is not in general compatible with the symmetries of string theory \cite{Blanco-PilladoIglesiasSiegel07}, as this would imply that $\langle X\rangle=X_{\rm classical}$, and this is not possible when states are invariant under spacelike worldsheet translations, see comments below (\ref{eq:X^+cov}).}. One of the most important features of these vertex operators is that they have a classical interpretation -- what we mean by a state with a classical interpretation has been explained in the opening lines of Sec.~\ref{SCS}. The \emph{rms} transverse distance from the center of mass to an arbitrary point on the string, see (\ref{eq:rms r}), is arbitrary and specified by the magnitude of the polarization tensors $\lambda_n^i$, $\bar{\lambda}_n^i$ -- when $|\lambda_n|^2\gg1$ these strings are macroscopic with expectation values evolving according to the classical equations of motion, and may therefore be identified with a toy model version of the macroscopic fundamental cosmic strings. A more realistic version would be the corresponding superstring construction with an appropriate compactification of the extra dimensions -- this is currently under investigation.

\subsection{Open string coherent states}

The open string coherent states (\ref{eq:DDF_coherent}) are constructed from a linear superposition of the open string mass eigenstates of Sec.~\ref{DDF}. The spacetime set-up we have in mind here corresponds to a vertex operator for an open string attached to a single D$p$-brane or two parallel D$p$-branes (of the same dimensionality), the so called $p$-$p$ string vertex operators NN and DD. The construction of the more general $p$-$p'$ vertex operators with possibly mixed boundary conditions ND and DN would also be interesting, see e.g.~\cite{Hashimoto96}. We have concentrated on strings with excitations within the D-brane worldvolume (i.e.~polarization tensors with non-zero components in directions parallel to the brane), the corresponding transverse excitations which have the interpretation of ripples of the brane being related to these via T-duality \cite{NairShapereStromingerWilczek87,Polchinski96}. Apart from these, there are also open strings with excitations in both the transverse and tangent directions relative to the brane. 

We have also provided a one-to-one correspondence between every open string covariant coherent state vertex operator, the corresponding lightcone gauge description and finally the classical solutions to which these vertex operators correspond to. We computed the angular momentum and mass of these states and showed that there is agreement between these three descriptions, see (\ref{eq:Jcorr}) and (\ref{eq:Jcorrlong}).

\subsection{DLCQ closed string coherent states}

The closed string coherent states we have considered are composed of two copies of the open string. The construction (\ref{eq:DDF_coherent_closed}), of Sec.~\ref{CSLCB}, with the corresponding lightcone gauge expression \footnote{A similar expression has appeared already in the literature, e.g.~\cite{ChialvaDamour06}.} (\ref{eq:DDF_coherent_closed_lcg}), is only consistent in a spacetime with \emph{lightlike compactification}, $X^-\sim X^-+2\pi R^-$, see Fig.~\ref{fig:nullcyl}. The normal ordered expression has been given in (\ref{eq:DDF_coherent_cov_closed_transverse}) for the case of traceless polarization tensors. Although these states are presumably not phenomenologically relevant (at least if they are interpreted as cosmic strings because lightlike compactification breaks 4-dimensional Lorentz invariance), they serve as a good starting point for the more refined closed string coherent state construction of Sec.~\ref{CSNCB}.

The lightlike compactified coherent states, nevertheless, have many interesting features and may have other applications: lightlike compactification also known as Discrete Lightcone Quantization (DLCQ) \cite{PauliBrodsky85,HellermanPolchinski99} of M-theory (which reduces to type IIA superstring theory when the radius of the 11$^{\rm th}$ dimension is taken to zero) has been conjectured \cite{Susskind97} to be equivalent to \emph{finite} N U(N) super Yang-Mills, see also \cite{DijkgraafVerlindeVerlinde97,Sen98,Seiberg97} and \cite{GrignaniOrlandPaniakSemenoff00,Semenoff00, Semenoff02} \footnote{DS would like to thank Sanjaye Ramgoolam for a very interesting discussion on the Matrix Model -- string theory correspondence.}. Therefore, there should be a one-to-one correspondence of the (superstring version of the) DLCQ vertex operators of the current paper, to the U(N) super Yang-Mills spectrum of states. A concise overview of these developments can be found in \cite{Das09}. Although the present article is specific to the bosonic string, many of these results go through to the superstring as we hope to show in a forthcoming article. The DLCQ coherent states have been shown to have certain perhaps surprising features: even though $X^-\sim X^-+2\pi R^-$ the expectation value is single-valued: $\langle X^-(\sigma+2\pi,\tau)\rangle=\langle X^-(\sigma,\tau)\rangle$ with all spacetime components being non-trivially consistent with the classical evolution, $\partial_z\partial_{\bar{z}}\langle X^{\mu}(z,\bar{z})\rangle=0$, see (\ref{eq:<X>_coherent_lc closed}), (\ref{eq:<X^->_coherent_lc closed}) and (\ref{eq:N=Nbar}). This presumably implies that lightlike compactification is a quantum-mechanical effect which is invisible at the classical level -- it may be interesting to understand what the corresponding implications are.

There are certain subtleties here, related to whether the vertex operators are invariant under spacelike worldsheet shifts or not: when vertex operators are invariant under such shifts, the expectation value $\langle X^{\mu}(z,\bar{z})\rangle$ cannot satisfy the classical equations of motion non-trivially \cite{Blanco-PilladoIglesiasSiegel07}. This seems to be a gauge dependent issue that is not related to whether vertex operators have a classical interpretation or not. For example in lightcone gauge, lightlike compactification breaks the invariance under spacelike worldsheet shifts (while preserving conformal invariance) and this is why the expectation values are compatible with the equations of motion (\ref{eq:<X>_coherent_lc closed}) and (\ref{eq:<X^->_coherent_lc closed}). Indeed, for every classical solution to the equations of motion there is a lightlike compactified coherent state with expectation values consistent with these equations of motion. These are subtle issues and have been explained in great detail in Sec.~\ref{CSLCB}. For example, the covariant gauge version of the coherent state (\ref{eq:DDF_coherent_closed}) is invariant under spacelike worldsheet shifts and so does not satisfy the equations of motion non-trivially: there is only the zero mode contribution (\ref{eq:<Xpm>}) with a similar expression for the transverse indices. 

We suggest that states with a classical interpretation that \emph{are} invariant under spacelike worldsheet shifts should satisfy the equation (\ref{eq:class_defn_}), which may be interpreted as a definition of classicality for such states. In fact, this definition is relevant for most states with a classical interpretation: all states in lightcone or covariant gauge in a spacetime without lightlike compactification are invariant under such shifts, whether or not they have a classical interpretation. Static gauge on the other hand breaks the invariance under shifts and so instead the definition $\langle X\rangle=X_{\rm cl}$ is appropriate.

Another interesting feature is the mass-shell constraint, which is identical to the usual expression for non-compact spacetimes, $m^2=2(N+\bar{N}-2)/\alpha'$, but with $N$ not necessarily equal to $\bar{N}$ (without breaking conformal invariance): the radius of compactification, $R^-$, does not appear in this expression. Furthermore, there is a rather curious dependence of the \emph{total} zero mode momentum on $R^-$, see (\ref{eq:massshell}).

Finally, as a consistency check we have also shown that the covariant vertex operator (\ref{eq:DDF_coherent_closed}) and the lightcone gauge state (\ref{eq:DDF_coherent_closed_lcg}) have identical angular momenta in all spacetime directions which is in agreement with the corresponding classical computation, see (\ref{eq:Jcorr_closed_long}) and (\ref{eq:Jcorr_closed}). This, together with the fact that there is a one-to-one correspondence between the covariant and lightcone gauge states, supports our conjecture that the lightcone gauge states (\ref{eq:DDF_coherent_closed_lcg}) and the covariant vertex operators (\ref{eq:DDF_coherent_closed}) are different manifestations of the same states and therefore share identical interactions.

\subsection{Minkowski space closed string coherent states}

Consistency in the above closed string coherent state construction led to the requirement of a lightlike compactification of spacetime, which led us to identify these vertex operators with DLCQ coherent state vertex operators. Our main objective has been to construct covariant coherent state vertex operators that may be identified with the fundamental cosmic strings, and therefore the requirement of a lightlike compactification is possibly too constraining. In Sec.~\ref{CSNCB} we have shown that, with an appropriate projection, closed string coherent states can consistently be embedded in a spacetime without lightlike compactification: starting from the DLCQ coherent states we project out the lightlike winding modes and end up with a vertex operator (\ref{V_0A}) that satisfies the definition of a coherent state, see Sec.~\ref{CS}, and has a classical interpretation. The corresponding normal ordered vertex operator is given in (\ref{V_0P}) for the case of traceless polarization tensors. By projecting out the winding states, translation invariance is restored in both lightcone and covariant gauges and so the relevant definition of classicality is (\ref{eq:class_defn_}), which as we have shown (\ref{eq:<XX>cov}) is satisfied by the projected states.

\subsection{Outlook}

An immediate application for the coherent state vertex operators is in fundamental cosmic string evolution: it is likely that these are then the correct vertex operators for the description of cosmic strings and it is now possible to search for discrepancies between the classical computations and the string theory predictions. Here the coherent states are useful not only because they correspond to an exact perturbative description of an arbitrarily excited macroscopic cosmic string, but because gravitational backreaction which is almost always neglected in the classical computations is automatically taken into account in string perturbation theory. In a forthcoming article we hope to present the first such computation of the gravitational radiation from cosmic string loops, including the effects of gravitational backreaction.

A particularly interesting set-up is the gravitational radiation from strings with cusps which classically have been shown \cite{DamourVilenkin00,DamourVilenkin01} to lead to strong signals that may be detected in the gravitational wave experiments Advanced LIGO and LISA. It is likely \cite{O'CallaghanChadburnGeshnizjaniGregoryZavala10} that the effect of extra dimensions can play a significant role in the damping of the cusp signal, although it is also important to better understand how the size of the extra dimensions constrains the statistically favorable configurations of long strings. Cusps are likely to be a generic feature of string with junctions as well \cite{DavisRajamanoharanSakellariadou08}. Recent evidence  \cite{BinetruyBoheHertogSteer10} also suggests that for string loops with junctions the kink signal plays a more significant role than does the gravitational wave signature from cusps, although one might expect the number of loops with junctions to be smaller than the number of loops without junctions. It might be that gravitational backreaction plays a significant role in all these computations \cite{QuashnockSpergel90}, especially close to cusps and kinks on cosmic strings and therefore it is very important to carry out the corresponding string theory computations and check that there is agreement. In any case, given the quantum nature of fundamental cosmic strings, it is important to check that the evolution is predominantly classical and that quantum effects are small.

Another interesting avenue is the comparison of mass eigenstates and coherent states. A number of decay rate computations of mass eigenstate vertex operators have been carried out, see e.g.~\cite{MitchellTurokWilkinsonJetzer89,MitchellSunborgTurok90,IengoRusso02,IengoRusso03,ChialvaIengoRusso03,ChialvaIengo04,Iengo06,GutperleKrym06,Chialva09}, although explicit results have been limited to vertices on the leading trajectory (i.e.~first harmonics only excited), where for example one does not expect to find non-degenerate cusps. At the qualitative level these are in line with one's geometrical classical expectation: mass eigenstate vertex operators corresponding classically to rotating circular loops are more stable than vertex operators corresponding to collapsed rotating loops for example \cite{ChialvaIengoRusso03}, thus showing that these states do share at least certain characteristics of the classical evolution. However, the spectrum of gravitational radiation from mass eigenstates does not match the corresponding classical computation \cite{Iengo06}. It will be interesting to determine how the mass eigenstate amplitude computations compare with the corresponding coherent state vertex operator computations.

Finally, we mention also an analogy with standard point particle quantum mechanics. An important feature of harmonic oscillator coherent states is that in the presence of interactions an initial coherent state, $|\psi(0)\rangle$, remains a coherent state when the Hamiltonian is linear in the operators of the Heisenberg-Weyl group, $H_4$, e.g. $a$, $a^{\dagger}$, $\mathds{1}$ and $a^{\dagger}a$ with $[a^{\dagger},a]=1$. That is to say, if $\hat{H}(t)=\hbar \omega a^{\dagger}a+j(t)a^{\dagger}+j^*(t)a$ and $j(t)\neq0$, the solution to the Schrodinger equation, $i\partial_t|\psi\rangle=\hat{H}(t)|\psi\rangle$, reads \cite{ZhangFengGilmore90}, $|\psi(t)\rangle = \exp(\lambda(t)a^{\dagger}-\lambda^*(t)a)|0\rangle e^{-i\eta(t)}$, with $\lambda(t)=-i e^{-i\omega t}\int_0^td\tau e^{i\omega \tau}j^*(\tau)$ and $\eta(t) = \frac{1}{2}\omega t+\int_0^t d\tau {\rm Re}[j(\tau)\lambda(\tau)]$. Therefore, in the presence of interactions the resulting state is a coherent state for all $t$, in accordance with the above statement.  It is conceivable that this remains true in string theory, i.e.~that coherent states evolve into coherent states at least at weak coupling, and it would be interesting to establish whether this is indeed the case. In the cosmic string context this is related to the question of what the final state of a radiating cosmic string is, or whether interactions preserve the classical nature of cosmic strings, questions that can be addressed using the coherent state vertex operators that we have constructed.

The developments presented here are expected to lead to greater insight into the observational prospects of cosmic strings, and in a wider sense of string theory.


\begin{acknowledgements}
The authors are gratefully indebted to Joe Polchinski for providing crucial insight and suggestions which ultimately made the coherent state construction in non-compact spacetime possible. DS would also like to thank KITP, UC Santa Barbara, for hospitality during the initial stages of this research and Joe Polchinski for interesting him in this project. Both authors have benefitted from discussions with Diego Chialva. DS would also like to thank David Bailin, Jose Blanco-Pillado, Edmund Copeland, Robert Dijkgraaf, Paolo Di Vecchia, Kevin Falls, Jo\~ao Penedones, Sanjaye Ramgoolam, Oliver Rosten, Xavier Siemens, Andrew Strominger, Arkady Tseytlin, Tanmay Vachaspati, Eric Verlinde, Steven Weinberg, Edward Witten and especially Ashoke Sen for very helpful discussions and suggestions.
\end{acknowledgements}

\appendix
\section{Closed String Conventions}\label{C}

Consider a worldsheet cylinder with coordinates $0\leq \sigma\leq 2\pi$ and $-\infty<\tau<\infty$, and the identification $\sigma\sim\sigma+2\pi$. We usually work in the conformally equivalent coordinates on the complex plane, $z=e^{-i(\sigma+i\tau)}$ and $\bar{z}=e^{i(\sigma-i\tau)}$, where the string at asymptotic infinity $\tau=-\infty$ is mapped to a point at the origin. Unless otherwise noted we work in the coordinate system ($z,\bar{z}$). States are then specified by local functionals on the complex plane, $V(0,0)$, which by translation invariance is shifted to some generic point, $V(z,\bar{z})$. We use a Euclidean signature worldsheet unless  specified otherwise, where $\tau=(\tau)_{\rm Euclidean}=i(\tau)_{\rm Minkowski}$. For easy reference we note that $i\partial_{\sigma}\equiv z\partial_z-\bar{z}\partial_{\bar{z}}$, $\partial_{\tau}\equiv z\partial_z+\bar{z}\partial_{\bar{z}}$ and $2\tau=\ln|z|^2$. It is sometimes useful to work in the coordinate system, $w=\sigma+i\tau$ and $\bar{w}=\sigma-i\tau$ with $\partial_w\equiv\frac{1}{2}(\partial_{\sigma}-i\partial_{\tau})$ and $\partial_{\bar{w}}\equiv\frac{1}{2}(\partial_{\sigma}+i\partial_{\tau})$.

\subsection{Closed String Mode Expansion}

Recall that for the closed string the mode expansion for the position operator reads,
\begin{equation}\label{X lc}
\begin{aligned}
X^{\mu}(z,\bar{z})&=\hat{x}^{\mu}-i\frac{\alpha'}{2}\,\hat{p}_{\rm L}^{\mu}\ln z-i\frac{\alpha'}{2}\,\hat{p}_{\rm R}^{\mu}\ln \bar{z}\\
&+i\Big(\frac{\alpha'}{2}\Big)^{1/2}\sum_{n\neq0}\frac{1}{n}\,\big(\alpha_n^{\mu}\,z^{-n}+\tilde{\alpha}_n^{\mu}\,\bar{z}^{-n}\big),
\end{aligned}
\end{equation}
with $\hat{x}^{\mu}=\hat{x}_{\rm L}^{\mu}+\hat{x}_{\rm R}^{\mu}$, total momentum $\hat{p}^{\mu}=\frac{1}{2}\big(\hat{p}_{\rm L}^{\mu}+\hat{p}_{\rm R}^{\mu}\big)$, and winding vector $\hat{w}^{\mu}=\frac{1}{2}\big(\hat{p}_{\rm L}^{\mu}-\hat{p}_{\rm R}^{\mu}\big)$. If we define $\dslash z=dz/(2\pi)$, $\alpha_0^{\mu}=\sqrt{\frac{\alpha'}{2}}\hat{p}_{\rm L}^{\mu}$ and $\tilde{\alpha}_0^{\mu}=\sqrt{\frac{\alpha'}{2}}\hat{p}_{\rm R}^{\mu}$, the dimensionless mode expansion operators are given by \cite{Polchinski_v1},
\begin{equation*}
\begin{aligned}
&\alpha_n^{\mu}=\sqrt{\frac{2}{\alpha'}}\oint \dslash z\,\partial X^{\mu}\,z^n,\\ &\tilde{\alpha}_n^{\mu}=-\sqrt{\frac{2}{\alpha'}}\oint \dslash \bar{z}\,\bar{\partial} X^{\mu}\,\bar{z}^n,
\end{aligned}
\end{equation*}
with $(\alpha_n^{\mu})^{\dagger}=\alpha_{-n}^{\mu}$ and the zero modes are given by \cite{Polchinski88},
$$
\hat{x}^{\mu}=\oint \Big(\frac{d z}{2\pi iz}-\frac{d \bar{z}}{2\pi i\bar{z}}\Big)X^{\mu}(z,\bar{z}),
$$
$$
\hat{p}^{\mu}=\frac{1}{\alpha'}\oint \Big(\dslash z\,\partial X^{\mu}-\dslash \bar{z}\,\bar{\partial} X^{\mu}\Big).
$$
The angular momentum operator reads,
$$
\hat{J}^{\mu\nu} =\frac{2}{\alpha'}\oint \Big(\dslash z X^{[\mu}\partial X^{\nu]}- \dslash \bar{z} X^{[\mu}\bar{\partial} X^{\nu]}\Big),
$$
the integrals being along a spacelike curve, e.g.~$|z|^2=1$,  and $a^{[\mu\nu]}=\frac{1}{2}(a^{\mu\nu}-a^{\mu\nu})$. 

\subsection{Operator Products and Commutators}

Recall that for two operators $$A=\oint dz\,a(z), \qquad  B=\oint dw\, b(w),$$ there exists the interpretation, see e.g.~\cite{DiFrancescoMatheuSenechal97},
\begin{equation}\label{eq:[A,B]}
\begin{aligned}
&[A,B]\cong A\cdot B= \oint_0 dw\oint_w dz \,a(z) \cdot b(w),\\
&[A,b(w)]\cong A\cdot b(w)=\oint_w dz\, a(z)\cdot b(w),
\end{aligned}
\end{equation}
the dot denoting an operator product expansion (OPE), where for a free scalar contractions are taken with respect to the propagator,
$$
\big\langle X^{\mu}(z,\bar{z})X^{\nu}(w,\bar{w})\big\rangle = -\frac{\alpha'}{2}\,\eta^{\mu\nu}\ln |z-w|^2.
$$
Formally factorizing the position operator according to $X^{\mu}(z,\bar{z})=X^{\mu}(z)+X^{\mu}(\bar{z})$ then leads to the standard commutation relations,
\begin{equation}\label{eq:[aa]XdXxp}
\begin{aligned}
&\big[\alpha^{\mu}_n,\alpha^{\nu}_m\big]=n\eta^{\mu\nu}\delta_{n+m,0},\quad [x^{\mu},p^{\nu}]=i\eta^{\mu\nu},\\
&\big[X^{\mu}(z),\partial_{\tau} X^{\nu}(z')\big]=\eta^{\mu\nu}\delta(\sigma-\sigma'),
\end{aligned}
\end{equation}
and similarly for the corresponding antiholomorphic quantities. 

\subsection{Closed String DDF Operators and Covariant Commutators}

The relevant components of the DDF operators are defined according to,
\begin{equation}
\begin{aligned}
&A_n^i=\sqrt{\frac{2}{\alpha'}}\oint \dslash z\,\partial X^ie^{inq\cdot X(z)},\\
&\bar{A}_n^i=-\sqrt{\frac{2}{\alpha'}}\oint \dslash \bar{z}\,\bar{\partial} X^ie^{inq\cdot X(\bar{z})}.
\end{aligned}
\end{equation}
The spacetime vector $q^{\mu}$ is transverse to the spacelike indices $i$, and $q^2=0$. These satisfy an oscillator algebra, 
\begin{equation}\label{eq:[A,A]}
\big[A_n^i,A_m^j\big]\cong n\delta^{ij}\delta_{n+m,0},\quad{\rm and}\quad \big[\bar{A}_n^i,\bar{A}_m^j\big]\cong n\delta^{ij}\delta_{n+m,0}
\end{equation}
from which it follows that $(A_n^i)^{\dagger}=A_{-n}^i$.  We define a vacuum according to, $\alpha_{n>0}^{\mu}\cdot e^{ip\cdot X(z,\bar{z})}\cong0$ and $A_{n>0}^i\cdot e^{ip\cdot X(z,\bar{z})}\cong0$ with,
$$
p^2=\frac{4}{\alpha'},\qquad p\cdot q=\frac{2}{\alpha'},\qquad{\rm and}\qquad q^2=0.
$$
From the above definition of the commutators we learn that,
\begin{equation*}
\begin{aligned}
&\big[\alpha_m^{\mu},A^i_n\big] = m\delta^{\mu, i}B^n_m+n\sqrt{\frac{\alpha'}{2}}q^{\mu}D_{m,n}^i,\\
&\big[\alpha_{\ell}^{\mu},B^n_m\big] = n\sqrt{\frac{\alpha'}{2}}q^{\mu}B^n_{m+\ell},\\
&\big[\alpha_{\ell}^{\mu},D_{m,n}^i\big] = \ell\delta^{\mu, i} B^n_{m+\ell}+n\sqrt{\frac{\alpha'}{2}}q^{\mu}D_{m+\ell,n}^i,\\
&\big[\alpha_m^{\mu},E^{n}_{\ell}\big] = m\sqrt{\frac{\alpha'}{2}}q^{\mu}B^{n}_{m+\ell}-n\sqrt{\frac{\alpha'}{2}}q^{\mu}E^{n}_{m+\ell},
\end{aligned}
\end{equation*}
where following \cite{CornalbaCostaPenedonesVieira06} we have defined,
\begin{equation*}
\begin{aligned}
&B^{n}_{m} = \oint \frac{\dslash z}{iz}\,z^m\,e^{inq\cdot X(z)},\\
&D^i_{m,n}=\sqrt{\frac{2}{\alpha'}}\oint \dslash zz^m\partial X^ie^{inq\cdot X(z)},\\
&E^n_m=\oint \dslash z z^mq\cdot \partial Xe^{inq\cdot X(z)}.
\end{aligned}
\end{equation*}
From these commutators and $(\alpha_n^{\mu})^{\dagger}=\alpha_{-n}^{\mu}$, $(A_n^i)^{\dagger}=A_{-n}^i$, it follows that $(B^n_m)^{\dagger}=B^{-n}_{-m}$, $(D^i_{m,n})^{\dagger}=D^i_{-m,-n}$ and $(E^n_m)^{\dagger}=E^{-n}_{-m}$. In addition we learn that,
\begin{equation*}
\begin{aligned}
&\big[A^i_{\ell},D_{m,n}^j\big] = \ell\delta^{ij}E^{\ell+n}_m,\\
&\big[D^i_{-\ell,n},D^j_{\ell,-m}\big]=\delta^{ij}\big(nE^{n-m}_0-\ell B^{n-m}_0\big),
\end{aligned}
\end{equation*}
with $\big[B^r_{\ell},D_{m,n}^i\big] =\big[A_n^i,B^{\ell}_m\big]= \big[A_n^i,E^{\ell}_m\big]=0$ and $[B^n_m,B^{\ell}_r]=[E^n_m,E^{\ell}_r]=[B^n_m,E^{\ell}_r]=0$.

On the chiral half of a (tachyonic) vacuum state, $e^{ip\cdot X(z)}$, one can readily compute the operator products, 
\begin{subequations}\label{eq:opes BDE}
\begin{align}
&B^{-n}_{\phantom{a}m}\cdot e^{ip\cdot X(z)}\cong S_{n-m}(nq;z)\,e^{i(p-nq)\cdot X(z)},\phantom{\Big|}\\
&D_{m,-n}^i\cdot e^{ip\cdot X(z)}\cong H^i_{n-m}(nq;z)e^{i(p-nq)\cdot X(z)},\phantom{\Big|}\\
&E^{-n}_{m}\cdot e^{ip\cdot X(z)}\cong \sqrt{\frac{\alpha'}{2}}q\cdot H_{n-m}(nq;z)e^{i(p-nq)\cdot X(z)},\phantom{\Big|}\label{eq:EeipXc}
\end{align}
\end{subequations}
where the polynomials $S_{n-m}(nq;z)$ and  $H^{i}_{n-m}(nq;z)$ have been defined below and we have made use of the Taylor expansion, $e^{-inq\cdot X(w)}=\sum_{a=0}^{\infty}(w-z)^aS_a{(nq;z)}e^{-inq\cdot X(z)}.$ Note that in (\ref{eq:EeipXc}) we have extended the definition of $H^i_{n-m}(nq;z)$, to include also longitudinal indices, $H^{\mu}_{n-m}(nq;z)$, without changing the form of the polynomial.

\subsection{Gauge Invariant Position Operator}

The position operator is not a gauge invariant quantity, $[L_n,X^{\mu}(z,\bar{z})]\neq0$, and so cannot be inserted into covariant path integrals. It is sometimes useful to have a operator that is gauge invariant that does in many respects have the properties of a position operator and we discuss this next. Motivated by the isomorphism of the algebras satisfied by $\alpha_n^{\mu}$ and $A_n^i$ and by the fact that $[L_n,A_m^i]=0$, let us by direct analogy to (\ref{X lc}) define the following position-like gauge invariant operator for the transverse indices \cite{GebertKoepsellNicolai97,GebertNicolai97},
\begin{equation}\label{eq:curly X}
\begin{aligned}
\mathsf{X}^i(z,\bar{z})=\hat{\mathsf{x}}^i&-i\frac{\alpha'}{2}\,\hat{p}^i\ln |z|^2\\
&+i\Big(\frac{\alpha'}{2}\Big)^{1/2}\sum_{n\neq0}\frac{1}{n}\,\big(A_n^i\,z^{-n}+\bar{A}_n^i\,\bar{z}^{-n}\big).
\end{aligned}
\end{equation}
Here $\hat{p}^i=A_0^i=\alpha_0^i$ and $\mathsf{x}^i=\frac{\alpha'}{2}q_{\mu}J^{i\mu} $ with the lightlike vector $q^{\mu}$ and the angular momentum operator $J^{i\mu}$ as defined above. On account of (\ref{eq:[A,B]}) one finds,
\begin{equation}
\begin{aligned}
&\big[A^i_n,A^j_m\big]=n\delta^{ij}\delta_{n+m,0},\qquad  [\mathsf{x}^i,p^j]=i\delta^{ij}\\
&\big[\mathsf{X}^i(z),\partial_{\tau} \mathsf{X}^j(z')\big]=\delta^{ij}\delta(\sigma-\sigma'),
\end{aligned}
\end{equation}
in direct analogy to (\ref{eq:[aa]XdXxp}). Unlike the standard position operator, $X(z,\bar{z})$, however, the quantity (\ref{eq:curly X}) is gauge invariant given that the DDF operators and the zero modes $\hat{\mathsf{x}}^i$ and $\hat{p}^i$ commute with the Virasoro generators, $\big[L_n,\mathsf{X}^i(z,\bar{z})\big]=0$, and $[L_n,\hat{\mathsf{x}}^i]=[L_n,\hat{p}^i]=0,$ for all $n\in\mathbb{Z}$ (and similarly for $\bar{L}_n$), and therefore define sensible operators that may be inserted into covariant path integrals. 

In fact, $\mathsf{X}^i(z,\bar{z})$ can in some sense be thought of as the covariant version of the lightcone quantity $X^i(z,\bar{z})$: the $A_n^i$ reduce to the $\alpha_n^i$ when one restricts to lightcone gauge in which case (\ref{eq:curly X}) reduces to (\ref{X lc}). We can use $q^{\mu}$ to define a lightcone time $q\cdot X(z,\bar{z})=-i\ln|z|^2$ to find that (at least classically),
\begin{equation*}
\begin{aligned}
A_n^i\big|_{\rm l.c.} &= \sqrt{\frac{2}{\alpha'}}\oint \dslash z \,\partial X^i(z)e^{inq\cdot X(z)}\Big|_{\rm l.c.}\\
&=\sqrt{\frac{2}{\alpha'}}\oint \dslash z\,\partial X^i(z)\, z^n\\
&=\alpha_n^i,
\end{aligned}
\end{equation*}
where we have formally factorized $q\cdot X(z,\bar{z})$ into $q\cdot X(z)=-i\ln z$ and $q\cdot X(\bar{z})=-i\ln\bar{z}$. We hence deduce that at least at the classical level,
$$\big(\mathsf{X}^i(z,\bar{z})-\mathsf{x}^i\big)\big|_{\rm l.c.}=X^i(z,\bar{z})-x^i.$$ 
We \emph{conjecture} that this be elevated to a quantum-mechanical statement:
\begin{equation}\label{eq:<cov>=<lc>}
\big\langle V\big|F\big(\mathsf{X}^i(z,\bar{z})-\mathsf{x}^i\big)\big|V\big\rangle_{\rm cov}=\big\langle V\big|F\big(X^i(z,\bar{z})-x^i\big)\big|V\big\rangle_{\rm lc},
\end{equation}
for some well behaved functional $F(A)$ of the argument $A$. Here by $|V\rangle_{\rm cov}\cong V(z,\bar{z})$ we mean the covariant vertex operator (\ref{eq:DDF state}),
\begin{equation}\label{eq:DDF state apx}
\begin{aligned}
V(z,\bar{z})=C&\xi_{ij\dots,kl\dots}\\
&A^{i}_{-n_1} A^j_{-n_2}\dots\bar{A}_{-\bar{n}_1}^k\bar{A}_{-\bar{n}_2}^l\dots\,e^{ip\cdot X(z,\bar{z})},
\end{aligned}
\end{equation}
and $|V\rangle_{\rm lc}$ represents the corresponding lightcone gauge state (\ref{eq:lc state}),
\begin{equation}\label{eq:lc state apx}
\begin{aligned}
|V\rangle_{\rm lc}=C&\xi_{ij\dots,kl\dots}\\
&\alpha^{i}_{-n_1}\alpha_{-n_2}^{j}\dots\tilde{\alpha}_{-\bar{n}_1}^{k}\tilde{\alpha}_{-\bar{n}_2}^{l}\dots |0,0;p^+,p^i\rangle.
\end{aligned}
\end{equation}
The expression (\ref{eq:<cov>=<lc>}) follows from the isomorphism of lightcone (in terms of the $\alpha_n^i,\tilde{\alpha}_n^i$) and covariant states (in terms of the $A_n^i,\bar{A}_n^i$), the isomorphism of the lightcone gauge and gauge invariant position operators, the fact that the states (\ref{eq:DDF state apx}) and (\ref{eq:lc state apx}) have the same mass and angular momenta, the isomorphism of the corresponding oscillator algebras and finally from out main conjecture that the lightcone and covariant states, (\ref{eq:lc state apx}) and  (\ref{eq:DDF state apx}), share identical correlation functions (provided these are gauge invariant).

For example, (\ref{eq:<cov>=<lc>}) implies that the expectation value of the gauge invariant position operator in some covariant state tells us about the position expectation value of the \emph{lightcone gauge description} of this covariant state.

\subsection{Closed String Polynomials}\label{SP}

Elementary Schur polynomials \cite{Kac90} are defined by the generating series \footnote{Elementary Schur polynomials, $S_m({\bf x})$, are not to be confused with the Schur polynomials, $S_{\lambda}({\bf x})$. Given a partition $\lambda=\{\lambda_1\geq\lambda_2\geq \dots\}$ these are related however, $S_{\lambda}({\bf x})=\det (S_{\lambda_i-i+j}({\bf x}))_{1\leq i,j,\leq|\lambda|}$.}, $\sum_{m=0}^{\infty}S_m(a_1,\dots,a_m)z^m=\exp \sum_{n=1}^{\infty}a_n\,z^n
$ and read explicitly:
\begin{subequations}\label{eq:S_m(a)}
\begin{align}
S_m(a_1,\dots,a_m)&=\sum_{k_1+2k_2+\dots+mk_m=m}\frac{a_1^{k_1}}{k_1!}\dots\frac{a_m^{k_m}}{k_m!}\label{eq:S_m(a)a}\\
&=-i\oint_0\dslash w\,w^{-m-1} \exp \sum_{s=1}^{m}a_sw^s\label{eq:S_m(a)b}
\end{align}
\end{subequations}
with $\dslash w\equiv dw/(2\pi)$, $S_0=1$ and $S_{m<0}=0$. When $a_s=-\tfrac{1}{s!}inq\cdot\partial^sX(z)$, with $q^{\mu}$ defined in  (\ref{eq:pq cond_closed}) we write $S_m(nq;z)\equiv S_m(a_1,\dots,a_m)$. For instance, when $a_s=-\tfrac{1}{s!}inq\cdot\partial^sX(z)$ or $a_s=-\tfrac{1}{s!}inq\cdot\bar{\partial}^sX(\bar{z})$,
\begin{subequations}\label{eq:SSbardfn apx}
\begin{align}
S_m(nq;z)&=\oint_0 \frac{dw}{2\pi iw}\,w^{-m} \exp\Big( -inq\cdot \sum_{s=1}^{m}\frac{w^s}{s!}\partial_z^s X(z)\Big),\\
\bar{S}_m(nq;\bar{z})&=-\oint_0 \frac{d\bar{w}}{2\pi i\bar{w}}\,\bar{w}^{-m} \exp\Big( -inq\cdot \sum_{s=1}^{m}\frac{\bar{w}^s}{s!}\partial_{\bar{z}}^s X(\bar{z})\Big),
\end{align}
\end{subequations}
and when there is no ambiguity we shall write instead $S_m(nq)$ for the same object, and similarly for $\bar{S}_m(nq)$. The following Taylor series is useful, $$e^{-inq\cdot X(z)}=\sum_{a=0}^{\infty}z^aS_a{(nq;0)}e^{-inq\cdot X(0)}.$$ 

Elementary Schur polynomials,  $S_m$, are related to the complete Bell polynomials, $B_m$, according to, $S_m(a_1,a_2,\dots,a_m)=\frac{1}{m!}B_m(a_1,2!a_2,\dots,m!a_m).$ Properties of the latter have been studied in \cite{Riordan58,RomanRota78,WangWang09,Collins01, NoscheseRicci03}.

The following polynomials in $\partial^{\#}X$ and $\bar{\partial}^{\#}X$ are the fundamental building blocks in normal ordered covariant vertex operators and are recorded here for easy reference,
\begin{subequations}\label{eq:PPbardfn apx}
\begin{align}
P_n^i(z)&=\sqrt{\frac{2}{\alpha'}}\sum_{m=1}^n\frac{i}{(m-1)!}\, \partial^mX^i(z)S_{n-m}(nq;z),\label{eq:Pn apx}\\
\bar{P}_n^i(\bar{z}) &=\sqrt{\frac{2}{\alpha'}}\sum_{m=1}^n\frac{i}{(m-1)!}\,\bar{\partial}^mX^i(\bar{z})\bar{S}_{n-m}(nq;\bar{z}).\label{eq:Pnbar apx}
\end{align}
\end{subequations}
which when $\xi_{\dots i\dots }p^i$ is non-vanishing generalizes to
\begin{subequations}\label{eq:H_n apx}
\begin{align}
H_n^i(z)&\equiv \sqrt{\frac{\alpha'}{2}}p^i S_{n}(nq;z)+P_n^i(z),\phantom{\bigg|}\\ 
\bar{H}_n^i(\bar{z})&\equiv \sqrt{\frac{\alpha'}{2}}p^i \bar{S}_{n}(nq;\bar{z})+\bar{P}_n^i(\bar{z}).
\end{align}
\end{subequations}
When necessary we shall also note the argument of the Schur polynomials by writing $P_n^i(mq;z)$ and $H_n^i(mq;z)$ although usually $n=m$ which is why we have written instead $P_n(z)$ and $H_n(z)$. For vertex operators whose lightcone gauge representation is not traceless, $\xi_{\dots i\dots j\dots }\delta^{ij}\neq0$, the following polynomials appear,
\begin{subequations}\label{eq:HS apx}
\begin{align}
&\mathbb{S}_{m,n}(z) \equiv \sum_{r=1}^nrS_{m+r}(mq;z)S_{n-r}(nq;z),\phantom{\bigg|}\label{eq:HSa apx}\\
&\bar{\mathbb{S}}_{m,n}(\bar{z}) \equiv \sum_{r=1}^nr\bar{S}_{m+r}(mq;\bar{z})\bar{S}_{n-r}(nq;\bar{z}),\label{eq:HSb apx}
\end{align}
\end{subequations}
see (\ref{eq:xiA^g e^ipX1}). These polynomials have the properties, $S_0(nq;z)=\sqrt{\alpha'/2}\,q\cdot H_0(nq;z)=1$, and $H^i_0(nq;z)=\sqrt{\alpha'/2}\,p^i$ and vanish when the subscripts are negative. Explicitly, for the first few level numbers, $P_0^i(z)=0$, $P_1^i(z)=i\partial X^i(z)$, $P_2^i(z)=2\partial X^iq\cdot \partial X(z)+i\partial^2X^i(z)$, and so on, where we have taken $\alpha'=2$ for simplicity; also, $S_0(Nq)=1$, $S_1(Nq)=-iNq\cdot \partial X$, $S_2(Nq) = 2(Nq\cdot i\partial X)^2-Nq\cdot i\partial^2X$,\dots

Also the following elementary Schur polynomials also appear in the final section,
\begin{equation}\label{eq:S_n(n,z) and bar{S}_m(m,zbar)}
\begin{aligned}
&\mathcal{S}_{n-\ell}(n,z)=\\
&\qquad \oint \frac{du}{2\pi i u}\,u^{-(n-\ell)}\exp\Big(\!\!-n\sum_{s=1}^{\infty}\frac{u^s}{s!}\,\partial_z^sG(z,w)\Big)\\
&\bar{\mathcal{S}}_{m-\ell}(m,\bar{z})=\\
&\qquad -\oint \frac{d\bar{u}}{2\pi i \bar{u}}\,\bar{u}^{-(m-\ell)}\exp\Big(\!\!-m\sum_{s=1}^{\infty}\frac{\bar{u}^s}{s!}\,\partial_{\bar{z}}^sG(z,w)\Big),
\end{aligned}
\end{equation}
obtained from the usual elementary Schur polynomials (\ref{eq:SSbardfn apx}) by the replacement:
$$
inq\cdot X(z)\rightarrow nG(z,w).
$$

\section{Open String Conventions}\label{O}

We label the spacetime directions tangent to the D$p$-brane by lower case latin letters from the beginning of the alphabet, $X^a$, with $a=0,\dots,p$, and directions transverse to the brane by upper case latin letters from the middle of the alphabet, $X^I$, with $I=p+1,\dots 25$.  In lightcone \emph{coordinates} and assuming the associated lightcone directions satisfy Neumann boundary conditions we may define, $$X^{\pm}=\tfrac{1}{\sqrt{2}}\big(X^0\pm X^{p}\big).$$ This is necessary \cite{Giddings96} in order to establish the correspondence between covariant and lightcone gauge: recall that in lightcone gauge $X^+= 2\alpha' p^+\tau_{\rm M}$ (with $\tau\equiv\tau_{\rm Euclidean}=i\tau_{\rm Minkowski}\equiv i\tau_{\rm M}$), which is compatible with Neumann and not Dirichlet boundary conditions, see (\ref{eq:ND BC's apx}). A general spacetime direction is as always labelled by Greek lower case letters, $X^{\mu}$. In summary,
\begin{equation*}
\begin{aligned}
&X^{a}=\{X^{\pm},X^A\}, \qquad{\rm  with}\qquad A=1,\dots,p-1,\\
&X^{i}\,=\{X^A,X^I\}, \,\qquad{\rm  with} \qquad I=p+1,\dots,25,\\
&X^{\mu}=\{X^{\pm},X^i\},
\end{aligned}
\end{equation*}
with the scalar product of two general vectors in components being, 
$U^{\mu}V_{\mu}=-U^-V^+-U^+V^-+U^AV^A+U^IV^I.$
The directions, $X^A$, therefore satisfy Neumann boundary conditions, whereas directions transverse to the brane, $X^I$, satisfy Dirichlet boundary conditions. In the Euclidean worldsheet coordinate \footnote{Our conventions are mostly in agreement with Polchinski \cite{Polchinski_v1}.} $z=e^{-i(\sigma+i\tau)}$, $\bar{z}=e^{i(\sigma-i\tau)}$ with $\sigma\in [0,\pi]$ and $\tau\in(-\infty,\infty)$, (considering only the case of NN and DD strings) Neumann and Dirichlet boundary conditions read respectively, 
\begin{equation}\label{eq:ND BC's apx}
N:\,\,\partial_{\sigma}X^a|_{\partial \Sigma_{1,2}}=0\quad {\rm and} \quad D:\,\,\partial_{\tau}X^I|_{\partial \Sigma_{1,2}}=0.
\end{equation}
It is useful to note furthermore that, $\partial_{\sigma}=i(\bar{z}\bar{\partial}-z\partial)$ and $\partial_{\tau}=\bar{z}\bar{\partial}+z\partial$. In the $(z,\bar{z})$ coordinates the open string physical worldsheet, $\Sigma$, is conformally mapped to the upper half plane with the identification, $z\sim \bar{z}$. The fixed point of this identification (the real line, $z=\bar{z}$) defines the open string boundaries, 
\begin{equation}
\begin{aligned}
&\partial \Sigma_1 \equiv \{z\,|\,z=e^{\tau},-\infty<\tau<\infty\},\\\
&\partial\Sigma_2 \equiv \{z\,|\,z=-e^{\tau},-\infty<\tau<\infty\}.
\end{aligned}
\end{equation}

\subsection{Open String Mode Expansion}

In the open string conventions, the general solution to the equations of motion, $\partial\bar{\partial}X^{\mu}=0$, is given by $X^{\mu}(z,\bar{z})=X^{\mu}(z)+X^{\mu}(\bar{z})$, with
\begin{equation*}
\begin{aligned}
X^{\mu}(z) = x^{\mu}_{\rm L}-i\alpha'p_{\rm L}^{\mu}\ln z+i\sqrt{\frac{\alpha'}{2}}\sum_{n\neq0}\frac{1}{n}\,\frac{\alpha_n^{\mu}}{z^n},\\
X^{\mu}(\bar{z}) = x^{\mu}_{\rm R}-i\alpha'p_{\rm R}^{\mu}\ln \bar{z}+i\sqrt{\frac{\alpha'}{2}}\sum_{n\neq0}\frac{1}{n}\,\frac{\tilde{\alpha}_n^{\mu}}{\bar{z}^n},
\end{aligned}
\end{equation*}
and the momentum is half that of the closed string, $\alpha_0^{\mu}=\sqrt{2\alpha'}\hat{p}_{\rm L}^{\mu}$, $\tilde{\alpha}_0^{\mu}=\sqrt{2\alpha'}\hat{p}_{\rm R}^{\mu}$. If we define the total momentum and winding vectors respectively by,
\begin{equation}\label{eq:pw apx}
p^{\mu}=\frac{1}{2}(p_{\rm L}+p^{\mu}_{\rm R})\qquad{\rm and}\qquad w^{\mu}=\frac{1}{2}(p_{\rm L}^{\mu}-p_{\rm R}^{\mu}),
\end{equation}
it follows that the boundary conditions (\ref{eq:ND BC's apx}) require,
\begin{subequations}\label{eq:wapa}
\begin{align}
w^a = 0,\qquad \alpha_n^a + \tilde{\alpha}_n^a = 0,\\
p^I = 0,\qquad \alpha_n^a - \tilde{\alpha}_n^a = 0,
\end{align}
\end{subequations}
reflecting the fact that open strings cannot wind in the Neumann directions and that the centre of mass momentum in the transverse directions vanishes. Therefore, the string mode expansions take the form,
\begin{equation}\label{eq:X open ND apx}
\begin{aligned}
&NN:\qquad X^{\pm}(z,\bar{z}) = x^{\pm}-i\alpha'p^{\pm}\ln |z|^2\\
&\hspace{3cm}+i\sqrt{\frac{\alpha'}{2}}\sum_{n\neq0}\frac{\alpha_n^{\pm}}{n}\Big(\frac{1}{z^n}+\frac{1}{\bar{z}^n}\Big),\\
&NN:\qquad X^A(z,\bar{z}) = x^A-i\alpha'p^A\ln |z|^2\\
&\hspace{3cm}+i\sqrt{\frac{\alpha'}{2}}\sum_{n\neq0}\frac{\alpha_n^A}{n}\Big(\frac{1}{z^n}+\frac{1}{\bar{z}^n}\Big),\\
&DD:\qquad X^I(z,\bar{z}) = x^I-i\alpha'w^I\ln\frac{z}{\bar{z}}\\
&\hspace{3cm}+i\sqrt{\frac{\alpha'}{2}}\sum_{n\neq0}\frac{\alpha_n^I}{n}\Big(\frac{1}{z^n}-\frac{1}{\bar{z}^n}\Big),
\end{aligned}
\end{equation}
with the two string endpoints located respectively at (switching back to a Minkowski worldsheet, $\tau=\tau_{\rm E} = i\tau_{\rm M}$),
\begin{equation*}\label{eq:X open ND boundary}
\begin{aligned}
&X^a(z,\bar{z})|_{\partial\Sigma_1} = x^a+(2\alpha')p^a\tau_{\rm M}+i\sqrt{2\alpha'}\sum_{n\neq0}\frac{\alpha_n^a}{n}e^{-in\tau_{\rm M}},\\
&X^I(z,\bar{z})|_{\partial\Sigma_1}=x^I,
\end{aligned}
\end{equation*}
and
\begin{equation*}
\begin{aligned}
&X^a(z,\bar{z})|_{\partial\Sigma_2} = x^a+(2\alpha')p^a\tau_{\rm M}+i\sqrt{2\alpha'}\sum_{n\neq0}(-1)^n\frac{\alpha_n^a}{n}e^{-in\tau_{\rm M}},\\
&X^I(z,\bar{z})|_{\partial\Sigma_2}=x^I-(2\alpha')w^I\pi.
\end{aligned}
\end{equation*}
With the definition $\dslash z=dz/(2\pi)$, the dimensionless mode expansion operators are as in the closed string \cite{Polchinski_v1},
\begin{equation*}
\begin{aligned}
&\alpha_n^{\mu}=\sqrt{\frac{2}{\alpha'}}\oint \dslash z\,\partial X^{\mu}\,z^n,\\
&\tilde{\alpha}_n^{\mu}=-\sqrt{\frac{2}{\alpha'}}\oint \dslash \bar{z}\,\bar{\partial} X^{\mu}\,\bar{z}^n,
\end{aligned}
\end{equation*}
with $(\alpha_n^{\mu})^{\dagger}=\alpha_{-n}^{\mu}$, and using the open string constraints (\ref{eq:wapa}) one may work with the holomorphic quantity, $\alpha_n^{\mu}$, only. The zero modes and angular momentum are given by \cite{GSW1},
\begin{equation*}
\begin{aligned}
&\hat{x}^{\mu}=\oint \Big(\frac{d z}{2\pi iz}-\frac{d \bar{z}}{2\pi i\bar{z}}\Big)X^{\mu}(z,\bar{z}),\\
&\hat{p}^{\mu}=\frac{1}{\alpha'}\oint \dslash z\,\partial X^{\mu},\\
&\hat{J}^{\mu\nu} =\frac{2}{\alpha'}\oint \dslash z X^{[\mu}\partial X^{\nu]},
\end{aligned}
\end{equation*}
and we have used the doubling trick \cite{Polchinski_v1} so that the integrals are along a spacelike curve, e.g.~$|z|^2=1$,  and $a^{[\mu\nu]}=\frac{1}{2}(a^{\mu\nu}-a^{\mu\nu})$. The physical worldsheet is in the upper half plane -- one identifies antiholomorphic quantities in the upper half plane with holomorphic quantities in the lower half plane and therefore one may just as well work with holomorphic quantities only in the full complex plane. For example, $\hat{p}^{\mu}=\frac{1}{2\alpha'}\int_{C_+}\big(\dslash z\,\partial X^{\mu}-\dslash \bar{z}\,\bar{\partial} X^{\mu}\big)=\frac{1}{2\alpha'}\big(\int_{C_+}+\int_{C_-}\big)\dslash z\,\partial X^{\mu}$ and $\int_{C_+}+\int_{C_-}=\oint$, so that $C_+$ represents an open spacelike contour in the upper half (stretching from $\sigma=0$ to $\pi$), $C_-$ represents the corresponding quantity in the lower half plane (stretching from $\sigma=\pi$ to $2\pi$), and $C$ represents a closed contour, $C=C_-\cup C_+$.

\subsection{Open String DDF Operators and Vertex Operators}

The relevant propagators on the upper half plane are,
\begin{equation}\label{eq:<XX>_op apx}
\begin{aligned}
&N:\quad\big\langle X^+(z,\bar{z})X^-(w,\bar{w})\big\rangle \\
&\hspace{1.5cm}= \frac{\alpha'}{2}\Big(\ln|z-w|^2+ \ln |z-\bar{w}|^2\Big),\\
&N:\quad\big\langle X^A(z,\bar{z})X^B(w,\bar{w})\big\rangle \\
&\hspace{1.5cm}= -\frac{\alpha'}{2}\delta^{AB}\Big(\ln|z-w|^2+ \ln |z-\bar{w}|^2\Big),\\
&D:\quad\big\langle X^I(z,\bar{z})X^J(w,\bar{w})\big\rangle\\
&\hspace{1.5cm} = -\frac{\alpha'}{2}\delta^{IJ}\Big(\ln|z-w|^2-\ln |z-\bar{w}|^2\Big),\\
\end{aligned}
\end{equation}
for the Neumann (N) or Dirichlet (D) directions respectively, with the normalization convention $\partial_z\partial_{\bar{z}}G(z,w) = -\pi\alpha^{\prime}\delta^2(z-w),$ and $G(z,w)=\langle X(z,\bar{z})X(w,\bar{w})\rangle$. 

To construct vertex operators we now distinguish between excitations tangent or transverse to the brane respectively,
\begin{equation}\label{eq: DDF A op apx}
\begin{aligned}
&A_n^A=\sqrt{\frac{2}{\alpha'}}\oint \dslash z\,\partial X^A(z)e^{inq\cdot X(z,\bar{z})},\\
&A_n^I=\sqrt{\frac{2}{\alpha'}}\oint \dslash z\,\partial X^I(z)e^{inq\cdot X(z,\bar{z})},
\end{aligned}
\end{equation}
and these act on the open string vacuum, $e^{ip\cdot X(z,\bar{z})}$, which is restricted to the real axis, $z=\bar{z}$. This procedure gives rise to vertex operators of the form,
\begin{equation}\label{eq:DDF state op apx}
V(z,\bar{z})=C\xi_{ij\dots}A^i_{-n_1} A^j_{-n_2}\dots\,e^{ip\cdot X(z,\bar{z})},
\end{equation}
as explained in the main text. Self-contractions are subtracted using the correlation functions (\ref{eq:<XX>_op apx}). The integrands of the DDF operators are to be restricted to the real axis, $z=\bar{z}$, and only after the normal ordering has been carried out are we to analytically continue the integrand in the complex plane so as to perform the contour integrations shown in (\ref{eq: DDF A op apx}). At this point the integrations should all be analytic in $z$.

Given that open string vertex operators live on the boundary of the worldsheet it is sometimes useful to represent them as holomorphic functions of a single variable, $z$. In the main text we concentrate on open string vertex operators with excitations in the directions tangent to the D$p$-brane, and so it is possible to construct vertex operators using instead of (\ref{eq: DDF A op apx}) the DDF operator,
\begin{equation}\label{eq: DDF A op apx2}
A_n^A=\sqrt{\frac{2}{\alpha'}}\oint \dslash z\,\partial X^A(z)e^{inq\cdot X(z)},
\end{equation}
with the corresponding vertex operators given by,
\begin{equation}\label{eq:DDF state op apx2}
V(z,\bar{z})=C\xi_{AB\dots}A^A_{-n_1} A^B_{-n_2}\dots\,e^{ip\cdot X(z)},
\end{equation}
in which case to obtain the normal ordered expression, the self-contractions are to be subtracted using the propagator,
\begin{equation}
N:\qquad\big\langle X^a(z)X^b(w)\big\rangle = -(2\alpha')\eta^{ab}\ln (z-w),
\end{equation}
which follows from (\ref{eq:<XX>_op apx}) by restricting the worldsheet arguments to the real axis. To carry out the contour integrations shown in (\ref{eq: DDF A op apx2}) we analytically continue in $z$ around the real axis and the contour is to contain the vacuum.

On a Minkowski signature worldsheet the DDF integrals are along the boundary of the worldsheet which is coincident with the D$p$-brane. 
The vacuum momenta $p^{\mu}$ and null vectors $q^{\mu}$ are restricted to lie within the D-brane worldvolume, see (\ref{eq:wapa}), and the $q^{\mu}$ are transverse to the DDF operators: $$q^A=q^I=p^I=0.$$ The onshell constraints for the open string are,
\begin{equation}\label{eq:pq cond_open apx}
p^2=\frac{1}{\alpha'},\qquad p\cdot q=\frac{1}{2\alpha'},\qquad {\rm and}\qquad q^2=0,
\end{equation}
so as to ensure that the vertex operators (\ref{eq:DDF state op apx}) are onshell with mass spectrum $m^2=-(p-Nq)^2=(N-1)/\alpha'$ as appropriate for open strings. The contractions appearing in (\ref{eq:pq cond_open apx}) are with respect to all spacetime indices $\mu$. 

\subsection{Open String Covariant Commutators}

In direct analogy to the closed string case above we learn that,
\begin{equation*}
\begin{aligned}
&\big[\alpha_m^{\mu},A^i_n\big] = m\delta^{\mu, i}B^n_m+n\sqrt{2\alpha'}q^{\mu}D_{m,n}^i,\\
&\big[\alpha_{\ell}^{\mu},B^n_m\big] = n\sqrt{2\alpha'}q^{\mu}B^n_{m+\ell},\\
&\big[\alpha_{\ell}^{\mu},D_{m,n}^i\big] = \ell\delta^{\mu, i} B^n_{m+\ell}+n\sqrt{2\alpha'}q^{\mu}D_{m+\ell,n}^i,\\
&\big[\alpha_m^{\mu},E^{n}_{\ell}\big] = m\sqrt{2\alpha'}q^{\mu}B^{n}_{m+\ell}-n\sqrt{2\alpha'}q^{\mu}E^{n}_{m+\ell},
\end{aligned}
\end{equation*}
where we have defined,
\begin{equation*}
\begin{aligned}
&B^{n}_{m} = \oint \frac{\dslash z}{iz}\,z^m\,e^{inq\cdot X(z)},\\
&D^i_{m,n}=\sqrt{\frac{2}{\alpha'}}\oint \dslash zz^m\partial X^ie^{inq\cdot X(z)},\\
&E^n_m=\oint \dslash z z^mq\cdot \partial Xe^{inq\cdot X(z)}.
\end{aligned}
\end{equation*}
From these commutators and $(\alpha_n^{\mu})^{\dagger}=\alpha_{-n}^{\mu}$, $(A_n^i)^{\dagger}=A_{-n}^i$, it follows that $(B^n_m)^{\dagger}=B^{-n}_{-m}$, $(D^i_{m,n})^{\dagger}=D^i_{-m,-n}$ and $(E^n_m)^{\dagger}=E^{-n}_{-m}$. In addition we learn that,
\begin{equation*}
\begin{aligned}
&\big[A^i_{\ell},D_{m,n}^j\big] = \ell\delta^{ij}E^{\ell+n}_m,\\
&\big[D^i_{-\ell,n},D^j_{\ell,-m}\big]=\delta^{ij}\big(nE^{n-m}_0-\ell B^{n-m}_0\big),\end{aligned}
\end{equation*}
and, $\big[B^r_{\ell},D_{m,n}^i\big] =\big[A_n^i,B^{\ell}_m\big]= \big[A_n^i,E^{\ell}_m\big]=0$ and $[B^n_m,B^{\ell}_r]=[E^n_m,E^{\ell}_r]=[B^n_m,E^{\ell}_r]=0$.

On the chiral half of a (tachyonic) vacuum state, $e^{ip\cdot X(z)}$, one can readily compute the operator products, 
\begin{subequations}\label{eq:opes BDE}
\begin{align}
&B^{-n}_{\phantom{a}m}\cdot e^{ip\cdot X(z)}\cong S_{n-m}(nq;z)\,e^{i(p-nq)\cdot X(z)},\phantom{\Big|}\\
&D_{m,-n}^i\cdot e^{ip\cdot X(z)}\cong H^i_{n-m}(nq;z)e^{i(p-nq)\cdot X(z)},\phantom{\Big|}\\
&E^{-n}_{m}\cdot e^{ip\cdot X(z)}\cong \sqrt{2\alpha'}q\cdot H_{n-m}(nq;z)e^{i(p-nq)\cdot X(z)},\phantom{\Big|}\label{eq:EeipX}
\end{align}
\end{subequations}
where the polynomials $S_{n-m}(nq;z)$ and  $H^{i}_{n-m}(nq;z)$ have been defined below and we have made use of the Taylor expansion, $e^{-inq\cdot X(w)}=\sum_{a=0}^{\infty}(w-z)^aS_a{(nq;z)}e^{-inq\cdot X(z)}.$ Note that in (\ref{eq:EeipX}) we have extended the definition of $H^i_{n-m}(nq;z)$, to include also longitudinal indices, $H^{\mu}_{n-m}(nq;z)$, without changing the form of the polynomial.

\subsection{Open String Polynomials}

In the open string sections of the main text we give explicit results for normal ordered vertex operators with excitations in the directions, $A=1,\dots,p-1$, tangent to the D$p$-brane. The various polynomials that appear in the open string analogous to (\ref{eq:PPbardfn apx}), (\ref{eq:SSbardfn apx}), (\ref{eq:H_n apx}) and (\ref{eq:HS apx}) of the closed string are in holomorphic language given respectively by,
\begin{equation*}
\begin{aligned}
&S_N(nq;z)=\oint_0 \frac{dw}{2\pi iw}\,w^{-N} \exp\Big( -inq\cdot \sum_{s=1}^{m}\frac{w^s}{s!}\partial_z^s X(z)\Big),\\
&H_N^A(z) \equiv \sqrt{2\alpha'}p^A S_{N}(Nq;z)+P_N^A(z),\\
&P_N^A(z) \equiv\sqrt{\frac{2}{\alpha'}}\sum_{m=1}^{N}\frac{i}{(m-1)!}\, \partial^mX^A(z)S_{N-m}(Nq;z),\\
&\mathbb{S}_{m,n}(z) \equiv \sum_{r=1}^nrS_{m+r}(mq;z)S_{n-r}(nq;z),\phantom{\bigg|}
\end{aligned}
\end{equation*}
and further properties and examples for $N=0,1$ and 2 of these are given in Appendix \ref{C}. The $\alpha'=2$ results there correspond to $\alpha'=1/2$ results here.


\bibliographystyle{apsrev}
\bibliography{SPI}

\end{document}